\documentclass[acmtog]{acmart}

\renewcommand\footnotetextcopyrightpermission[1]{}

\AtBeginDocument{%
  }

\usepackage{algorithm}
\usepackage{algpseudocode}
\usepackage{graphicx}
\usepackage{array}
\usepackage{mwe}
\usepackage{placeins}
\usepackage{subcaption}

\begin{document}

\title{Geometric Shape Optimization for Limbless Locomotion}

\author{Utpal Khanal}
\orcid{1234-5678-9012}
\affiliation{%
  \institution{Indian Institute of Technology Palakkad}
  \city{Palakkad}
  \state{Kerala}
  \country{India}
}
\email{utpalkhanal8@gmail.com}

\author{Avirup Mandal}
\affiliation{%
  \institution{Indian Institute of Technology Palakkad}
  \city{Palakkad}
  \state{Kerala}
  \country{India}
}
\email{mandal.avirup@gmail.com}

\renewcommand{\shortauthors}{Khanal and Mandal}

\begin{abstract}
    The simulation of locomotion in limbless, deformable organisms remains a challenging problem across computer graphics, soft robotics, and computational modeling. In this work, we present a novel differential-geometric framework for modeling the motion of slender soft bodies, such as snakes. The body is represented as a three-dimensional parametric curve using a Fourier–Chebyshev polynomial basis. Motion is computed by solving an optimization problem that determines the interaction between the curve and its environment by estimating polynomial coefficients. To ensure physically plausible and non-self-intersecting behavior, bending and torsional energy terms are incorporated into the formulation. The resulting curve is then used to drive a surface representation via interpolation, enabling realistic visualization analogous to skinning techniques. We evaluate the proposed approach across a range of complex scenarios and parameter settings to demonstrate its robustness and versatility. Comparative analysis with state-of-the-art methods indicates that our approach achieves improved simulation quality and generates more physically realistic motion. 
\end{abstract}


\begin{teaserfigure}
  \centering  
  \includegraphics[width = \textwidth]{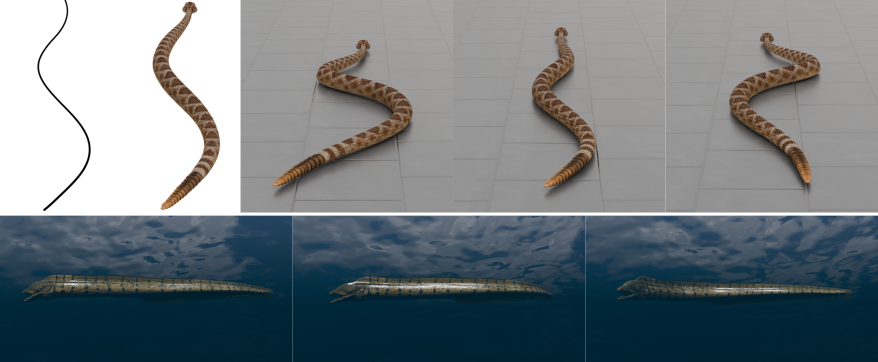}
  \caption{\emph{Top row}: The leftmost panel shows how the snake’s body is mapped onto a specified curve. The following panels illustrate its time-dependent evolution and movement pattern on solid terrain. \emph{Bottom row}: Temporal locomotion patterns of an eel moving through a fluid medium.}
  \Description{Teaserimage}
  \label{fig:teaser}
\end{teaserfigure}

\received{20 February 2007}
\received[revised]{12 March 2009}
\received[accepted]{5 June 2009}

\maketitle

\thispagestyle{empty}
\pagestyle{empty}

\section{Introduction}
Limbless soft-body locomotion has remained a prominent and challenging area of interest in graphics, soft-robotics, and computational modeling over the decades~\cite{kuznetsov1967fishsnake}~\cite{hirose1993biorobot}~\cite{transeth2009snakesurvey}. It is a field of study that explores the movement of flexible, deformable, limbless structures, often inspired by biological organisms such as snakes, worms, octopuses, and slugs. These organisms can move without dedicated limbs. Instead, they rely on their highly deformable bodies to interact dynamically with their surroundings. Limbless soft-body locomotion has garnered attention for its potential to enable soft robots that are resilient, adaptable, and capable of navigating complex and hazardous terrain. Unlike traditional rigid-body robots, soft robots can deform in response to environmental constraints. It enables complex maneuvers and applications in areas where traditional robots or humans fall short~\cite{laschi_soft_robot}. The mathematical models for such locomotion have a wide range of applications, including medical robotics, advanced robotic grippers, the visual effects and design industries, educational tools, and custom robots for hazardous environments, such as nuclear power plants.

A soft body achieves locomotion through continuous shape deformation, including bending, twisting, and lifting motions, which collectively provide a highly adaptable and versatile mechanism of movement as shown in Figure~\ref{fig:teaser}. Soft robots and natural organisms that move this way can handle tight spaces, rough surfaces, and uneven terrain better than rigid machines. But this flexibility also makes soft-body movement difficult to model and control. Many earlier models ~\cite{Gray1946}~~\cite{kuznetsov1967fishsnake} simplify the body to a flat curve or assume a fixed motion pattern, like a simple wave. These simplified models support basic analysis, but they cannot capture the full three-dimensional motion that real soft bodies often require. They also limit the ability to explore new movement strategies, because the motion pattern is partly fixed by the model's assumptions. Linked snake motion~\cite{jing2013linksnake} or curve-based approximations~\cite{Alben2013}~\cite{Wang2014} are also studied for soft body motion. However, the soft-body motions generated by these works are jittery, excessively damped, and physically unrealistic. Moreover, they often undergo self-intersection or twisting. 

To overcome these restrictions, we model the soft body as a smooth curve in three-dimensional space whose shape is controlled directly by its curvature and torsion~\cite{pinkallgross2024}. These two quantities completely determine the local geometry of the body, and by integrating the Frenet-Serret equations~\cite{guggenheimer1963differential}, we can rebuild the entire 3D shape at each moment. This approach to modeling gives us full control over shape deformation without relying on predefined templates or motions, as in the works by ~\citet{gross2023shapechange}and ~\citet{soliman2024goflow}. It also separates the body's internal deformation from its global movement. Instead of imposing a forced trajectory on the body, we compute its translation and rotation from the balance of physical forces and torques at each time step. This ensures that the motion is physically meaningful and results directly from the body's interaction with the environment.

Our simulation model includes physical components that are important for realistic soft-body movement. The mass along the body need not be uniform; it can vary with arc length, allowing the model to represent soft bodies with heavier sections or tapered shapes as shown in Figure~\ref{fig:teaser}. Ground contact is handled with smooth, regularized equations that avoid abrupt jumps when the body touches or leaves the surface. This makes the simulation more stable and prevents unrealistic sticking or bouncing. The friction model is anisotropic, meaning the friction strength varies with the direction of motion. Forward, backward, and sideways motions may face different resistance, which is common in friction-based locomotion. The model also allows the body to move on an inclined plane, where gravity affects both the normal and sliding forces.

To measure how effective a given motion pattern is, we use a cost-of-locomotion metric based on dissipative power. The total power includes frictional power, strain power, and bending and torsional energy, all of which are calculated at each time step. This provides a clear measure of the power and energy required to move the body forward. As the curve varies smoothly over both space and time, we represent it using Chebyshev polynomials for the spatial dimension and a Fourier series for time. This gives the model sufficient flexibility to represent a wide range of smooth, complex 3D deformation patterns while keeping the number of parameters manageable.

Finding the most efficient motion pattern is challenging because the cost must be computed through the full state space. We employ two stochastic optimization methods: Finite-Difference Stochastic Approximation (FDSA) and Simultaneous Perturbation Stochastic Approximation (SPSA). In both methods, we incorporate momentum and adaptive learning rates to ensure stable, consistent parameter improvement.

All of these components together form a complete and general framework for studying soft-body locomotion in three dimensions. The differential-geometric formulation gives a compact and expressive way to describe body shape. The physical model includes realistic elements such as variable mass, inclined contact, and anisotropic friction. The cost metric gives a fair way to compare different movement patterns. And the methods like FDSA and SPSA provide a practical way to search for efficient movement strategies directly from simulation. This allows us to discover motion patterns that arise naturally from the physics of deformation and friction, rather than from predefined templates~\cite{soliman2024goflow}. The proposed model is capable of simulating various types of motion across different media.

Our contributions are summarized as follows:
\begin{itemize}
    \item We propose approximating a soft body as a continuous, differentiable curve embedded in 3D space using the Frenet–Serret formulation.
    \item We employ optimization techniques such as FDSA and SPSA to achieve fast, physically realistic simulations.
    \item We incorporate appropriate bending and torsional energy terms to prevent unrealistic configurations.
    \item We incorporate a novel penalty term for the net change in body orientation to ensure directional stability and energy constraints.
    \item We develop a novel strategy for handling non-uniform mass distribution.
\end{itemize}

The rest of the paper is organized as follows. After presenting the related works in Section~\ref{sec:relwork}, we delve into the technical details of the soft-body locomotion in Section~\ref{sec:technique}. Section~\ref{sec:opt} delves into the optimization framework, followed by the implementation details in Section~\ref{sec:discrete}. Next, we present our simulation results and compare them with the state of the art in Section~\ref{sec:result}, and conclude our paper in Section~\ref{sec:conclusion} by outlining future research directions.

\section{Related Work}\label{sec:relwork}
Research on limbless and soft-bodied locomotion has expanded steadily over the past two decades, generally following two major directions. The first focuses on physics-based models that analyze how the body interacts with the ground, often emphasizing frictional forces and contact mechanics~\cite{kuznetsov1967fishsnake}. The second uses geometry-based approaches that describe locomotion as motion through a space of shapes, highlighting how coordinated body deformations produce net movement~\cite{jing2013linksnake}~\cite{Alben2013}. Our work connects these two perspectives by directly controlling a full 3D curve through curvature and torsion, modeling anisotropic friction on sloped surfaces, and optimizing motion using stochastic methods.

\begin{algorithm}[h!tb]
\caption{Optimization of Curvature--Torsion Parameters for 3D Snake Locomotion}
\label{alg:snake_optimization}
\begin{algorithmic}[1]

\State Initialize parameter vector $\boldsymbol{\theta}_0$ \hfill [Eq.~\eqref{eq:theta}]

\While{ItrNo $<$ MaxItr}

    \For{each incremental time step $t_n+\Delta t$}
        \State Compute $\kappa$ and $\tau$ \hfill [Eq.~\eqref{eq:KappaAndTau}]
        \State Reconstruct local body shape, $\int_{0}^{1}\boldsymbol{T}(s,t)ds$\hfill [Sec.~\ref{subsec:fs}]
        \State Initialize $\boldsymbol{\psi} = \left[ x_0,\, y_0,\, z_0,\, \alpha,\, \beta,\, \sigma \right]^\top$ \hfill [Eq.~\eqref{eq:RigidBodyVar}]

        \While{$\mathcal{R}(\boldsymbol{\psi})  > \epsilon$ }
            \State Construct $R(\alpha,\beta,\gamma)$ \hfill [Sec.~\ref{subsec:fs}]
            \State Estimate global position $\boldsymbol{\gamma}(s,t)$ \hfill [Eq.~\eqref{eq:GlobalPosition}]
            \State Calculate $\boldsymbol{f}^{\textrm{ext}}$ \hfill [Sec.~\ref{sec:AllForces}]
            \State Compute $\mathcal{R}(\boldsymbol{\psi})$ residual \hfill [Sec.~\ref{subsec:force_balance}]
            \State Update $\boldsymbol{\psi}$ using a Newton step \hfill [Sec.~\ref{subsec: Calculate_RBVar}]
        \EndWhile
        \State Update global position $\boldsymbol{\gamma}(s,t)$ \hfill [Eq.~\eqref{eq:GlobalPosition}]
    \EndFor

    \State Compute $\langle P\rangle$, $\;\mathcal{E}{\tau}$, \& $\;\mathcal{E}{\kappa}$ \hfill [Sec.~\ref{sec:power} \& ~\ref{sec:intpower}]  
    \State Compute cost, $\mathcal{L}_T\left(\boldsymbol{\theta}_{t_n}\right)$ \hfill [Eq.~\eqref{eq:Cost}]
    \State Calculate $\nabla \mathcal{L}_T\left(\boldsymbol{\theta}_{t_n}\right)$ using FDSA/SPSA \hfill [Sec.~\ref{subsec:FDSA} \& ~\ref{subsec:SPSA}]
    \State $\hat{\boldsymbol{\theta}}_{t_n} \gets $ Update $\boldsymbol{\theta}_{t_n}$ using loss gradient \hfill [Sec.~\ref{subsec:Gradient Update}]
    \State $\hat{\hat{\boldsymbol{\theta}}}_{t_n} \gets$ Project $\hat{\boldsymbol{\theta}}_{t_n}$ onto admissible bounds \hfill 
    \State $\boldsymbol{\theta}_{t_n+\Delta t} \gets \hat{\hat{\boldsymbol{\theta}}}_{t_n}$
\EndWhile

\State \Return $\boldsymbol{\theta}^\star \gets \boldsymbol{\theta}_{t_{\textrm{MaxItr}}}$
\end{algorithmic}
\end{algorithm}

\subsection{Soft Body Locomotion}

A large amount of previous work models snakes and snake robots as inextensible curves that move across the ground under anisotropic Coulomb friction. An important early study by ~\citet{Hu2009} measured how real snakes interact with the ground and showed that their well-known lateral undulation can be accurately predicted using force balance. Their findings confirmed a basic idea, i.e., snakes benefit from having lower friction in the forward direction and higher friction sideways. This foundational approach continues to evolve~\cite{hu2009slithloc}, providing insights into the mechanics of locomotion across terrestrial, aquatic~\cite{shine2003seasnake}, and even aerial environments~\cite{socha2002glide}. Snakes, in particular, inspire scientists to develop bioinspired soft robotics~\cite{hirose1993biorobot}~\cite{transeth2009snakesurvey}~\cite{astley2015sidewind}. Their limbless design offers control, adaptability, and navigation advantages in complex and cluttered environments~\cite{fu2020robotmodel}~\cite{fu2022vertlatmotion}. The organic movements of deformable, limbless bodies have influenced the design of robotic systems~\cite{liljeback2012revrobo}. Later, \citet{Alben2013} developed an optimization framework for purely planar snake motion. In his approach, curvature waves are prescribed, the resulting forces and torques are computed, and the method searches for the most energy-efficient gait. This work clearly explained when snakes prefer retrograde waves (effective when sideways friction is high) versus direct waves (better when sideways friction is low), providing a useful reference for studying planar locomotion under different friction conditions. Moving on sloped surfaces introduces gravitational effects, making efficient locomotion more difficult. ~\citet{Wang2014} extended planar snake models to inclined surfaces by incorporating gravity and different friction values for forward, backward, and transverse motion. Beyond purely planar models, researchers have investigated how snakes lift parts of their bodies to reduce friction. ~\citet{Alben2022} developed a computational framework that combined 3D bending with lifting and optimized how snakes should time these motions. The work showed that lifting can significantly reduce energy loss, especially when the snake has good transverse friction but suffers high tangential drag. These results suggest that 3D motion is not just a geometric flourish; it directly helps manage frictional losses. However, the author uses a genetic algorithm for optimization, which results in unrealistic motion and physically implausible simulations, such as self-intersection.  

\subsection{Motion in Various Media}

Sidewinding is a unique and well-known way that many biological animals, like snakes, use to move in multiple planes. \citet{Astley2015} showed that this motion can be described simply --- it results from combining two waves, one in the horizontal plane and one in the vertical plane, that are offset by a quarter cycle. By changing the wave amplitudes or their phase difference, snakes can turn smoothly and move effectively on loose or unstable ground. On granular slopes, ~\citet{Marvi2014} demonstrated that both snakes and snake robots must carefully control how much of their body stays in contact with the ground. If too little of the body touches the surface, the animal or robot slips; if too much is in contact, the movement becomes inefficient. Their robotic experiments showed that maintaining a sufficiently long contact region is crucial when the substrate behaves like a fluid.

For loose materials such as sand, standard Coulomb friction models are often inaccurate. ~\citet{Li2013} developed terradynamics, a reduced-order method that sums resistive forces on small body segments. These forces depend on depth and orientation and give reasonably accurate predictions without needing expensive granular simulations (e.g., discrete element methods). These models have since been widely used to design gaits for robots on flowable substrates.

\subsection{Geometric Mechanics and Shape-Space Approaches}

A separate line of research developed geometric methods for understanding locomotion. ~\citet{Hatton2011} formulated locomotion as a mapping from a closed loop in a \textit{shape space} which represents the set of possible body deformations, to the resulting translation and rotation of the body. Using a mathematical construct known as the local connection, they showed that gaits can be optimized by evaluating area integrals in this shape space rather than performing full simulations. This geometric perspective has significantly influenced how robotics researchers design and analyze gaits, particularly in friction-dominated environments.

In computer graphics, mass-spring-based systems ~\citep{miller1988snakedynamics} and point-based physics simulations ~\citep{waszak2018limblessmov} have been developed to generate the motion of the bodies using applied forces. Recently, methods have been introduced to optimize the body's shape between transitions. It minimizes the internal energy of body configurations and produces smooth, optimized intermediate shapes between user-defined poses ~\citep{gross2023shapechange} and ~\citep{soliman2024goflow}. However, these methods require previously defined poses to estimate the motion.

\subsection{Optimization of Simulation-Based Locomotion}

Locomotion cost functions are usually obtained through simulation and may be noisy or non-differentiable. This makes it difficult to compute analytical gradients used for Newton's method. As a result, optimization methods that rely on gradient estimates rather than true derivatives are commonly used. Two influential approaches are finite-difference stochastic approximation (FDSA) and simultaneous perturbation stochastic approximation (SPSA). ~\citet{Spall1992} presented a comprehensive analysis of SPSA, and ~\citet{Bhatnagar2013} provided additional theoretical foundations and practical guidance. These methods are well-suited for simulation-based studies in which evaluating a single candidate gait requires full time-stepping and force-balance computations.
\section{Technique}\label{sec:technique}
We begin with a brief overview of a three-dimensional space curve and its key properties. We then introduce the optimization framework used to generate limbless motion.
\subsection{Space Curve}
The spinal trajectory of a limbless body can be mathematically approximated as a smooth vector-valued curve~\cite{lee2012diffgeo}~\cite{pinkallgross2024}
\begin{equation}\label{eq:sp_curve_gen}
    \boldsymbol{\gamma}:s\times t \longrightarrow \mathbb{R}^n, \;\; s \in [s_0, s_1] \subset \mathbb{R}, \; t \in [t_0, t_1] \subset \mathbb{R}
\end{equation}
such that the velocity vector $\displaystyle \frac{\partial \boldsymbol{\gamma}}{\partial s} \neq 0, \frac{\partial \boldsymbol{\gamma}}{\partial t} \neq 0, \; \forall s, t$. Here $s$ and $t$ denote arc length and time, respectively. 

The unit tangent field, $\displaystyle \boldsymbol{T}$, normal field, $\displaystyle \boldsymbol{N}$, curvature, $\kappa$ and torsion, $\tau$, of the curve $\gamma$ at a fixed time $t$, are given by~\cite{pinkallgross2024}~\cite{jost2017riemannian}
\begin{equation}\label{eq:sp_curve_quantities}
\begin{aligned}
    &\boldsymbol{T}:s \times t \longrightarrow \mathbb{S}^{n-1}, \; \boldsymbol{T} = \frac{\partial \boldsymbol{\gamma}}{\partial s} = \frac{\dot{\boldsymbol{\gamma}}}{|\dot{\boldsymbol{\gamma}}|}; \;\;
    \boldsymbol{N}:s \times t \longrightarrow \mathbb{S}^{n-1}, \; \langle\boldsymbol{N}, \boldsymbol{T} \rangle = 0\\
    &\kappa:s \times t \longrightarrow \mathbb{R}, \; \kappa = \frac{\partial \boldsymbol{T}}{\partial s}; \;\;
    \tau: s \times t \longrightarrow \mathbb{R}, \; \tau = \left\langle \frac{\partial\boldsymbol{N}}{\partial s}, \boldsymbol{T} \times \boldsymbol{N} \right\rangle
\end{aligned}
\end{equation}

Let us assume that $\boldsymbol{\gamma}(s,t)|_{s=0}$ corresponds to the tail and $\boldsymbol{\gamma}(s,t)|_{s=1}$ corresponds to the head of the body, approximated by a 3D curve. The shape of the curve is described in terms of two intrinsic quantities, the curvature $\kappa$, which specifies how sharply the body bends at each point, and the torsion $\tau$, which specifies how much the body twists out of the bending plane. A nonzero torsion introduces three-dimensional motion. Curvature and torsion control entirely the shape of the soft body, while the global position and orientation of the body must be specified separately.

\subsubsection{Frenet--Serret Equations}\label{subsec:fs}
\begin{figure}
    \centering
    \includegraphics[width=\columnwidth]{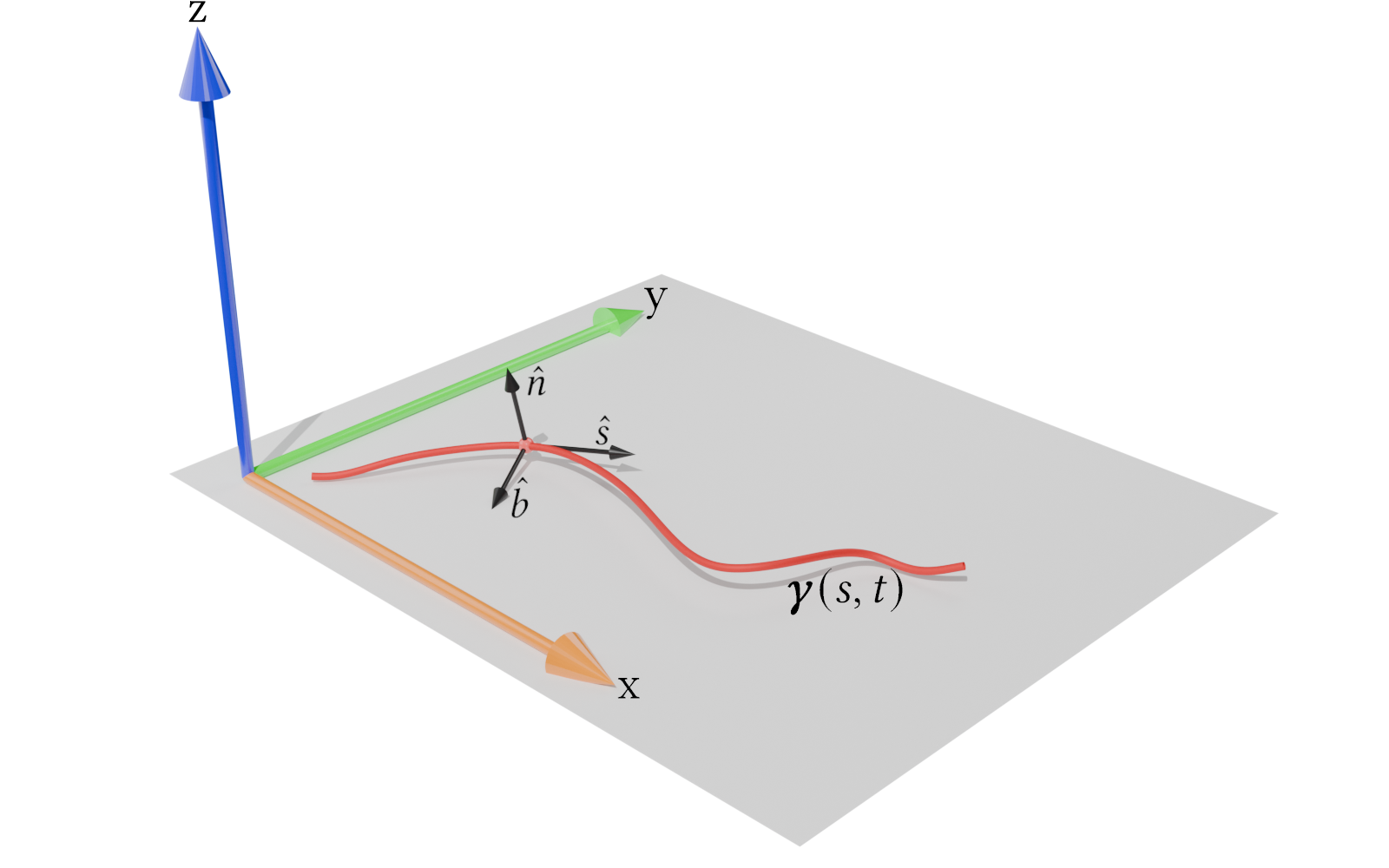}
    \caption{Representative configuration of the limbless body modeled as a three-dimensional space curve $\gamma(s,t)$ at a fixed time instant. The red curve denotes the body centerline, parameterized by arc length $s$, with $s=0$ corresponding to the tail and $s=1$ to the head. The global coordinate axes $(x,y,z)$ are shown, where the $x$--$y$ plane represents the ground surface and the $z$-axis denotes the vertical direction. At each point along the curve, the local Frenet--Serret frame $(\hat{s}, \hat{n}, \hat{b})$ defines the tangent, normal, and binormal directions, respectively. The shape of the curve is governed by the curvature $\kappa$, which controls bending, and the torsion $\tau$, which induces out-of-plane twisting, enabling three-dimensional locomotion.}
    \label{fig:frnet-serret}
\end{figure}

To describe the local geometry of the arc, we employ the Frenet-Serret frame~\cite{guggenheimer1963differential}, which provides a moving orthonormal basis attached to each point along the arc. As shown in Figure~\ref{fig:frnet-serret}, this frame consists of the unit tangent vector $\boldsymbol{T}$, the unit normal vector $\boldsymbol{N}$, and the unit binormal vector $\boldsymbol{B} = \boldsymbol{T}\times\boldsymbol{N}$, which together form a right-handed coordinate system that evolves smoothly along the arc length $s$. These vectors are governed by the Frenet-Serret equations~\cite{guggenheimer1963differential} as follows,

\begin{equation}\label{eq:fs}
\begin{bmatrix} 
    \frac{\partial\boldsymbol{T}}{\partial s} \\ \frac{\partial\boldsymbol{N}}{\partial s} \\ \frac{\partial\boldsymbol{B}}{\partial s} 
\end{bmatrix} = 
\begin{bmatrix} 
    0 & \kappa & 0 \\
    -\kappa & 0 & \tau \\
    0 & -\tau & 0    
\end{bmatrix}
\begin{bmatrix} 
    \boldsymbol{T} \\ \boldsymbol{N} \\ \boldsymbol{B}
\end{bmatrix}
\end{equation}
Therefore, curvature $\kappa$ and torsion $\tau$ are obtained by integrating the Frenet-Serret Equations~\eqref{eq:fs}. This provides a representation of the arc in a body-fixed frame, with
\begin{equation}\label{eq:init-pose}
    \begin{aligned}
        \textrm{Tail Position}&: \quad \boldsymbol{\gamma}(0,t) = \left[x(0,t), y(0,t), z(0,t)\right]\\
        \textrm{Orientation}&: \quad \left[\boldsymbol{T}(0,t), \boldsymbol{N}(0,t), \boldsymbol{B}(0,t)\right]\\
        &\quad \quad=\boldsymbol{R}_z\left(\alpha(t)\right)\boldsymbol{R}_y\left(\beta(t)\right)\boldsymbol{R}_x\left(\sigma(t)\right)
    \end{aligned}
\end{equation}
where $\displaystyle \left[\boldsymbol{R}_z\left(\alpha(t)\right), \boldsymbol{R}_y\left(\beta(t)\right), \boldsymbol{R}_x\left(\sigma(t)\right)\right] \text{ and } \left[\alpha(t),\beta(t),\sigma(t)\right]$ are the rotation matrices and Euler angles with the $z$-, $y$- and $x$- axes respectively.
Next, the global pose of the limbless body is obtained by applying the rigid-body transformation defined by the initial position and orientation in Equation~\ref{eq:init-pose}. Following the Equations~\eqref{eq:sp_curve_quantities}-~\eqref{eq:init-pose}, the final pose of the body is therefore expressed as
\begin{equation}\label{eq:GlobalPosition}
\begin{aligned}
    \boldsymbol{\gamma}(s,t) &= \boldsymbol{\gamma}(0,t) + \boldsymbol{R}_z\left(\alpha(t)\right)\boldsymbol{R}_y\left(\beta(t)\right)\boldsymbol{R}_x\left(\sigma(t)\right)\int_{0}^{1} \boldsymbol{T}(s,t)ds \\
    &= \boldsymbol{\gamma}(0,t) + \boldsymbol{R}(\alpha,\beta,\sigma) \int_{0}^{1} \boldsymbol{T}(s,t)ds \\
    &= \boldsymbol{\gamma}(0,t) + \boldsymbol{R}(\alpha,\beta,\sigma) \int_{0}^{1} d\boldsymbol{\gamma}
\end{aligned} 
\end{equation}
This formulation separates the description of the intrinsic shape of the soft body, governed by curvature and torsion, from its extrinsic motion, governed by the tail’s translation and orientation. Together, these elements provide a complete mathematical model for the limbless body's configuration at any given time. The parameter vector, $\displaystyle \boldsymbol{\psi}$, defined by the initial position and orientation of the tail, can be expressed as
\begin{equation}
\begin{aligned}
    \boldsymbol{\psi} = \left[ x(0,t),\, y(0,t),\, z(0,t),\, \alpha,\, \beta,\, \sigma \right]^\top \equiv \left[ x_0,\, y_0,\, z_0,\, \alpha,\, \beta,\, \sigma \right]^\top
    \label{eq:RigidBodyVar}
\end{aligned}
\end{equation}
which define the global position and orientation of the limbless body at each time step. These parameters are chosen so that the total residual force and torque acting on the body are zero, thereby ensuring mechanical equilibrium throughout the motion. Details of the residual calculation are presented in Section~\ref{subsec:force_balance}.



\begin{figure*}[tbp]
\centering

\vspace{5pt}

\noindent
\begin{minipage}[t]{0.14\textwidth}
\vspace{20pt}
\raggedright
\textbf{Case 1:}\\
$\mu_n/\mu_f = 1.2$ $\mu_b/\mu_f = 1.2$
\end{minipage}%
\hfill
\begin{minipage}[t]{0.85\textwidth}
\vspace{2pt}
\centering
\includegraphics[width=0.16\linewidth]{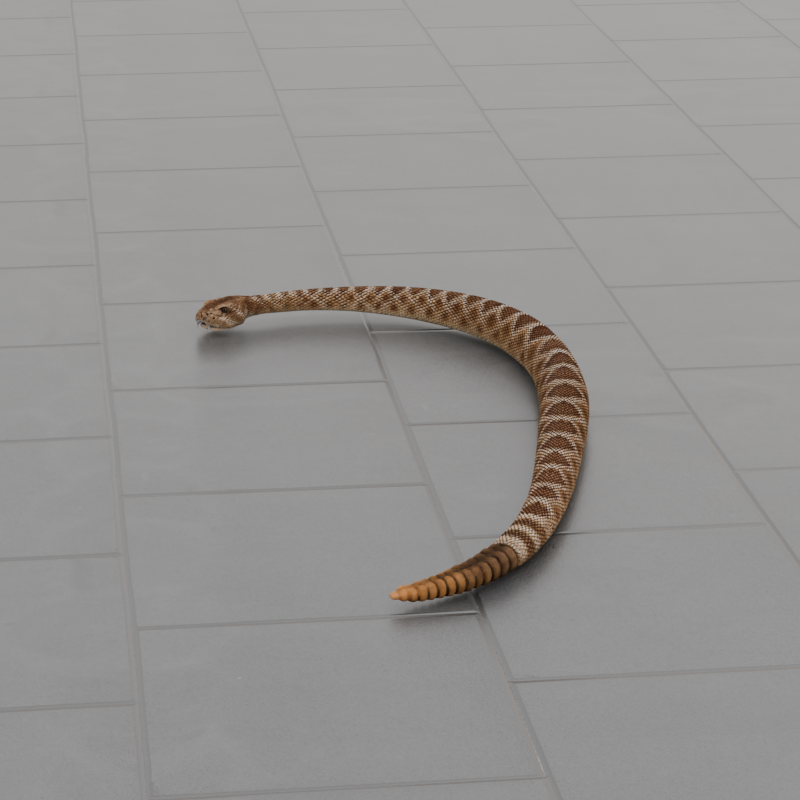}\hspace{2pt}%
\includegraphics[width=0.16\linewidth]{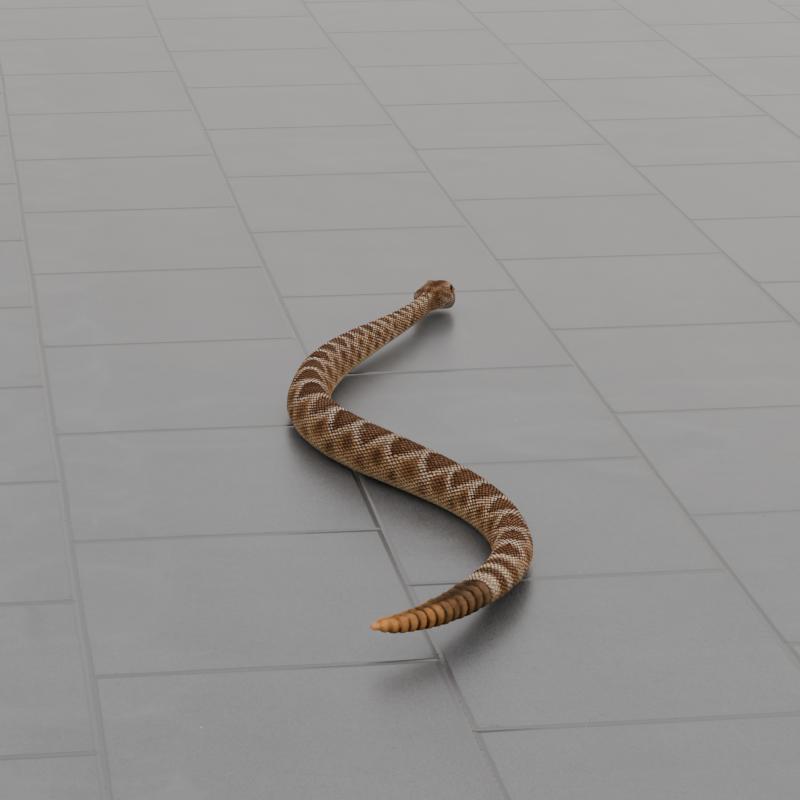}\hspace{2pt}%
\includegraphics[width=0.16\linewidth]{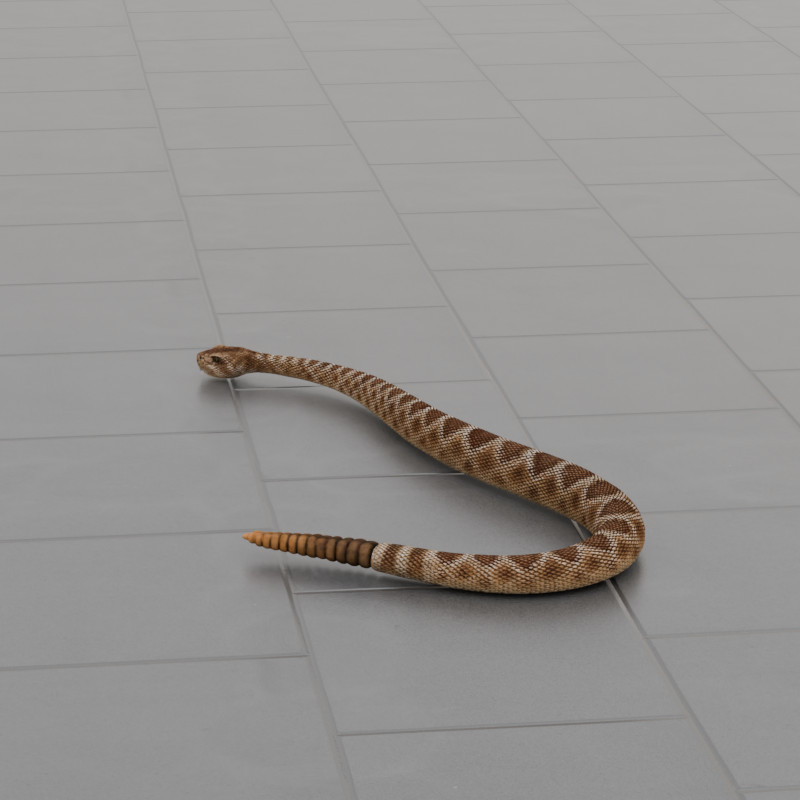}\hspace{2pt}%
\includegraphics[width=0.16\linewidth]{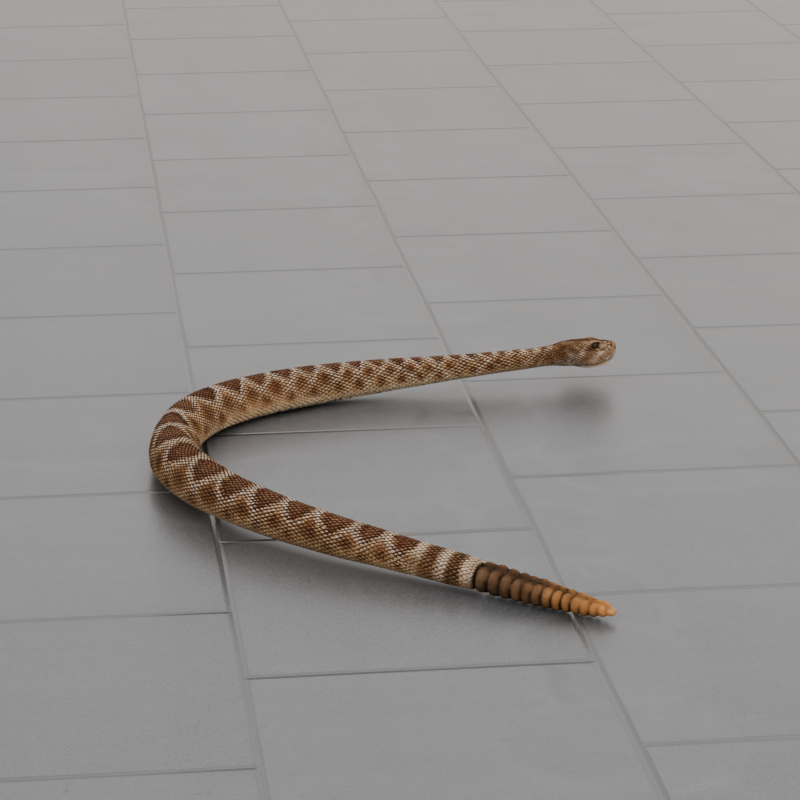}\hspace{2pt}%
\includegraphics[width=0.16\linewidth]{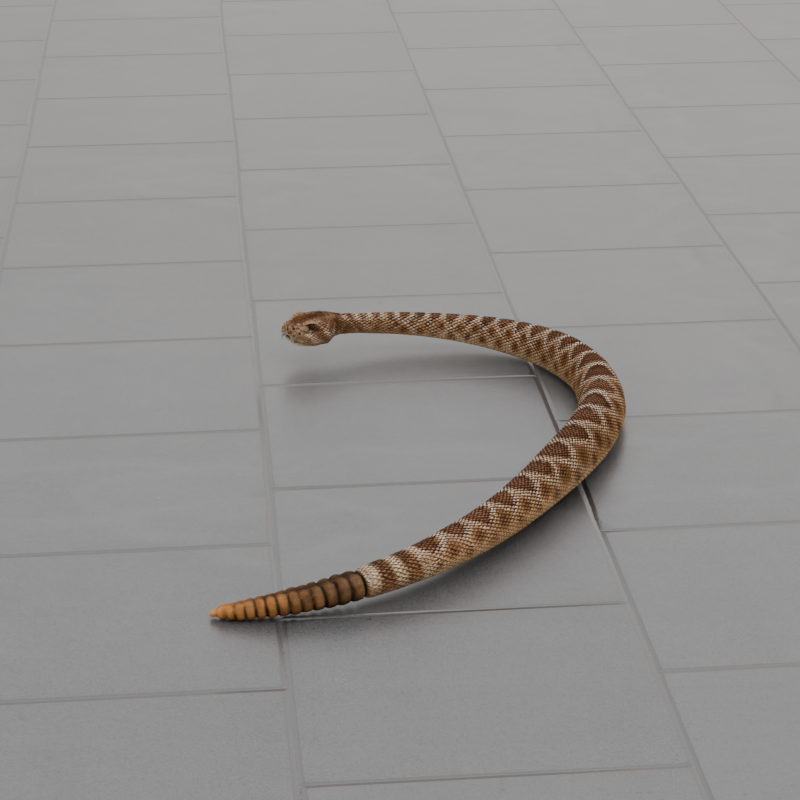}\hspace{2pt}%
\includegraphics[width=0.16\linewidth]{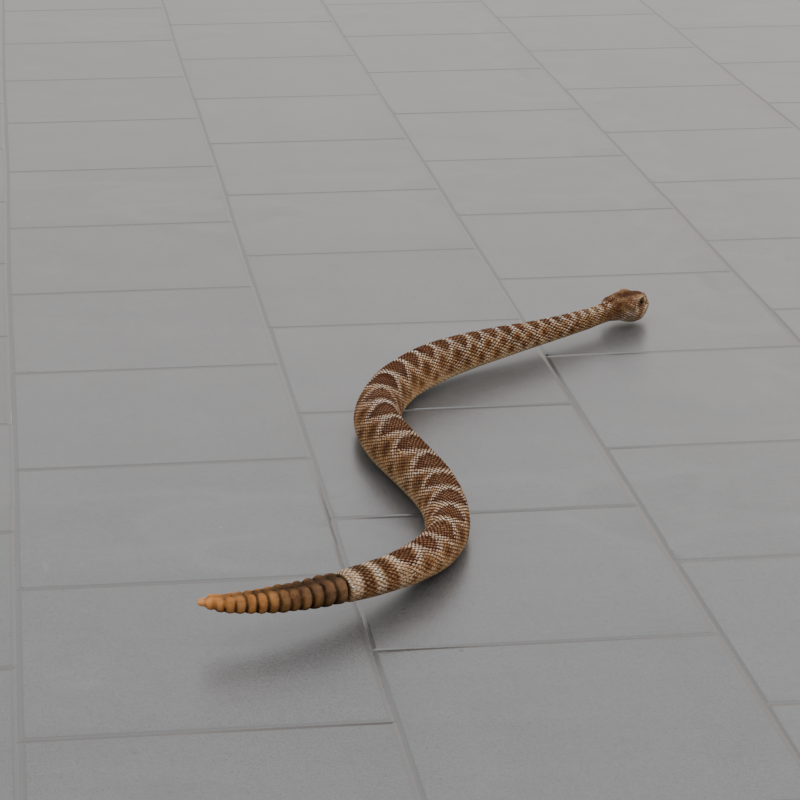}
\end{minipage}

\noindent
\begin{minipage}[t]{0.14\textwidth}
\vspace{20pt}
\raggedright
\textbf{Case 2:}\\
$\mu_n/\mu_f = 1$ $\mu_b/\mu_f = 5$
\end{minipage}%
\hfill
\begin{minipage}[t]{0.85\textwidth}
\vspace{2pt}
\centering
\includegraphics[width=0.16\linewidth]{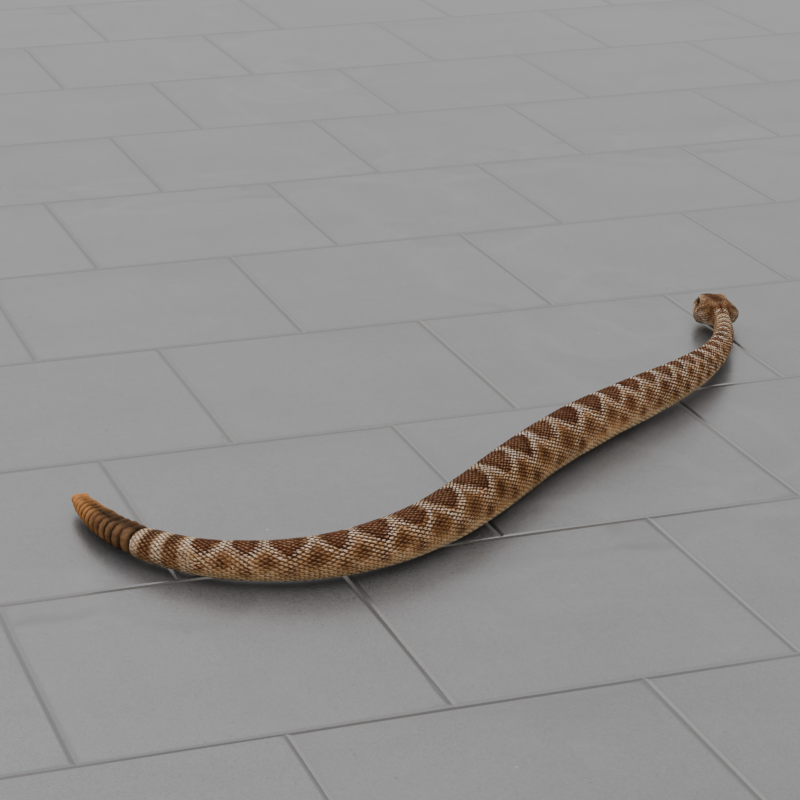}\hspace{2pt}%
\includegraphics[width=0.16\linewidth]{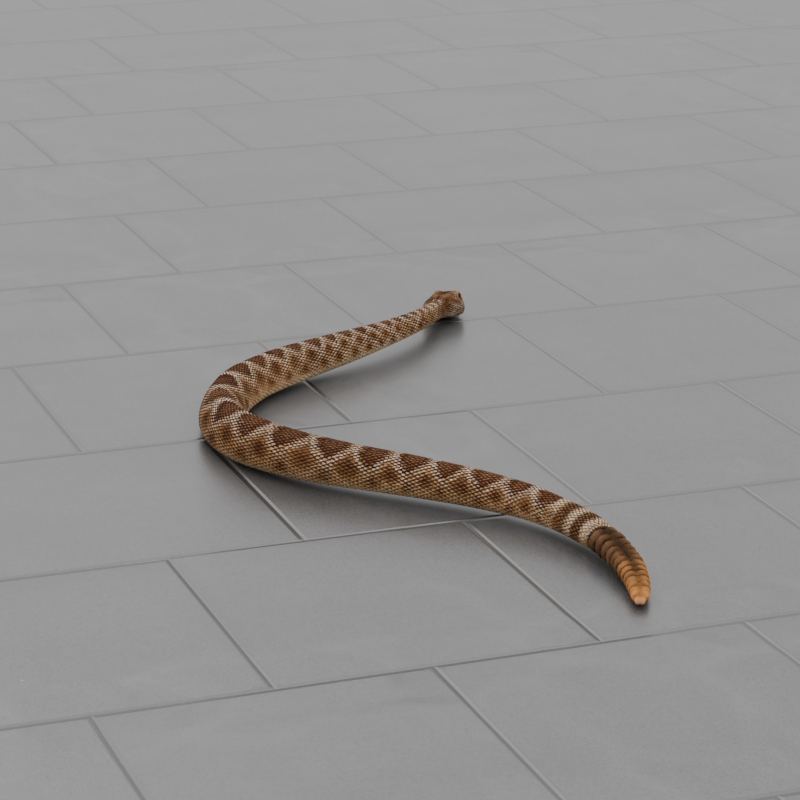}\hspace{2pt}%
\includegraphics[width=0.16\linewidth]{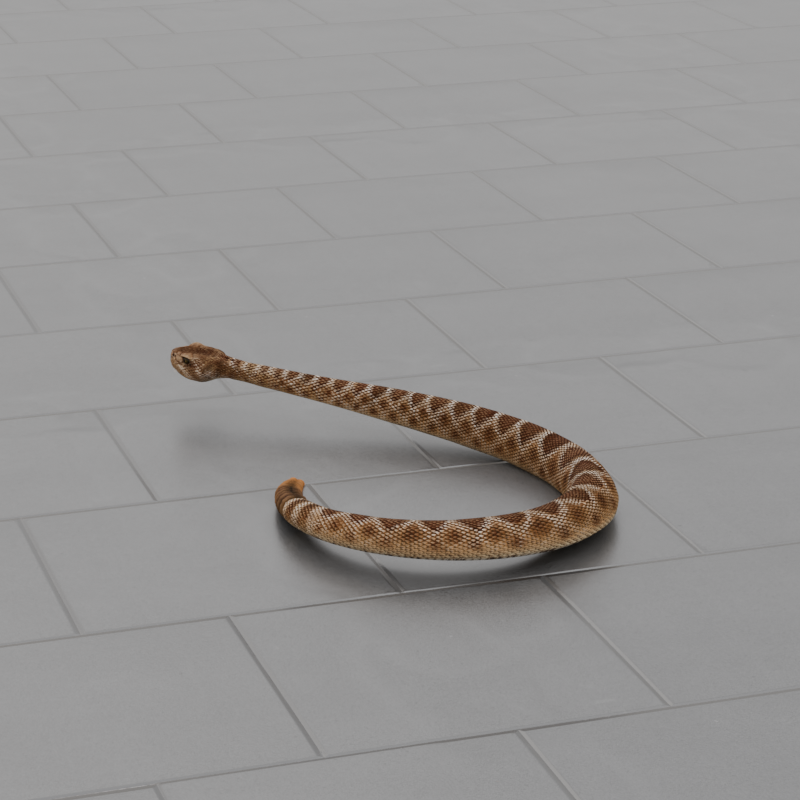}\hspace{2pt}%
\includegraphics[width=0.16\linewidth]{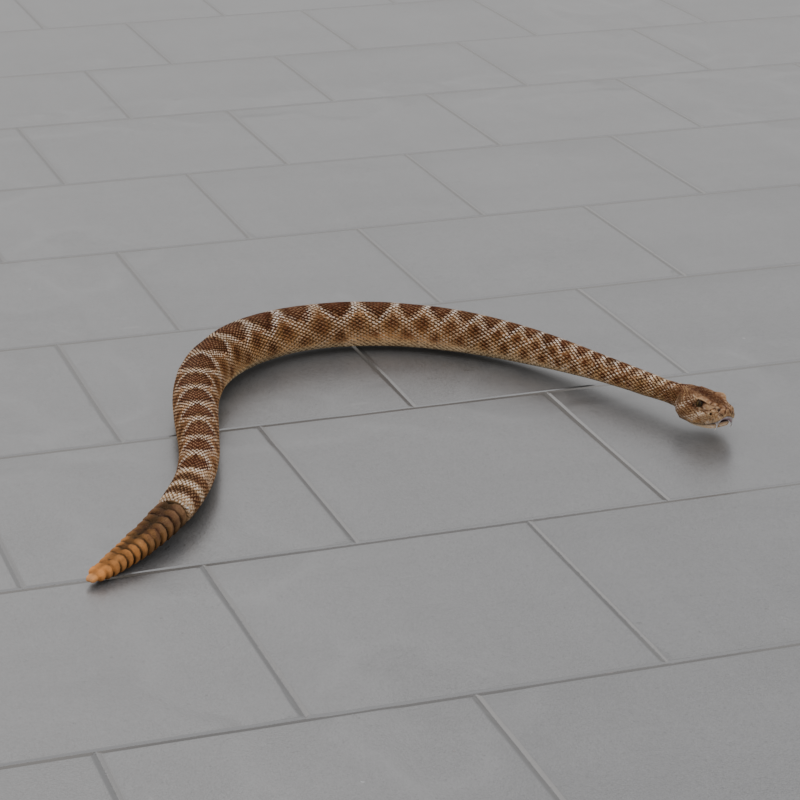}\hspace{2pt}%
\includegraphics[width=0.16\linewidth]{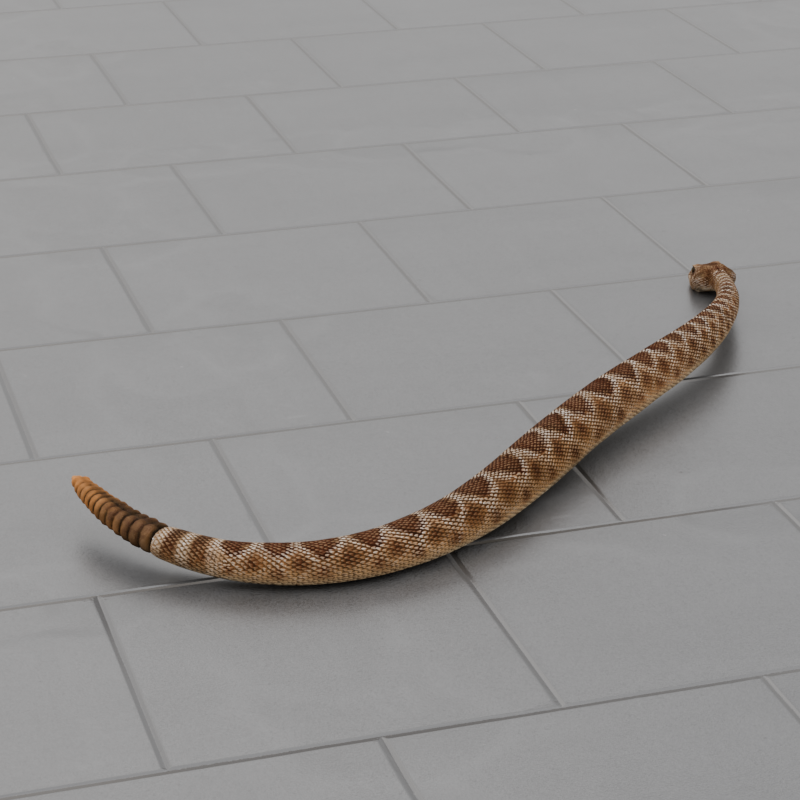}\hspace{2pt}%
\includegraphics[width=0.16\linewidth]{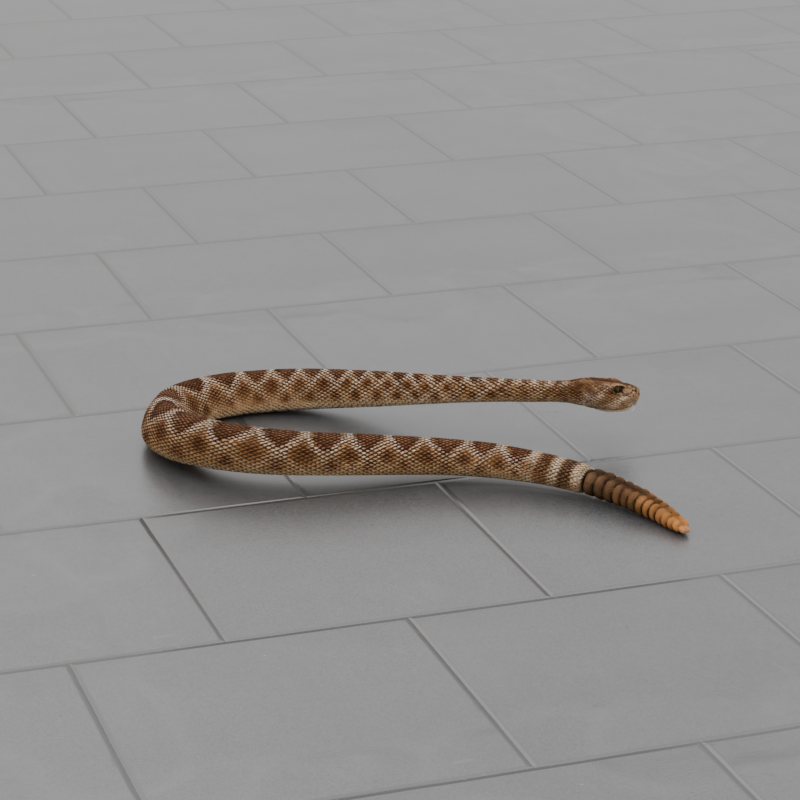}
\end{minipage}

\noindent
\begin{minipage}[t]{0.14\textwidth}
\vspace{20pt}
\raggedright
\textbf{Case 3:}\\
$\mu_n/\mu_f = 10$ $\mu_b/\mu_f = 1.2$ 
\end{minipage}%
\hfill
\begin{minipage}[t]{0.85\textwidth}
\vspace{2pt}
\centering
\includegraphics[width=0.16\linewidth]{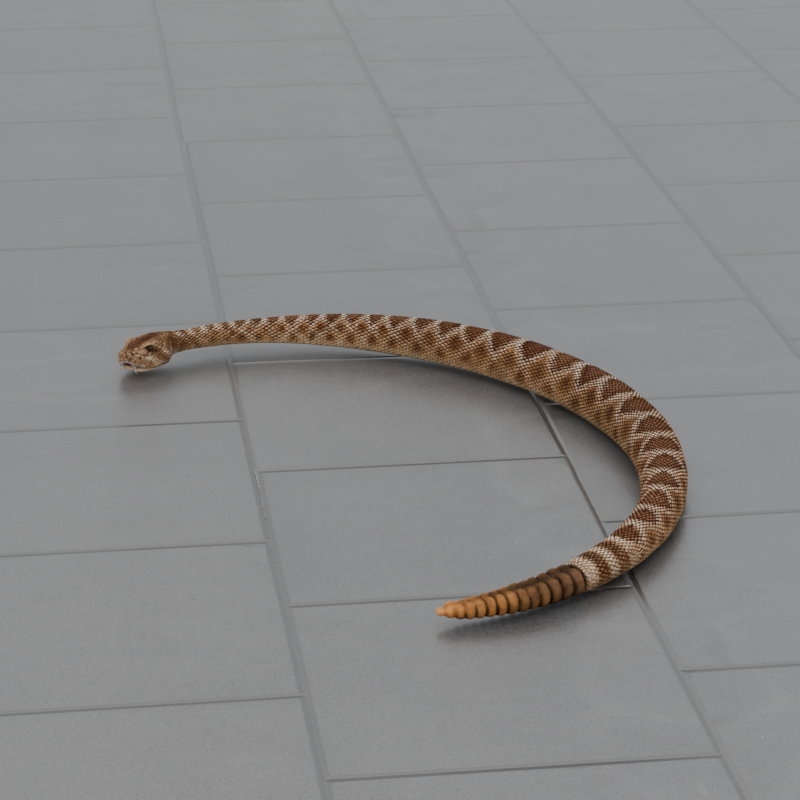}\hspace{2pt}%
\includegraphics[width=0.16\linewidth]{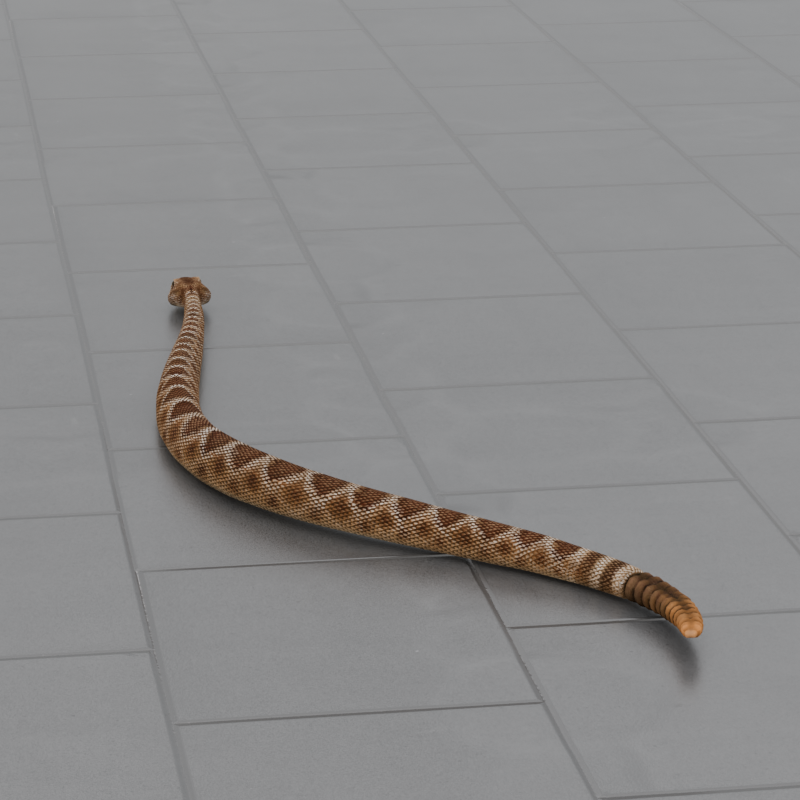}\hspace{2pt}%
\includegraphics[width=0.16\linewidth]{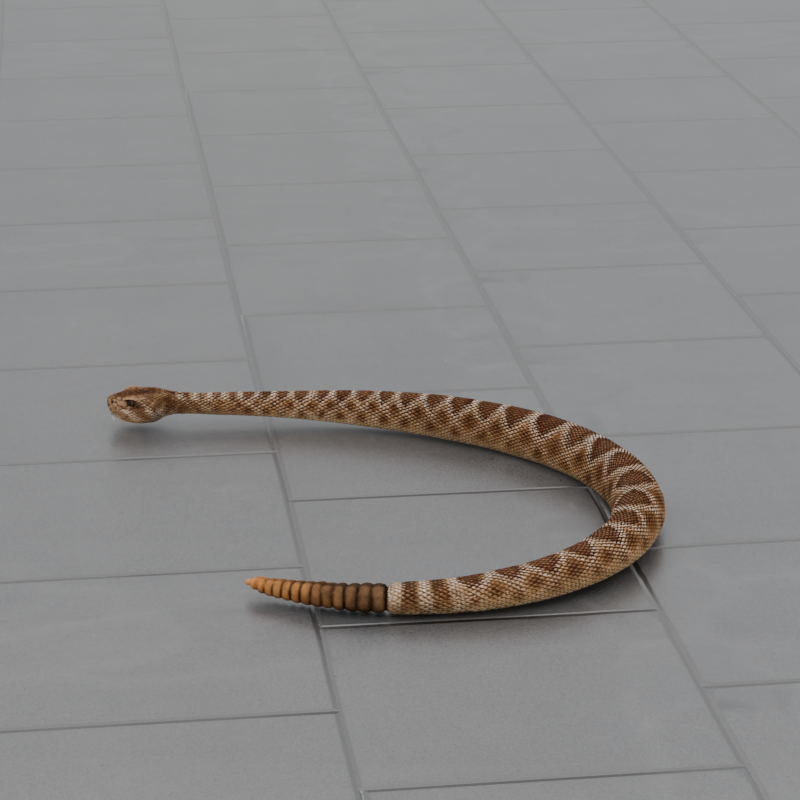}\hspace{2pt}%
\includegraphics[width=0.16\linewidth]{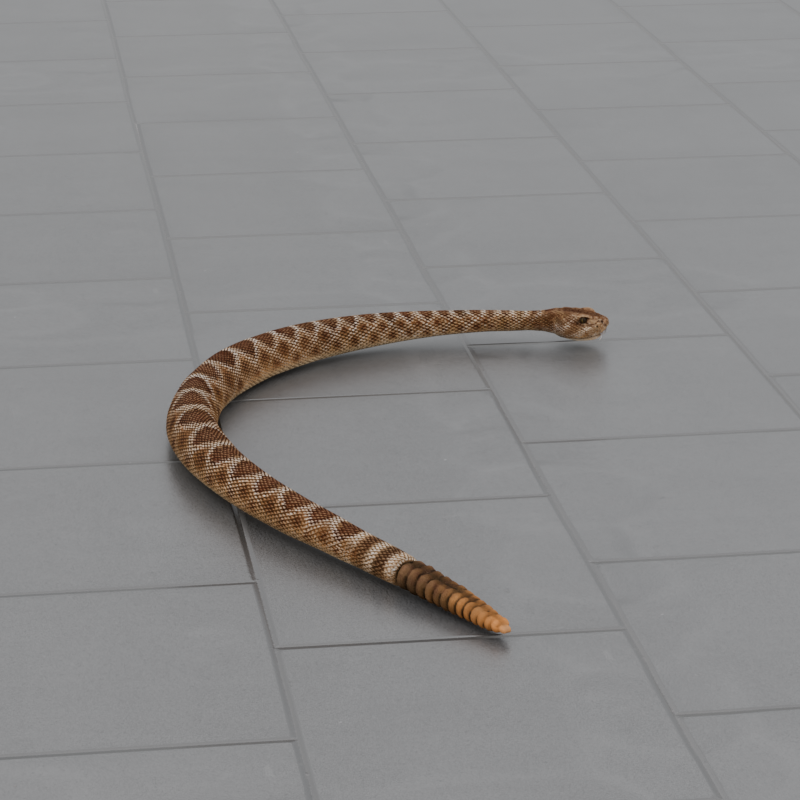}\hspace{2pt}%
\includegraphics[width=0.16\linewidth]{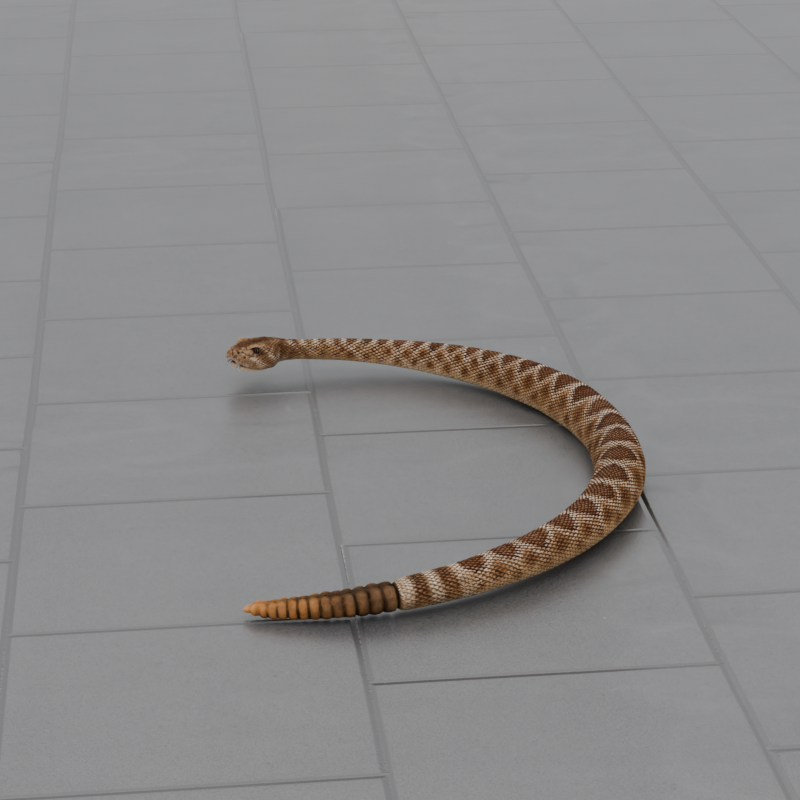}\hspace{2pt}%
\includegraphics[width=0.16\linewidth]{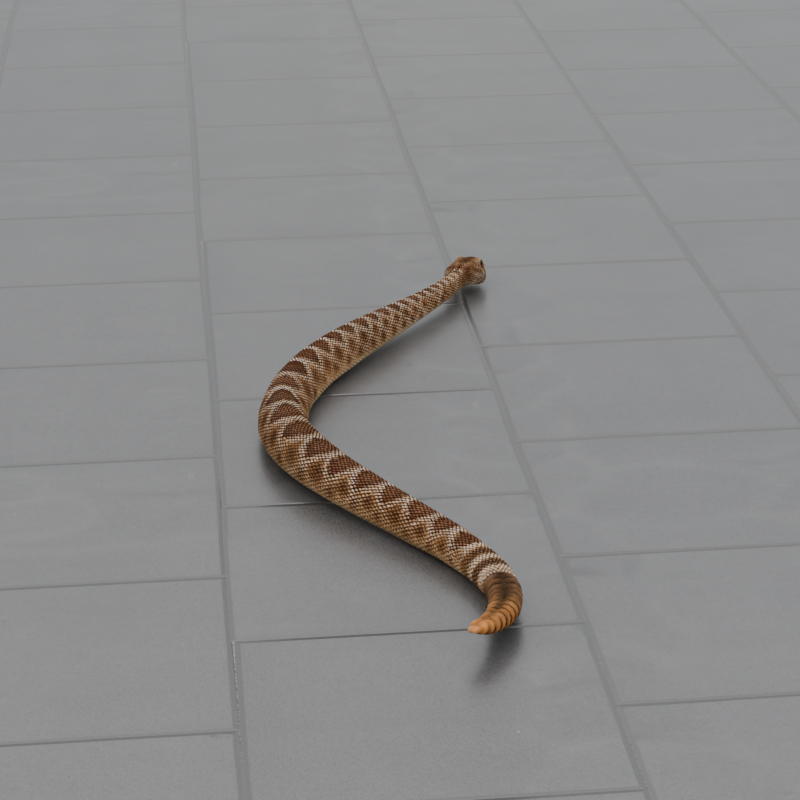}
\end{minipage}
\caption{Temporal evolution of motion, corresponding to an FDSA-optimized gait with uniform mass distribution and no bending energy term in the cost function. 
The coefficient bounds are $\max(A_{jk}) = 2$, $\max(B_{jk}) = 2$, $\max(C_{jk}) = 0.2$, and $\max(D_{jk}) = 0.2$.}
\label{fig:Img1}
\end{figure*}


\subsection{Non-uniform Mass Distribution}
\label{subsec:mass_distribution}

To represent spatial variation in body density, a continuous but non-uniform mass distribution is defined along the normalized arc length $\displaystyle s \in [0,1] $.  
The local mass per unit length is expressed as a smooth function $\displaystyle \rho(s)$, sampled from an analytic profile \( f(x) \):
\begin{equation}
\rho(s) = \mathcal{S}\left(f\left(\boldsymbol{\gamma}(s)\right)\right),
\label{eq:mass_function}
\end{equation}
where \( \mathcal{S}(\cdot) \) denotes a suitable scaling operation ensuring the physical consistency of the density field. This formulation enables modeling arbitrary spatial patterns without altering the numerical framework, including gradual tapering, mid-body concentration, or tail-heavy configurations.


The total body mass and instantaneous position of the center of mass (CM) are defined as the mass-weighted mean of the backbone coordinates
\begin{equation}\label{eq:mass_cm}
m_\text{t} = \int_0^1 \rho(s)ds, \;\; \boldsymbol{\gamma}_{CM}(s,t) = \frac{\int_0^1 \rho(s)\boldsymbol{\gamma}(s,t)ds}
{\int_0^1 \rho(s) ds}
\end{equation}

The net displacement of the center of mass is then used as the normalization factor in the total power expression given in Equation~\eqref{eq:cost_power}.  
This normalization ensures that the energetic cost is evaluated relative to the net translational motion of the body’s center of mass, providing a direct measure of locomotion efficiency that naturally accounts for the underlying mass distribution.

The distributed mass $\displaystyle \rho(s)$ is further incorporated into the \textit{contact} and \textit{gravitational} force components of the dynamic balance.  
In the contact model, the local ground-reaction force is scaled by $\displaystyle \rho(s)$, ensuring that heavier body segments exert proportionally stronger normal forces on the plane.

\subsection{Force Model}\label{sec:AllForces}
\subsubsection{Ground Contact Force Model with Inclined Angle}

 Ground contact generates vertical reaction forces that both prevent the limbless body penetration into the substrate and provide the necessary support for forward propulsion. 

To achieve smooth and numerically stable computations, a regularized ground contact model is employed, which expresses the vertical reaction force as a continuous function of the body’s height above the ground. Let \( z(s, t) \) denote the vertical coordinate of the spinal trajectory at arc length \( s \) and time \( t \). When a segment of the body lies above the ground (\( z > 0 \)), the support force decays smoothly, reflecting that the body is elevated and gradually losing contact with the substrate. Conversely, when the body penetrates slightly below the nominal ground plane (\( z \le 0 \)), a restoring force is generated that pushes the body upward.

The following continuous piecewise definition models this regularized behavior~\cite{Alben2022}
\begin{equation}
\boldsymbol{f}_{\text{contact}}(s, t) = \rho(s) \mathbf{g} \cos\alpha \hat{\mathbf{z}}
\begin{cases}
\text{exp}\left\{-\frac{z(s, t)}{d_w \cos\alpha }\right\}, & z(s,t)>0, \\
1 - \frac{z(s,t)}{d_w\cos\alpha }, & z(s, t) \leq 0,
\end{cases}
\label{eq:contact_force}
\end{equation}
where $\displaystyle \mathbf{g}$ is the gravitational acceleration, $\displaystyle \hat{\mathbf{z}}$ is the unit vector in the vertical direction $\displaystyle \alpha$ is the angle of inclination of the ground, and $d_w$ is a small regularization constant that smooths the contact force near the ground, preventing a discontinuous jump at $z=0$ and controlling its decay above the ground.

\subsubsection{Anisotropic Friction Model}
Unlike isotropic friction, which acts equally in all directions, the soft body locomotion occurs under anisotropic frictional conditions where resistance depends on the direction of motion relative to the body’s orientation. This directional dependence reflects the mechanical asymmetry of the ventral scales, which facilitates easier forward sliding along the body axis and higher resistance during backward or lateral motion. Let
\begin{equation}
\label{eq:velocityProj}
\boldsymbol{\gamma}'_{2D} = \frac{\partial \boldsymbol{\gamma}_{2D}(s,t)}{\partial t}  = 
\begin{bmatrix}
x'(s,t), y'(s,t)
\end{bmatrix}^\top
\end{equation}
denote the local velocity of the spinal trajectory projected onto the $\displaystyle xy$ plane. 
The local unit tangent and normal directions in this plane are defined as
\begin{equation}
\hat{\mathbf{s}}_{2D}(s,t) =
\begin{bmatrix}
s_x,  s_y
\end{bmatrix}^\top,
\quad
\hat{\mathbf{n}}_{2D}(s,t) =
\begin{bmatrix}
n_x, n_y
\end{bmatrix}^\top
\end{equation}
where $\displaystyle \hat{\mathbf{s}}_{2D}$ is the projection of the three-dimensional tangent vector $\displaystyle \hat{\mathbf{s}}$ onto the $\displaystyle xy$ plane, and $\displaystyle \hat{\mathbf{n}}_{2D} \equiv \hat{\mathbf{s}}_{2D}^{\perp}$ 
is obtained by rotating $\displaystyle \hat{\mathbf{s}}_{2D} $ by $\displaystyle 90^{\circ} $ counterclockwise about the vertical $\displaystyle \hat{\mathbf{z}}$ axis. 

The tangential and normal (transverse) velocity components are then obtained by projecting the planar velocity 
onto these local directions as
\begin{equation}
\begin{aligned}
    u_s(s,t) &= \boldsymbol{\gamma}'_{2D}(s,t)^\top \hat{\mathbf{s}}_{2D}(s,t) \\
    u_n(s,t) &= \boldsymbol{\gamma}'_{2D}(s,t)^\top \hat{\mathbf{n}}_{2D}(s,t)
\end{aligned}
\end{equation}
where $\displaystyle u_s$ and $\displaystyle u_n$ denote, respectively, the tangential and transverse components of the body’s motion. To ensure smooth numerical behavior when the velocity magnitude approaches zero, a regularized form of the local speed is defined as
\begin{equation}\label{eq:N}
N(s, t) = \sqrt{x'(s, t)^2 + y'(s, t)^2 + \delta^2},
\end{equation}
where $\displaystyle \delta >0$ is a small regularization constant introduced to prevent singularities in the friction computation as shown in Equation~\eqref{eq:friction_components} later. The tangential friction coefficient is direction-dependent and varies according to the sign of the tangential velocity, defined as
\begin{equation}
\mu_s =
\begin{cases}
    \mu_f, & u_s \ge 0 \quad \text{(forward motion)} \\
    \mu_b, & u_s < 0 \quad \text{(backward motion)}
\end{cases}
\label{eq:mu_s}
\end{equation}
where $\displaystyle \mu_f $ and $\displaystyle \mu_b $ are the forward and backward tangential friction coefficients, respectively, with $\displaystyle \mu_b > \mu_f $ reflecting the higher resistance encountered during backward sliding. The normal friction coefficient $\displaystyle \mu_n $ represents the transverse resistance.

Thus, the total planar frictional force per unit body length acting on the limbless body is given by
\begin{equation}\label{eq:friction_components}
    \mathbf{f}_{\text{fric}}(s,t) = -\lVert \mathbf{f}_{\text{contact}}(s, t) \rVert \frac{\mu_n u_n(s, t) \hat{\mathbf{n}}_{2D}(s,t)+\mu_s u_s(s,t) \hat{\mathbf{s}}_{2D}(s,t) }{N(s, t)} 
\end{equation}

where $\displaystyle \lVert \mathbf{f}_{\text{contact}}(s, t) \rVert$ represents the magnitude of the local normal contact force acting on the body. The coefficients $\displaystyle \mu_n$ quantify the anisotropic frictional coefficients in the transverse direction.

Equations~\eqref{eq:velocityProj}–\eqref{eq:friction_components} collectively define a continuous anisotropic friction model, where the magnitude and direction of resistive forces depend on both the body’s orientation and the direction of motion. As mentioned earlier, the denominator $\displaystyle N(s, t) $ provides a smooth normalization that ensures numerical stability even at near-zero velocities, avoiding abrupt transitions in frictional response. 

\subsubsection{Gravitational Force Model}

The gravitational contribution acts uniformly along the body in the vertical direction and is expressed as
\begin{equation}
\mathbf{f}_{\text{gravity}} = -\rho(s) \mathbf{g} \sin \alpha \hat{\mathbf{z}}
\label{eq:f_gravity}
\end{equation}
where \( \rho \) is the mass per unit length of the body, $\displaystyle \boldsymbol{g}$ denotes the gravitational acceleration, and \( \hat{\boldsymbol{z}} \) is the unit vector along the vertical axis. This term represents the body’s weight per unit length and acts downward, opposing the normal contact reaction exerted by the supporting surface. The gravitational force, therefore, provides the constant vertical load that balances the ground contact force during locomotion.


\begin{figure*}[tbp]
\centering

\vspace{5pt}

\noindent
\begin{minipage}[t]{0.14\textwidth}
\vspace{10pt}
\raggedright
\textbf{Case 1:}\\
{$\max(A_{jk}) = 1$}\\
{$\max(B_{jk}) = 1$}\\
{$\max(C_{jk}) = 0.1$}\\
{$\max(D_{jk}) = 0.1$}\\
\end{minipage}%
\hfill
\begin{minipage}[t]{0.85\textwidth}
\vspace{2pt}
\centering
\includegraphics[width=0.16\linewidth]{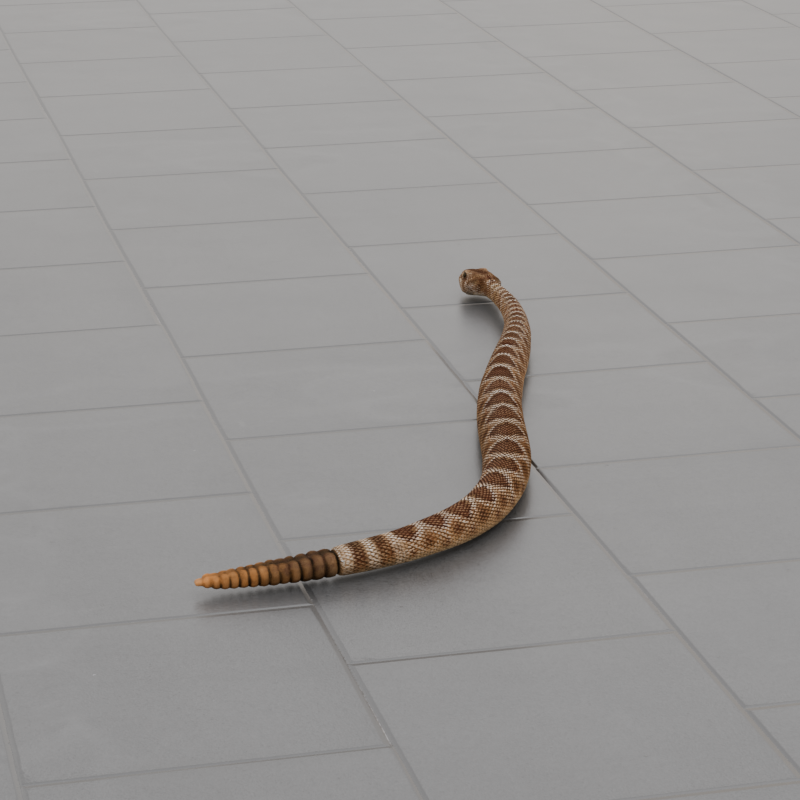}\hspace{2pt}%
\includegraphics[width=0.16\linewidth]{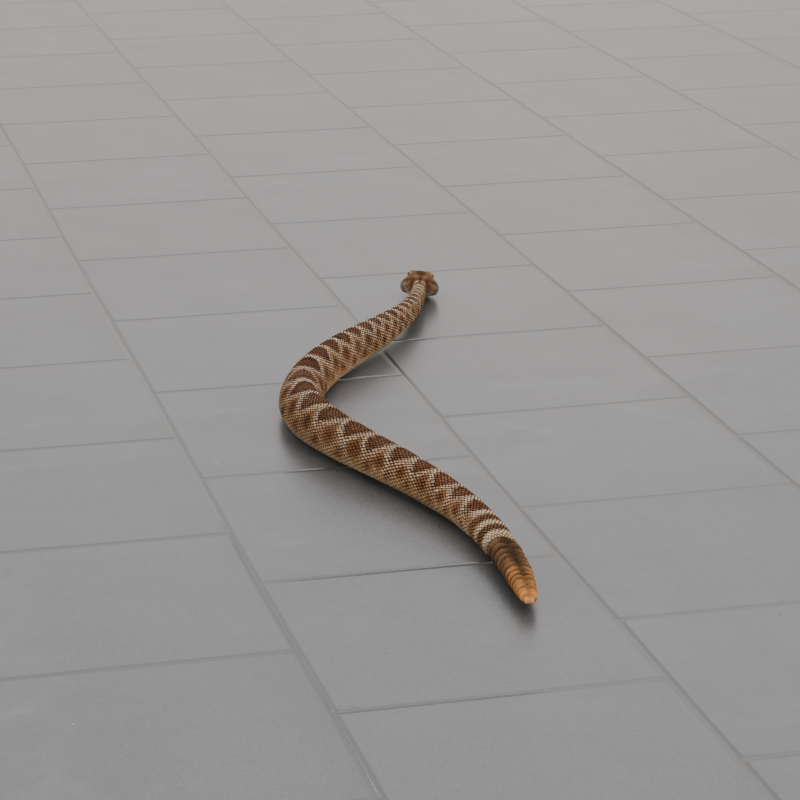}\hspace{2pt}%
\includegraphics[width=0.16\linewidth]{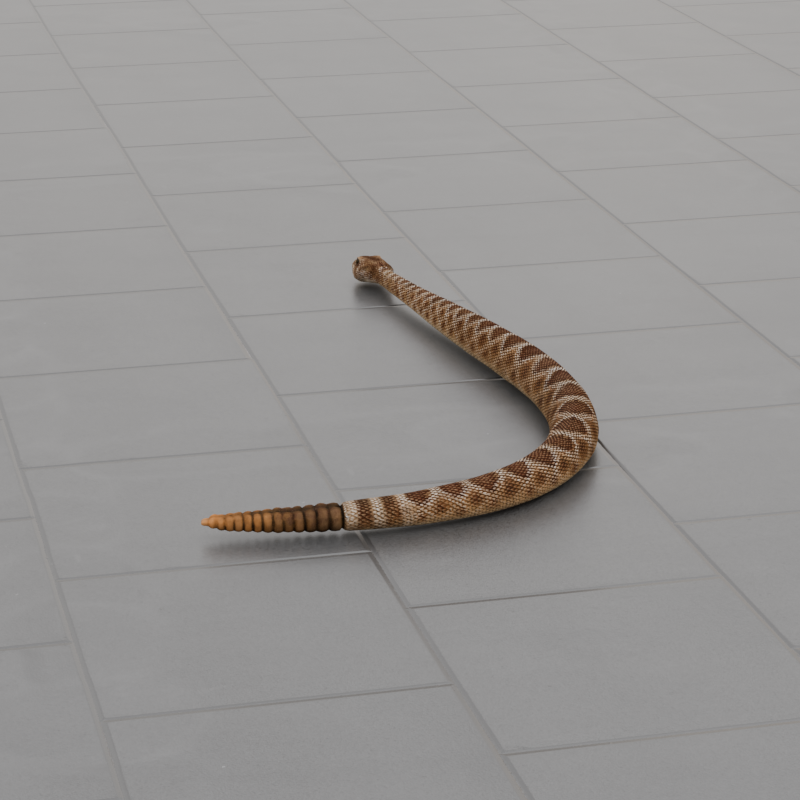}\hspace{2pt}%
\includegraphics[width=0.16\linewidth]{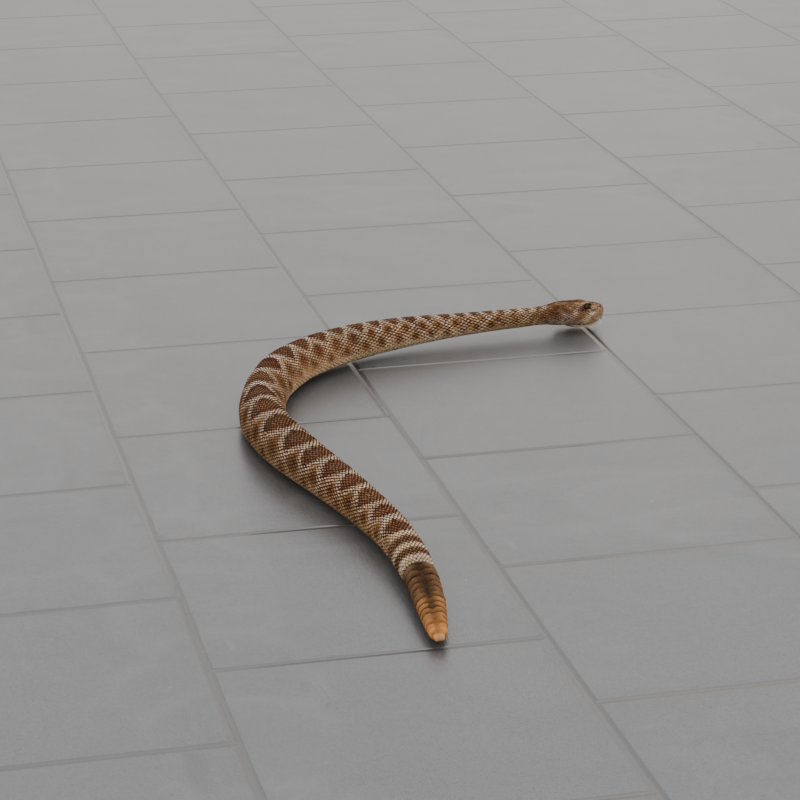}\hspace{2pt}%
\includegraphics[width=0.16\linewidth]{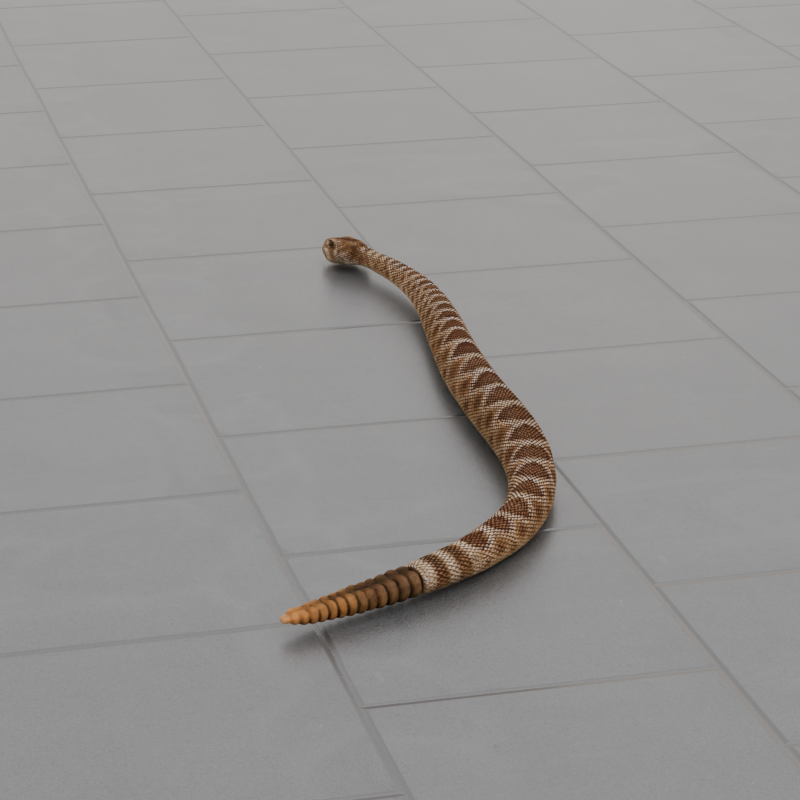}\hspace{2pt}%
\includegraphics[width=0.16\linewidth]{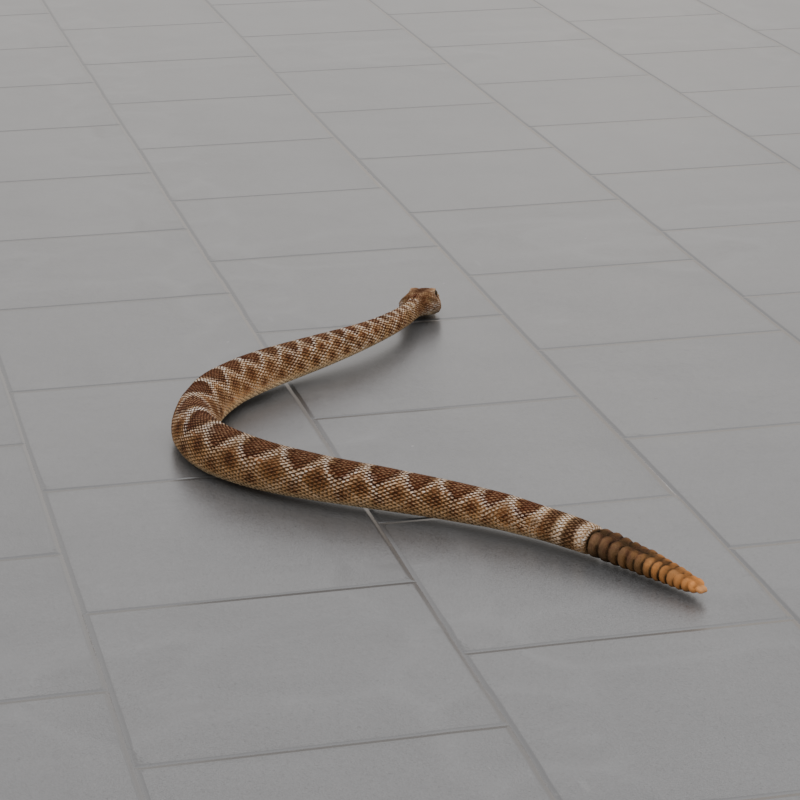}
\end{minipage}

\noindent
\begin{minipage}[t]{0.14\textwidth}
\vspace{10pt}
\raggedright
\textbf{Case 2:}\\
{$\max(A_{jk}) = 2$}\\
{$\max(B_{jk}) = 2$}\\
{$\max(C_{jk}) = 0.1$}\\
{$\max(D_{jk}) = 0.1$}\\
\end{minipage}%
\hfill
\begin{minipage}[t]{0.85\textwidth}
\vspace{2pt}
\centering
\includegraphics[width=0.16\linewidth]{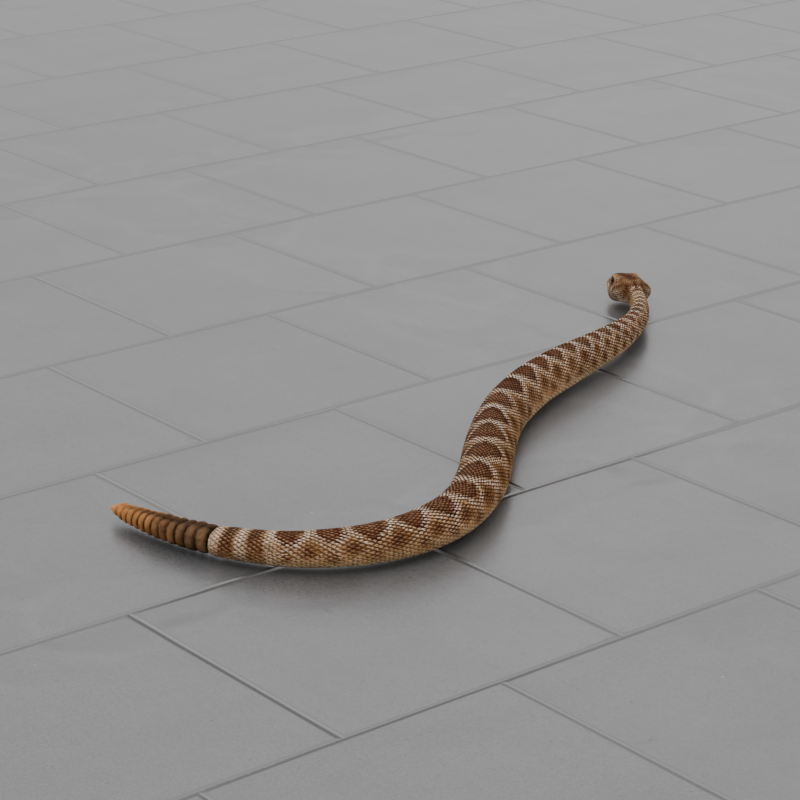}\hspace{2pt}%
\includegraphics[width=0.16\linewidth]{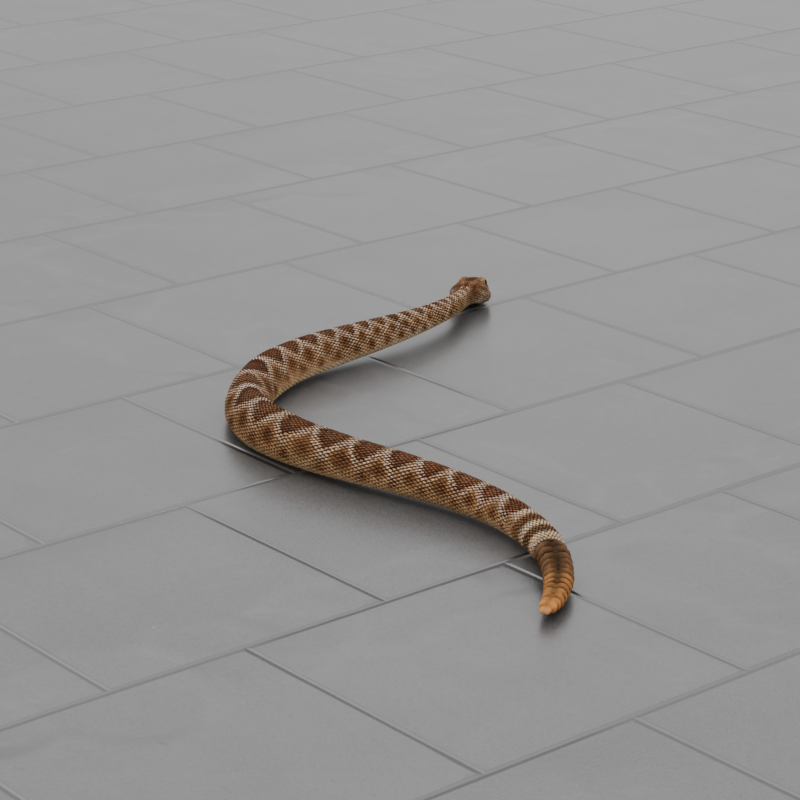}\hspace{2pt}%
\includegraphics[width=0.16\linewidth]{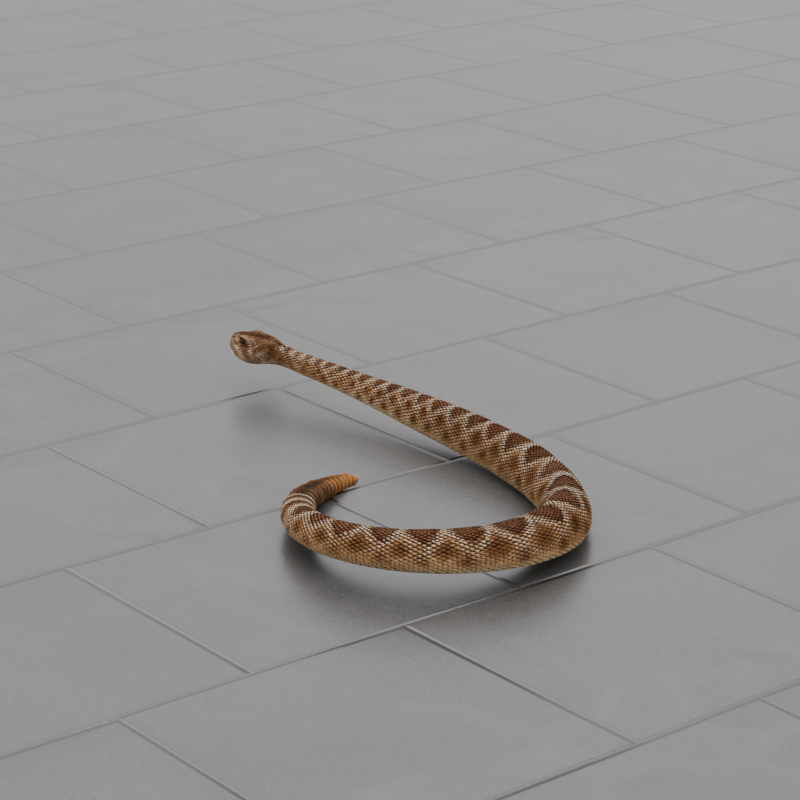}\hspace{2pt}%
\includegraphics[width=0.16\linewidth]{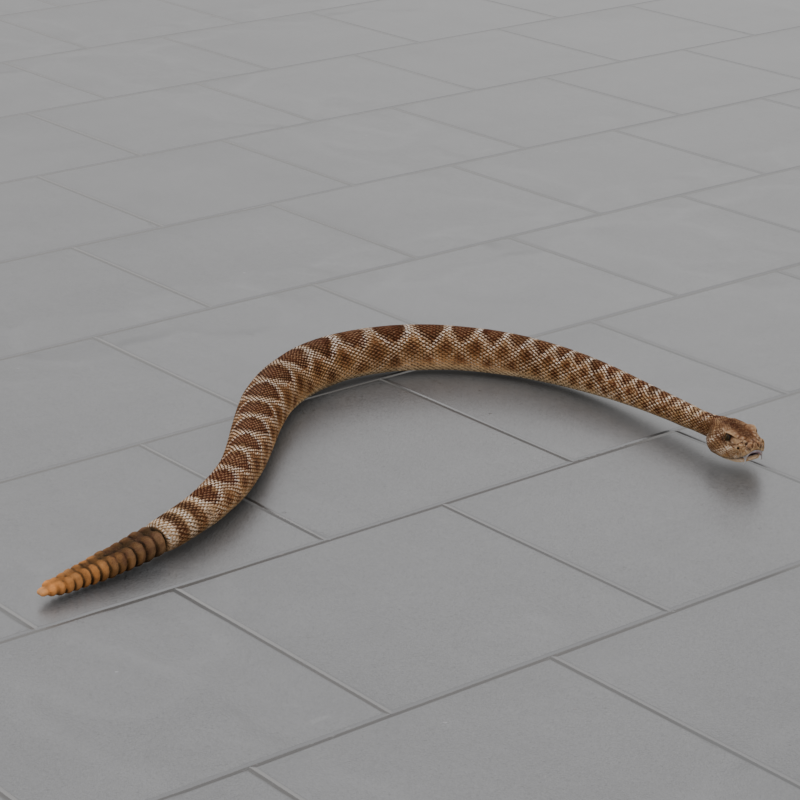}\hspace{2pt}%
\includegraphics[width=0.16\linewidth]{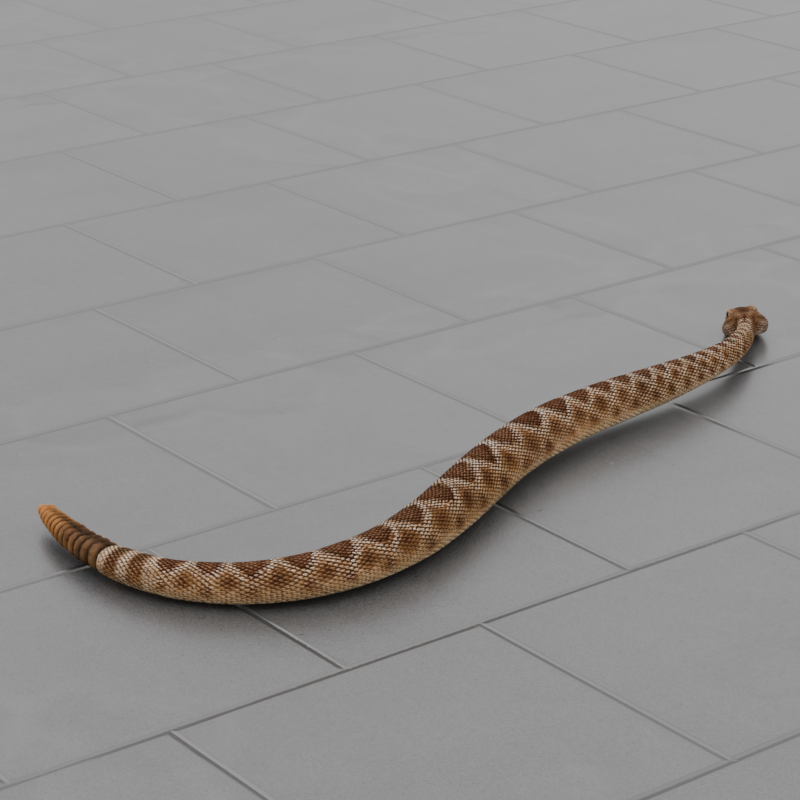}\hspace{2pt}%
\includegraphics[width=0.16\linewidth]{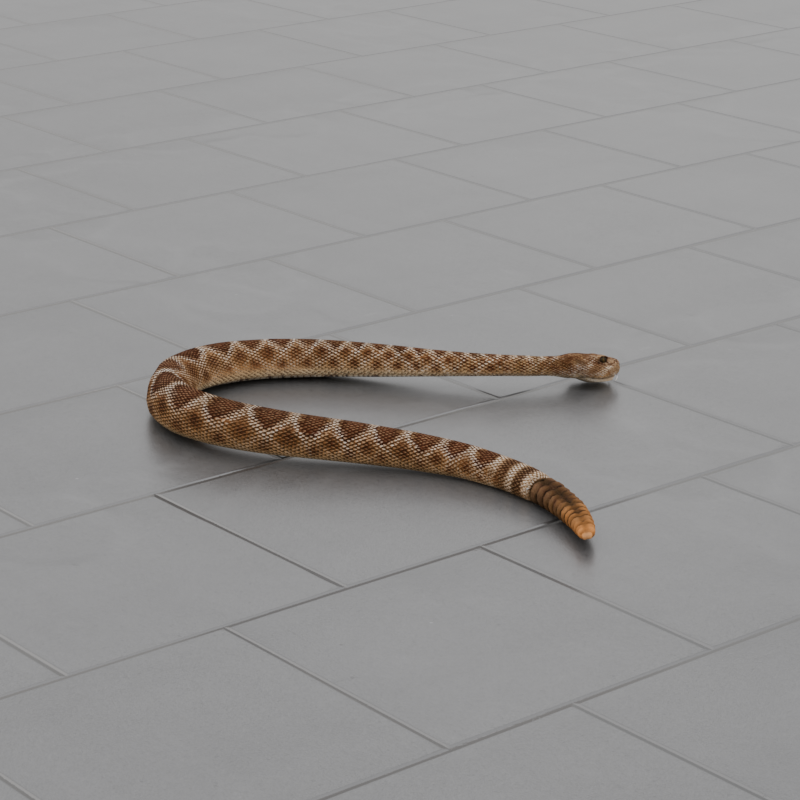}
\end{minipage}

\noindent
\begin{minipage}[t]{0.14\textwidth}
\vspace{10pt}
\raggedright
\textbf{Case 3:}\\
{$\max(A_{jk}) = 3$}\\
{$\max(B_{jk}) = 3$}\\
{$\max(C_{jk}) = 0.1$}\\
{$\max(D_{jk}) = 0.1$}\\
\end{minipage}%
\hfill
\begin{minipage}[t]{0.85\textwidth}
\vspace{2pt}
\centering
\includegraphics[width=0.16\linewidth]{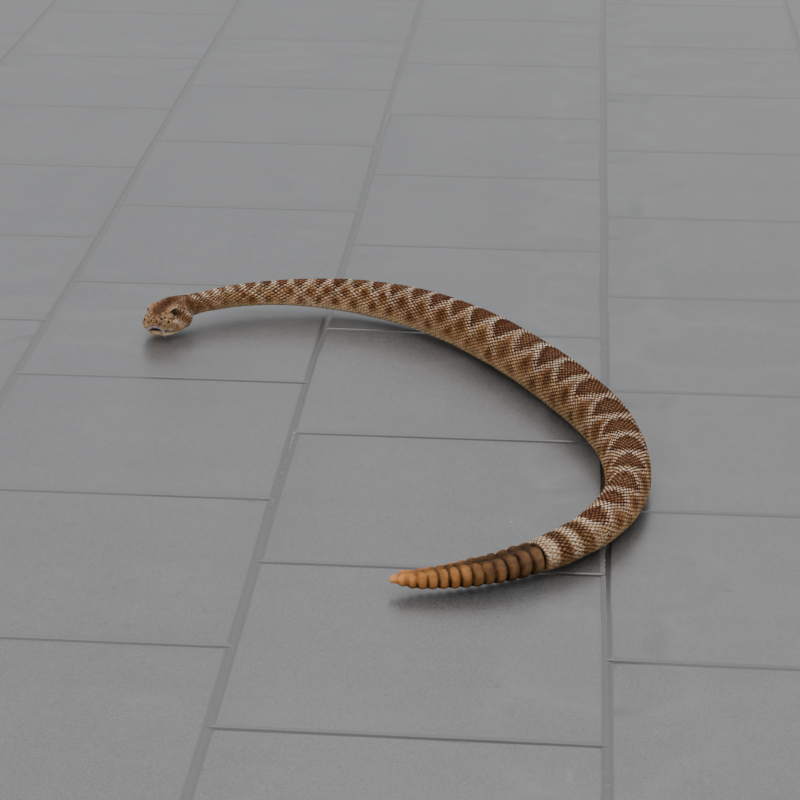}\hspace{2pt}%
\includegraphics[width=0.16\linewidth]{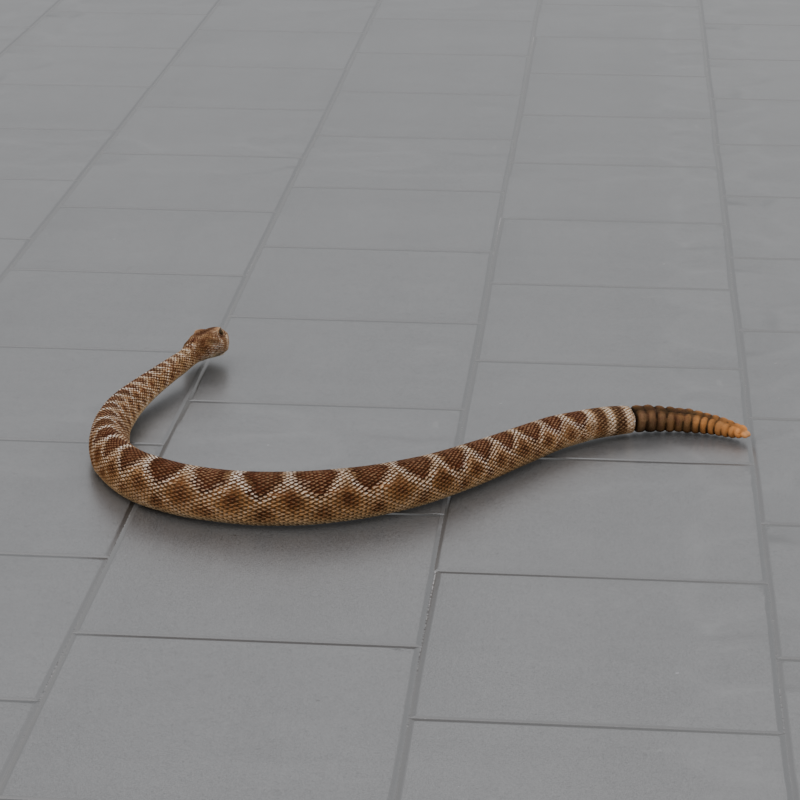}\hspace{2pt}%
\includegraphics[width=0.16\linewidth]{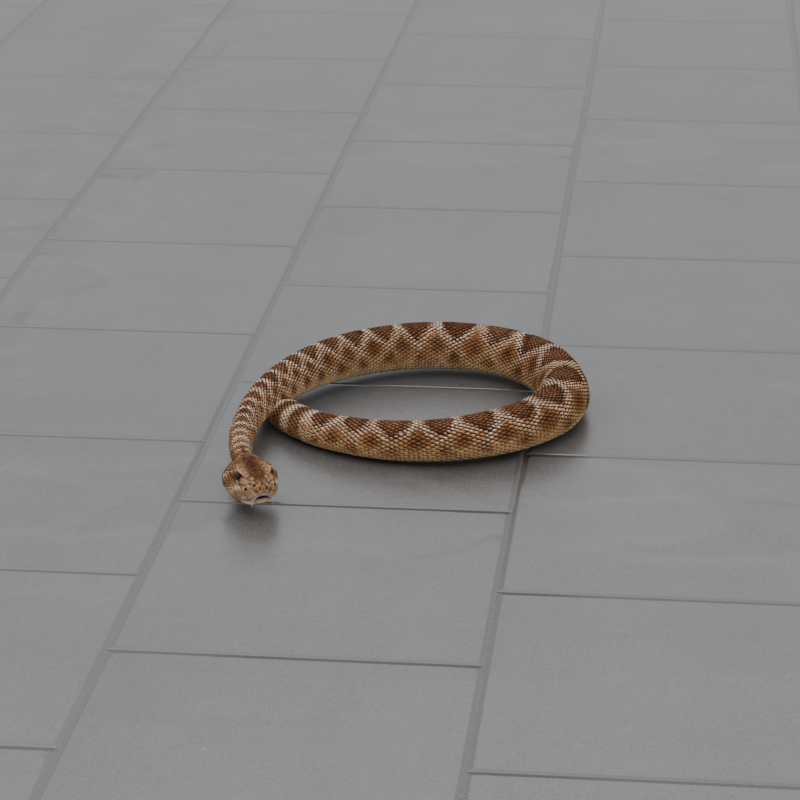}\hspace{2pt}%
\includegraphics[width=0.16\linewidth]{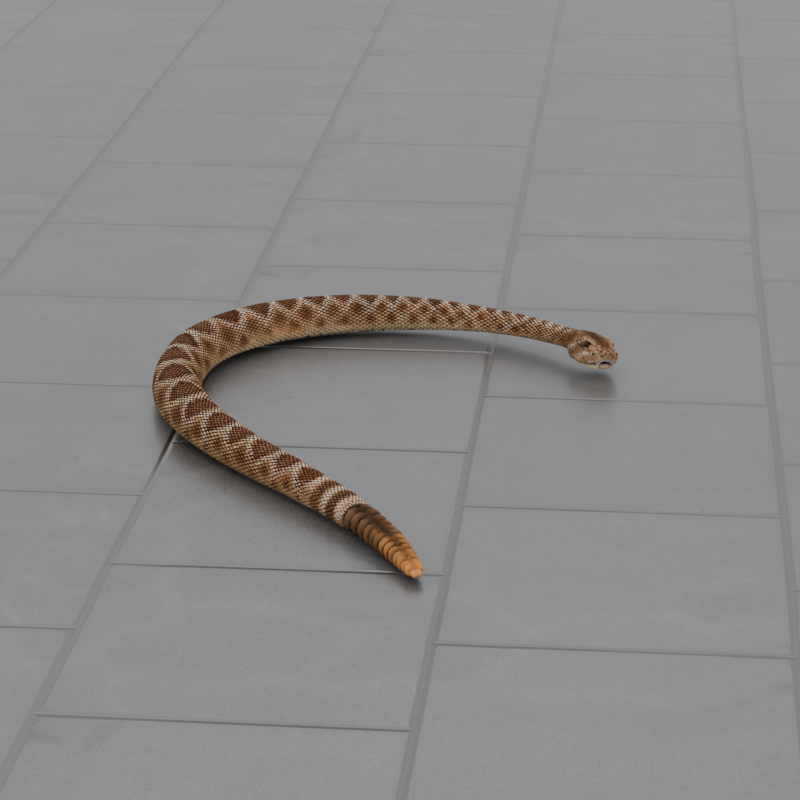}\hspace{2pt}%
\includegraphics[width=0.16\linewidth]{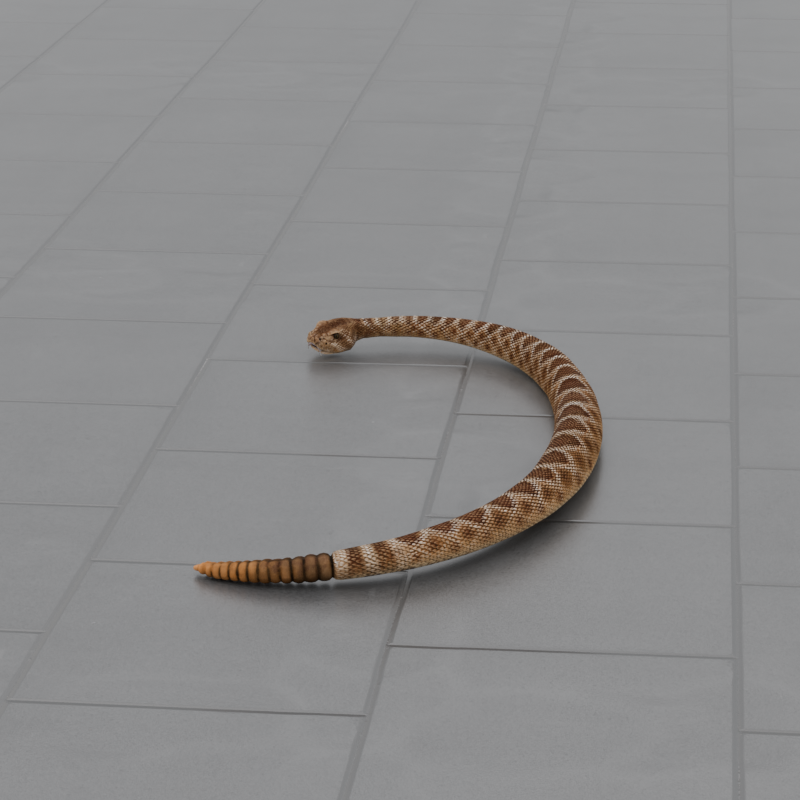}\hspace{2pt}%
\includegraphics[width=0.16\linewidth]{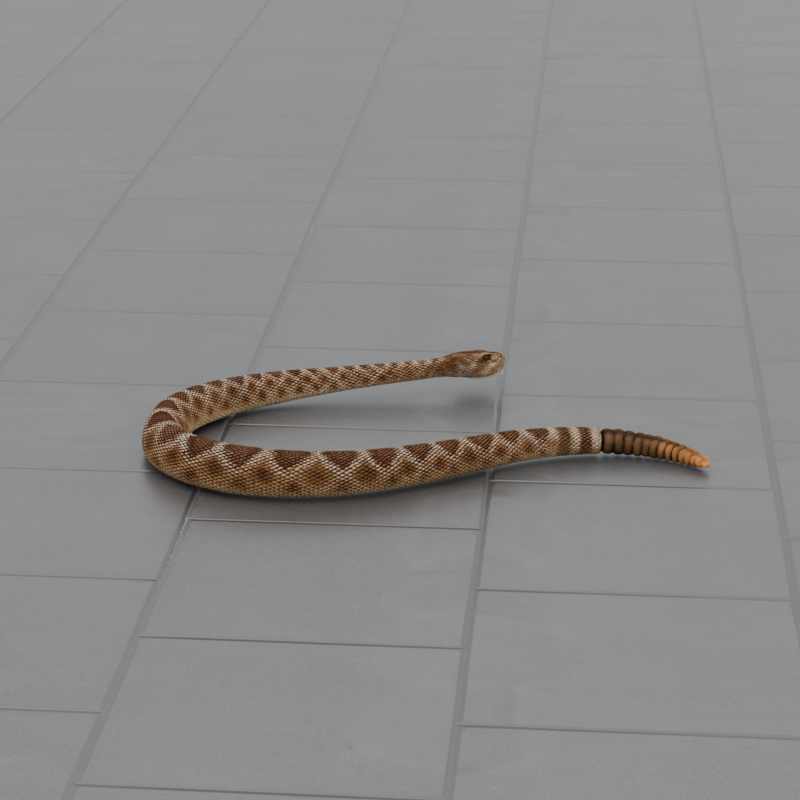}
\end{minipage}

\caption{Temporal evolution of motion, corresponding to an FDSA-optimized gait with friction ratios $\mu_n/\mu_f = 1$ and $\mu_b/\mu_f = 5$ with uniform mass distribution, and no bending energy term in the cost function.}
\label{fig:Img2}
\end{figure*}


\subsection{Force and Angular Momentum Residual Formulation}
\label{subsec:force_balance}

The motion of the limbless body is governed by the balance of linear and angular momentum over its entire body. The formulation is expressed in terms of residual quantities that measure the deviation from exact mechanical equilibrium.

Consider an infinitesimal segment of the limbless body with mass $dm$. This segment experiences an acceleration $\mathbf{a}(s,t)$, which is the acceleration of a point on the curve and arises from the time variation of the body configuration. In addition to inertial effects, the segment is subjected to external forces consisting of contact forces, gravitational forces, and frictional interactions with the ground. The total external force is expressed as
\begin{equation}
\mathbf{f}_{\text{ext}} = \mathbf{f}_{\text{contact}} + \mathbf{f}_{\text{gravity}} + \mathbf{f}_{\text{friction}}
\end{equation} 
The difference between the inertial force $dm \, \mathbf{a}(s,t)$ and the total external force represents the force imbalance acting on that segment. If the system were in perfect mechanical equilibrium, this imbalance would vanish everywhere. However, in the numerical formulation, this imbalance is integrated over the entire body to define a global residual of linear momentum, which is given by the equation as
\begin{equation}\label{eq:force_balance}
\int_{0}^{1} \left[ dm\,\mathbf{a}(s,t) - \mathbf{f}_{\text{ext}}(s,t) \right]\, ds = \epsilon_{r}, \;\; \textrm{s.t.}\;\; \epsilon_{r} \neq 0
\end{equation}
This residual represents the net unbalanced force acting on the limbless body along the three spatial directions. Similarly, the imbalance in angular momentum is obtained by computing the torque contribution of each segment about a reference origin. This is done by taking the cross product of the position vector $\boldsymbol{\gamma}(s,t)$ of the segment with the force imbalance and integrating it along the body, which is given by the equation as
\begin{equation}\label{eq:torque_balance}
\int_{0}^{1} \boldsymbol{\gamma}(s,t) \times \left[ dm\,\mathbf{a}(s,t) - \mathbf{f}_{\text{ext}}(s,t) \right]\, ds = \epsilon_{r}, \;\; \textrm{s.t.}\;\; \epsilon_{r} \neq 0
\end{equation}
The result is a residual of angular momentum, which quantifies the net unbalanced torque acting on the system. Together, these two quantities form a six-dimensional residual vector, with three components corresponding to linear-momentum imbalance and three to angular-momentum imbalance. This residual vector provides a complete measure of how far the current configuration is from satisfying global force and torque balance. The residual depends on a set of extrinsic parameters, denoted by $\displaystyle \boldsymbol{\psi}$ from Equation $\displaystyle \eqref{eq:RigidBodyVar}$, which describe the tail position of the curve. The physically consistent motion of the limbless body is obtained by finding the values of $\boldsymbol{\psi}$ for which this residual becomes zero. This formulation leads to a root-finding problem to determine the extrinsic parameters that make the net linear and angular momentum imbalances vanish, which is solved iteratively using a Newton-Raphson method, as described in Section~\ref{sec:extrinsic}.

\subsection{Power Consumption}\label{sec:power}
The efficiency of locomotion is determined by how much mechanical power the limbless body expends while moving. During this process, energy is continuously used in two ways: first, to overcome the resistive frictional forces of the ground; and second, to deform the body as it bends and straightens over time. These two contributions together determine the total power consumption and provide a measure of locomotion efficiency.

\subsubsection{Frictional Energy Dissipation}
The first contribution comes from the \textit{frictional work} performed by the limbless body against the substrate. As the body slides over the ground, each local segment experiences tangential and normal resistive forces that depend on its instantaneous velocity and direction of motion. Because friction is anisotropic, the resistance varies when the body moves in different directions, such as forward, backward, or sideways. Following the work by ~\citet{Alben2013}, the average rate of energy loss due to friction over time  $\displaystyle T$ is written as
\begin{equation}
    P_{\mathrm{fric}} =\frac{1}{T}\int_{0}^{T}\int_{0}^{L} \mathbf{f}_{\text{fric}}^\top \boldsymbol{\gamma}'ds dt, \;\;\; \frac{\boldsymbol{\gamma}'}{|\boldsymbol{\gamma}'|} = \frac{\partial \boldsymbol{\gamma}}{\partial t}
\end{equation}
where $\mathbf{f}_{\text{fric}}$ represents the local frictional force per unit length and $\boldsymbol{\gamma}'$ is the local velocity of the curve.

\subsubsection{Internal Viscous Dissipation}
The second contribution arises from \textit{internal viscous dissipation}, which accounts for the energy lost as the body bends and unbends. When the curvature $\displaystyle \kappa(s,t)$ changes over time, the body material undergoes internal strain, leading to viscous losses similar to those found in visco-elastic materials. This effect is modeled through a term proportional to the square of the curvature rate, with proportionality constant $\displaystyle \epsilon \mathcal{I}$, where $\displaystyle \epsilon$ is the effective extensional viscosity and $\displaystyle \mathcal{I}$ is the cross-sectional area moment of inertia. The corresponding viscous power is

\begin{equation}
\begin{aligned}
    P_{\mathrm{strain}} &= \frac{1}{T}\int_{0}^{T}\int_{0}^{L} \epsilon \mathcal{I} \left(\frac{\partial \kappa}{\partial t} \right)^2 ds dt \\
    &= \frac{1}{T}\int_{0}^{T}\int_{0}^{L} \epsilon \mathcal{I} \left(\kappa'\right)^2 ds dt, \;\;\; \kappa' = \frac{\partial \kappa}{\partial t}
\end{aligned}
\end{equation}

By combining these two effects, the external frictional losses and the internal viscous dissipation, the total average mechanical power is obtained as
\begin{equation}
    \langle P\rangle =\underbrace{\frac{1}{T}\int_{0}^{T}\int_{0}^{L}\mathbf{f}_{\mathrm{fric}}^\top \boldsymbol{\gamma}ds dt}_{\text{I}} + \underbrace{\frac{1}{T}\int_{0}^{T}\int_{0}^{L} \epsilon \mathcal{I} \left(\kappa'\right)^2ds dt}_{\text{II}}
\end{equation}
The term I quantifies the power used to overcome ground friction, while the term II measures the power dissipated internally as the body deforms.

\subsection{Torsion \& Bending Energy Model}\label{sec:intpower}
As discussed in the previous section, minimizing frictional power dissipation enhances the overall locomotion efficiency of the soft body. However, this objective alone does not impose any constraints on the body's geometric configuration, such as its shape or orientation in space. In the absence of additional regularization, we observed that the curve representing the body tends to exhibit excessive twisting and occasional self-intersection, resulting in physically inconsistent configurations.

To address these issues, we introduce two additional constraints, formulated as energy-minimization terms, that regularize the curve’s geometry. Specifically, we minimize the torsional energy, $\mathcal{E}{\tau}$, to suppress unwanted twisting, and the bending energy, $\mathcal{E}{\kappa}$, to prevent self-intersection along the body~\cite{pinkallgross2024}~\cite{yu2021rc}~\cite{yu2021rs}.

\begin{equation}\label{eq:torsion_energy}
    \mathcal{E}{\tau} = \frac{1}{2}\int_{0}^{1} \tau^2 ds = \frac{1}{2}\int_{0}^{1} \left\langle\left\langle \frac{\partial\boldsymbol{N}}{\partial s} \boldsymbol{T} \times \boldsymbol{N} \right\rangle, \left\langle \frac{\partial\boldsymbol{N}}{\partial s} \boldsymbol{T} \times \boldsymbol{N} \right\rangle\right\rangle ds
\end{equation}

\begin{equation}\label{eq:bend_energy}
    \mathcal{E}{\kappa} = \frac{1}{2}\int_{0}^{1} \kappa^2 ds = \frac{1}{2}\int_{0}^{1} \left\langle \frac{\partial\boldsymbol{T}}{\partial s}, \frac{\partial\boldsymbol{T}}{\partial s} \right\rangle ds
\end{equation}

\subsection{Curvature and Torsion Control}
 
The functions $\kappa(s,t)$ and $\tau(s,t)$ are expanded in terms of spatial basis functions and temporal basis functions. The coefficients of these expansions act as the tunable control variables. By adjusting them, the limbless body can explore a rich family of body deformations while maintaining smoothness and physical realizability.

The kinematics of the soft body and the power dissipation are completely described by parameterizing the curvature, $\displaystyle \kappa(s,t)$, and torsion, $\displaystyle \tau(s,t)$, using Chebyshev-Fourier series.

\begin{equation}
\begin{aligned}
    \kappa(s,t) &= \sum_{j=0}^{N_f-1} \sum_{k=0}^{N_c-1} \left[ A_{jk} \cos(2\pi jt) + B_{jk} \sin(2\pi jt) \right] T_k(s)\\
    \tau(s,t) &= \left[1 - e^{-\left(t/t_d\right)^2}\right]\cdot \\
    &\sum_{j=0}^{N_f-1} \sum_{k=0}^{N_c-1} \left[C_{jk} \cos(2\pi j t) + D_{jk} \sin(2\pi j t) \right] T_k(s)
\end{aligned}
\label{eq:KappaAndTau}
\end{equation}
where $\displaystyle t_d$ denotes a characteristic transient time scale that regulates the exponential onset of torsion, ensuring a gradual introduction of out-of-plane twisting during the initial phase of motion. The functions $T_k(s)$ are Chebyshev polynomials~\cite{gill1981practical} of the first kind of degree $\displaystyle k$. The integers $\displaystyle N_f$ and $\displaystyle N_c$ represent the number of temporal Fourier modes and spatial Chebyshev modes, respectively. The sets of coefficients $\displaystyle \{A_{jk}, B_{jk}, C_{jk}, D_{jk}\}$ fully determine the curvature and torsion profiles of the limbless body, with the constant terms $B_{0k}$ excluded to avoid redundancy.

\section{Optimization}\label{sec:opt}
To assess efficiency, the total power is normalized by the distance traveled by the center of mass $\displaystyle d_t$ over time and by the smallest friction coefficient $\displaystyle \mu_{\min}=\min(\mu_f,\mu_b,\mu_n)$. The total energetic cost that we need to minimize is then expressed as
\begin{equation}\label{eq:cost_power}
    \mathcal{L}_{P} = \frac{P_{\mathrm{fric}}}{d_t \mu_{\min}} + \zeta\frac{P_{\mathrm{strain}}}{d_t^{2}}
\end{equation}
where $\displaystyle \zeta$ is a user-defined damping coefficient. For body motions, gravity and normal contact force primarily cause minor vertical adjustments. These forces exchange potential energy as the body rises and falls slightly. As a result, their average contribution to the total mechanical work is negligible compared with the frictional and viscous components.


\begin{figure*}[tbp]
\centering

\vspace{5pt}

\noindent
\begin{minipage}[t]{0.14\textwidth}
\vspace{10pt}
\raggedright
\textbf{Case 1:}\\
{$\max(A_{jk}) = 1$}\\
{$\max(B_{jk}) = 1$}\\
{$\max(C_{jk}) = 0.1$}\\
{$\max(D_{jk}) = 0.1$}\\
\end{minipage}%
\hfill
\begin{minipage}[t]{0.85\textwidth}
\vspace{2pt}
\centering
\includegraphics[width=0.16\linewidth]{Figures/d1.png}\hspace{2pt}%
\includegraphics[width=0.16\linewidth]{Figures/d2.png}\hspace{2pt}%
\includegraphics[width=0.16\linewidth]{Figures/d3.png}\hspace{2pt}%
\includegraphics[width=0.16\linewidth]{Figures/d4.png}\hspace{2pt}%
\includegraphics[width=0.16\linewidth]{Figures/d5.png}\hspace{2pt}%
\includegraphics[width=0.16\linewidth]{Figures/d6.png}
\end{minipage}

\noindent
\begin{minipage}[t]{0.14\textwidth}
\vspace{10pt}
\raggedright
\textbf{Case 2:}\\
{$\max(A_{jk}) = 2$}\\
{$\max(B_{jk}) = 2$}\\
{$\max(C_{jk}) = 0.2$}\\
{$\max(D_{jk}) = 0.2$}\\
\end{minipage}%
\hfill
\begin{minipage}[t]{0.85\textwidth}
\vspace{2pt}
\centering
\includegraphics[width=0.16\linewidth]{Figures/b1.png}\hspace{2pt}%
\includegraphics[width=0.16\linewidth]{Figures/b2.png}\hspace{2pt}%
\includegraphics[width=0.16\linewidth]{Figures/b3.png}\hspace{2pt}%
\includegraphics[width=0.16\linewidth]{Figures/b4.png}\hspace{2pt}%
\includegraphics[width=0.16\linewidth]{Figures/b5.png}\hspace{2pt}%
\includegraphics[width=0.16\linewidth]{Figures/b6.png}
\end{minipage}

\noindent
\begin{minipage}[t]{0.14\textwidth}
\vspace{10pt}
\raggedright
\textbf{Case 3:}\\
{$\max(A_{jk}) = 3$}\\
{$\max(B_{jk}) = 3$}\\
{$\max(C_{jk}) = 0.3$}\\
{$\max(D_{jk}) = 0.3$}\\
\end{minipage}%
\hfill
\begin{minipage}[t]{0.85\textwidth}
\vspace{2pt}
\centering
\includegraphics[width=0.16\linewidth]{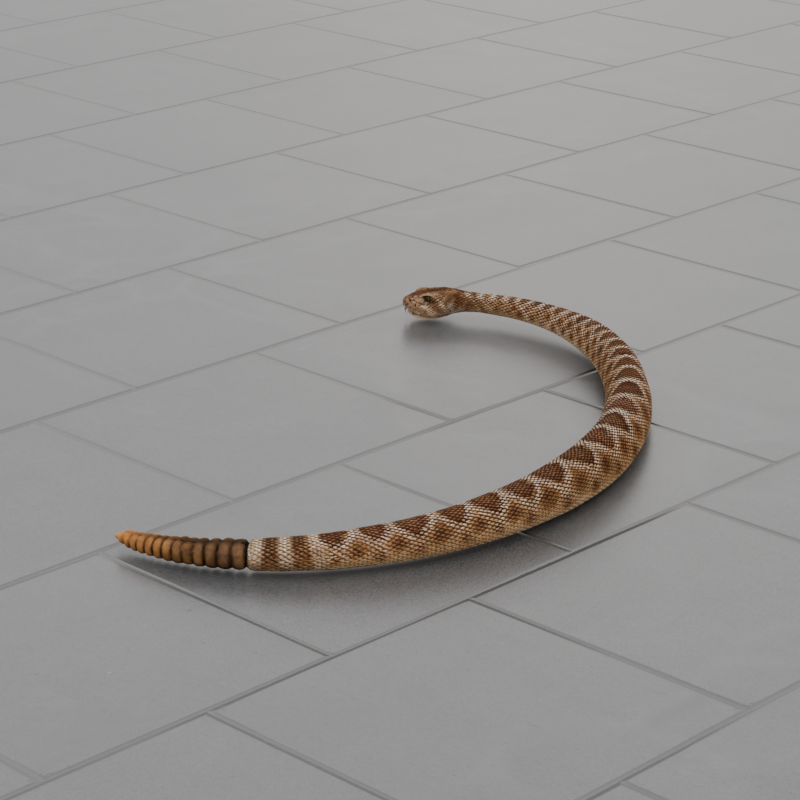}\hspace{2pt}%
\includegraphics[width=0.16\linewidth]{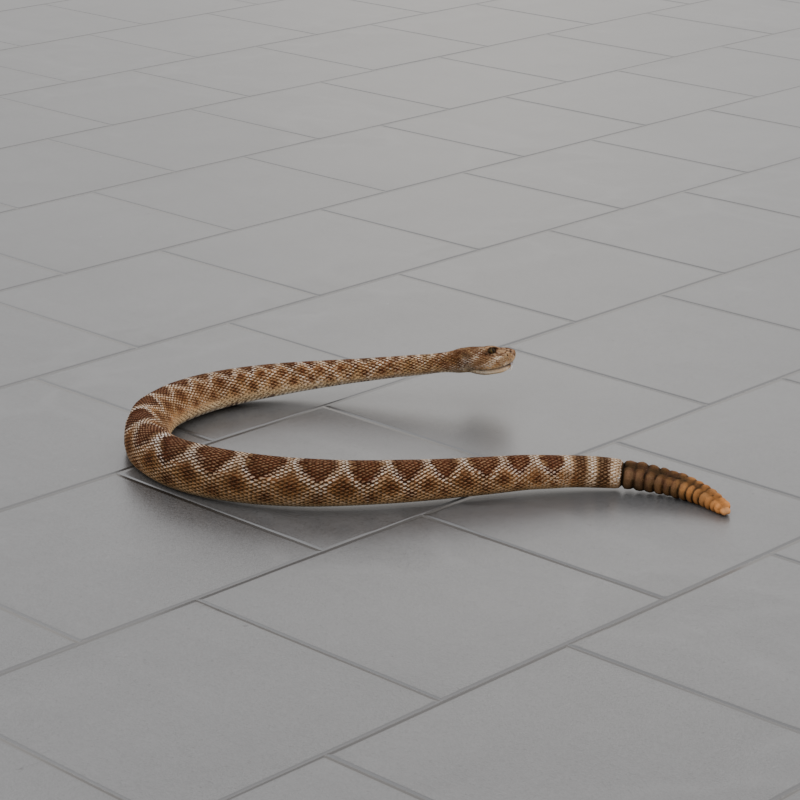}\hspace{2pt}%
\includegraphics[width=0.16\linewidth]{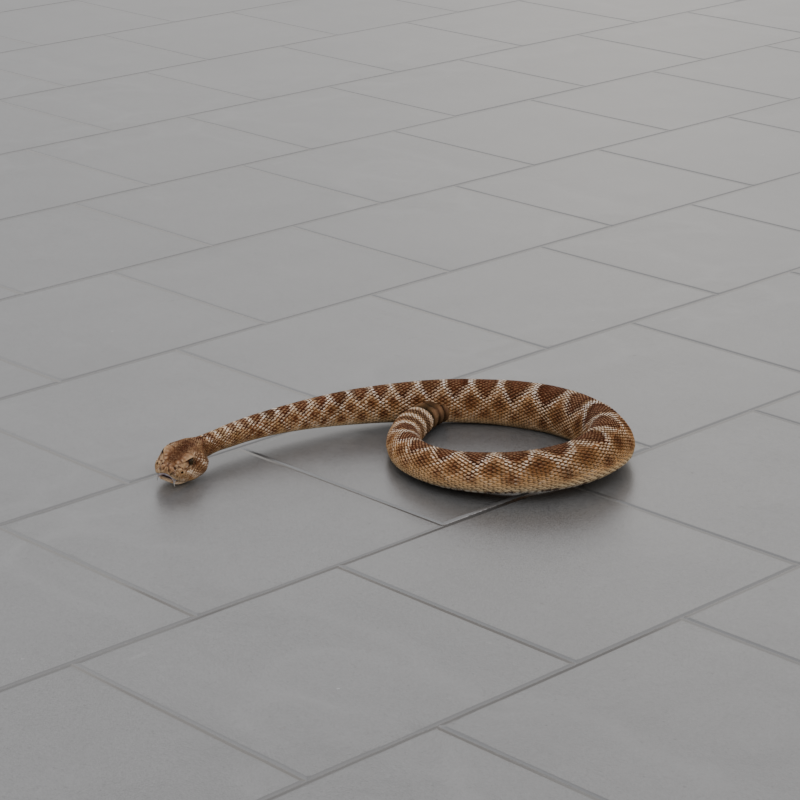}\hspace{2pt}%
\includegraphics[width=0.16\linewidth]{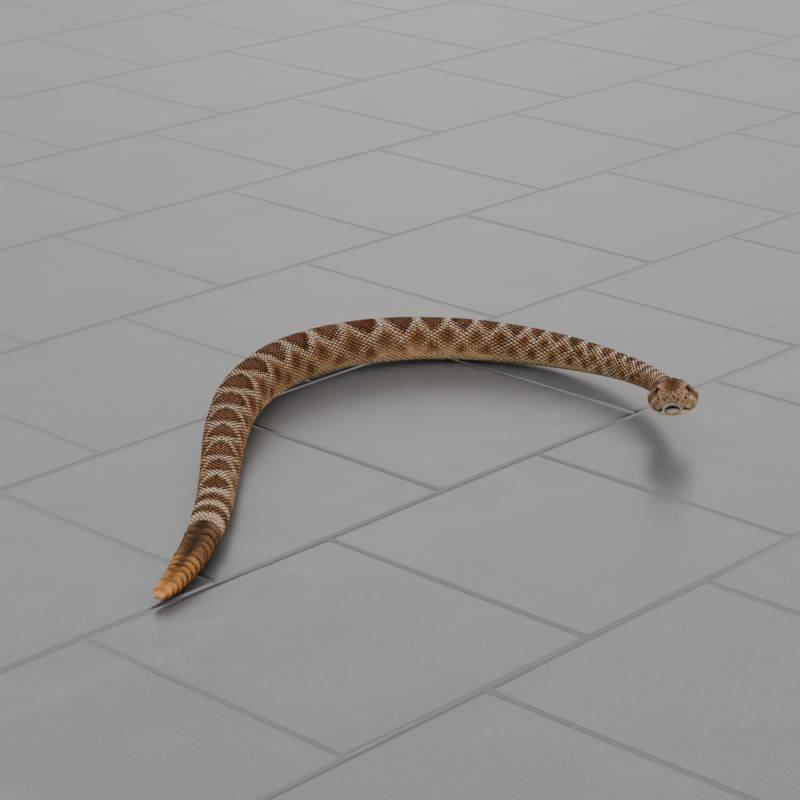}\hspace{2pt}%
\includegraphics[width=0.16\linewidth]{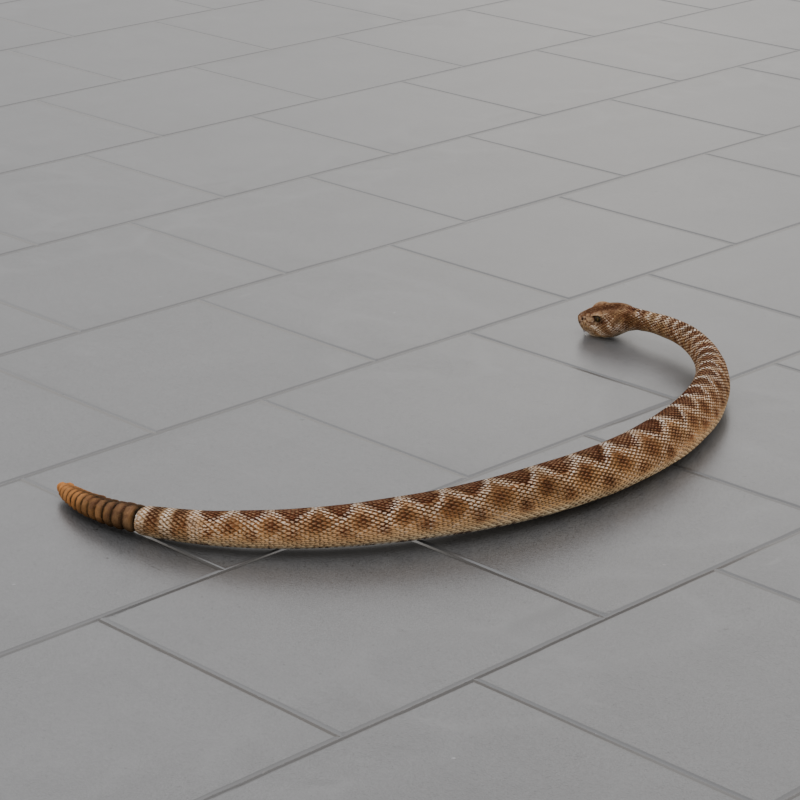}\hspace{2pt}%
\includegraphics[width=0.16\linewidth]{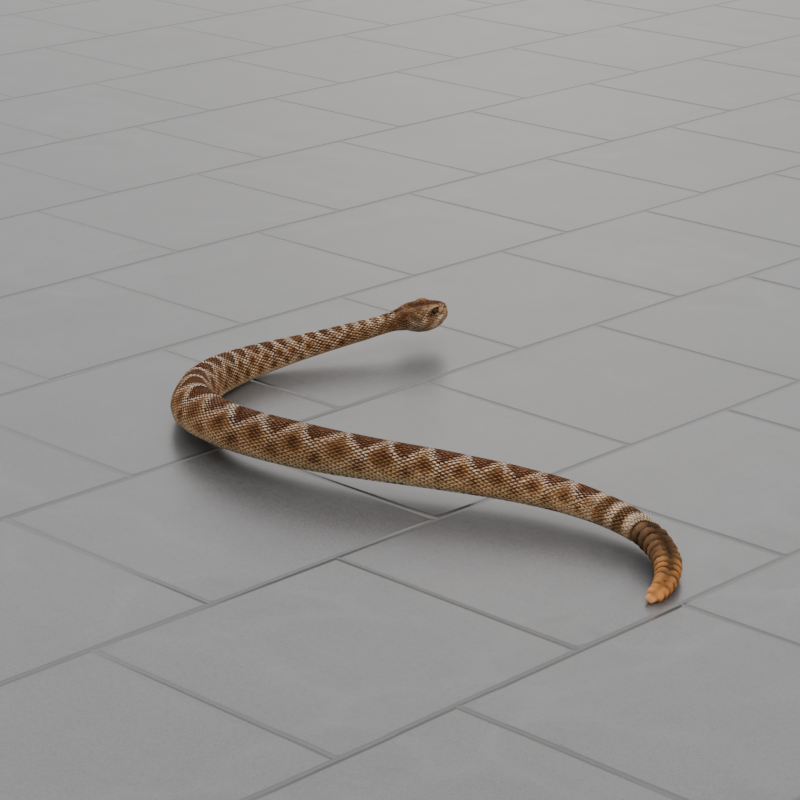}
\end{minipage}

\caption{Temporal evolution of motion, corresponding to an FDSA-optimized gait with friction ratios $\mu_n/\mu_f = 1$ and $\mu_b/\mu_f = 5$ with uniform mass distribution and no bending energy term in the cost function. 
}
\label{fig:Img3}
\end{figure*}



As previously defined, the curvature $\kappa(s,t)$ and torsion $\tau(s,t)$ are expressed as  Chebyshev–Fourier series, which collectively represent both spatial and temporal variations in the body shape. The coefficients of these expansions define the magnitude and phase of each deformation mode. Let's rearrange all these coefficients into a single parameter vector as
\begin{equation}
    \boldsymbol{\theta} =\left[A_{jk}, B_{jk}, C_{jk}, D_{jk}\right]^\top
    \label{eq:theta}
\end{equation}
which completely characterizes the body's intrinsic configuration. Each component of this vector represents the amplitude of one spatial–temporal mode contributing to the curvature or torsion field. The optimization aims to determine the specific set of coefficients $\boldsymbol{\theta}^{\ast}$ that yields the most energy-efficient gait while maintaining mechanical equilibrium and physical realism.

For each candidate parameter vector $\boldsymbol{\theta}$, the forward locomotion model computes the average mechanical power defined in Equation~\eqref{eq:cost_power}. In addition to minimizing energetic expenditure, an effective motion must also remain directionally stable. To achieve that, a penalty for rotation, $\displaystyle \delta \boldsymbol{\zeta}$, is imposed, which represents the net change in body orientation. A physically realistic and biologically meaningful gait should not contain any twist or self-intersection. Thus, the bending and torsion energies from Equations~\eqref{eq:bend_energy} \& ~\eqref{eq:torsion_energy} are added as a regularizer in the cost function. Incorporating everything, the optimization problem is formulated through the cost function
\begin{equation}
    \mathcal{L}_T\left(\boldsymbol{\theta}\right) = \mathcal{L}_P  + \lambda_1 \left(\delta \boldsymbol{\zeta}\right)^2 + \lambda_2 \mathcal{E}_{I}, \;\;  \mathcal{E}_{I} = \mathcal{E}_{\tau} + \mathcal{E}_{\kappa}
    \label{eq:Cost}
\end{equation}
where $\displaystyle \lambda_1, \, \lambda_2$ are Lagrange multipliers. The cost $\displaystyle \mathcal{L}_T\left(\boldsymbol{\theta}\right)$ is minimized to determine the optimized vector $\displaystyle \boldsymbol{\theta}^{\ast}$.

The coefficients that define the curvature and torsion must remain within prescribed bounds to preserve physical validity and prevent numerical instability. Excessive amplitudes in these coefficients would correspond to unrealistically sharp bends or twists of the body. To restrict the search space to feasible motions, each coefficient satisfies the constraint
\begin{equation}
    |A_{jk}|, |B_{jk}| \leq \kappa_{\max}, \quad |C_{jk}|, |D_{jk}| \leq \tau_{\mathrm{max}}w_j
\end{equation}
where $\displaystyle \kappa_{\max}$ and $\displaystyle \tau_{\mathrm{max}}$ denote the maximum allowable curvature and torsion amplitudes, respectively. The term $w_j = 1/j$ is a frequency-dependent weighting factor that suppresses higher-frequency oscillations, ensuring that the resulting deformations remain smooth. A projection operator, $\displaystyle \Pi(\boldsymbol{\theta})$, is applied after each update to enforce these bounds, guaranteeing that all parameters remain within the admissible region of the search space. The total cost, $\mathcal{L}_T\left(\boldsymbol{\theta}\right)$, is obtained from a non-linear simulation. The analytical derivative with respect to the parameters is extremely complex. We employ two numerical methods to estimate the gradient: Finite-Difference Stochastic Approximation (FDSA) and Simultaneous Perturbation Stochastic Approximation (SPSA).

\section{Implementation}\label{sec:discrete}
In this section, we present our novel discretization strategies and corresponding implementation details. 
\subsection{Finite-Difference Stochastic Approximation}\label{subsec:FDSA}
In this approach, each parameter is perturbed slightly in the positive and negative directions while keeping the other parameters fixed, and the change in cost is used to estimate the gradient
\begin{equation}
    \frac{\partial \mathcal{L}_T}{\partial \boldsymbol{\theta}} \approx \frac{\mathcal{L}_T\left(\boldsymbol{\theta} + \boldsymbol{\delta} \otimes \hat{\boldsymbol{\theta}}\right) - \mathcal{L}_T\left(\boldsymbol{\theta} - \boldsymbol{\delta} \otimes \hat{\boldsymbol{\theta}}\right)}{2\boldsymbol{\delta} \otimes \hat{\boldsymbol{\theta}}}
\end{equation}
If the size of $\displaystyle \boldsymbol{\theta}$ is $\displaystyle p$, then using the FDSA method, the cost function $\displaystyle \mathcal{L}_T$ needs to be perturbed twice for each of the $\displaystyle p$ parameters, requiring a total of $\displaystyle 2p$ evaluations per iteration.


\begin{figure*}[tbp]
\centering

\vspace{5pt}

\noindent
\begin{minipage}[t]{0.14\textwidth}
\vspace{10pt}
\raggedright
\textbf{Case 1:}\\
{$\max(A_{jk}) = 1$}\\
{$\max(B_{jk}) = 1$}\\
{$\max(C_{jk}) = 0.1$}\\
{$\max(D_{jk}) = 0.1$}\\
\end{minipage}%
\hfill
\begin{minipage}[t]{0.85\textwidth}
\vspace{2pt}
\centering
\includegraphics[width=0.16\linewidth]{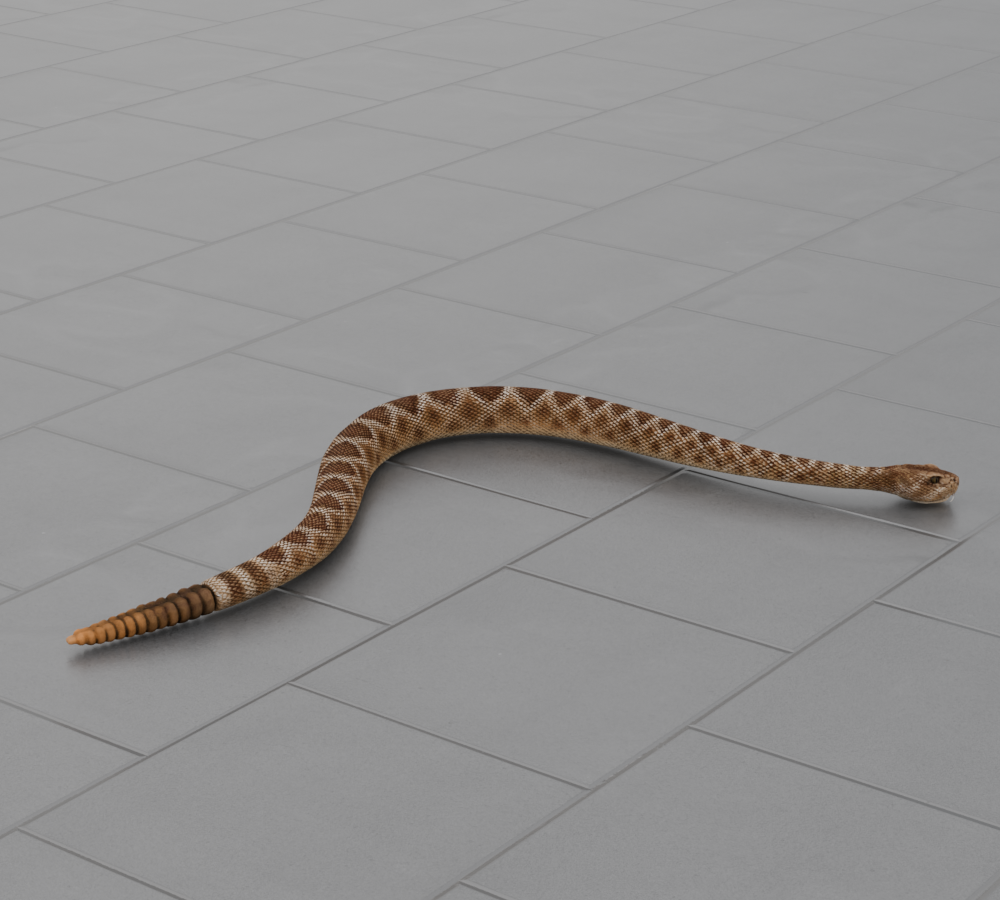}\hspace{2pt}%
\includegraphics[width=0.16\linewidth]{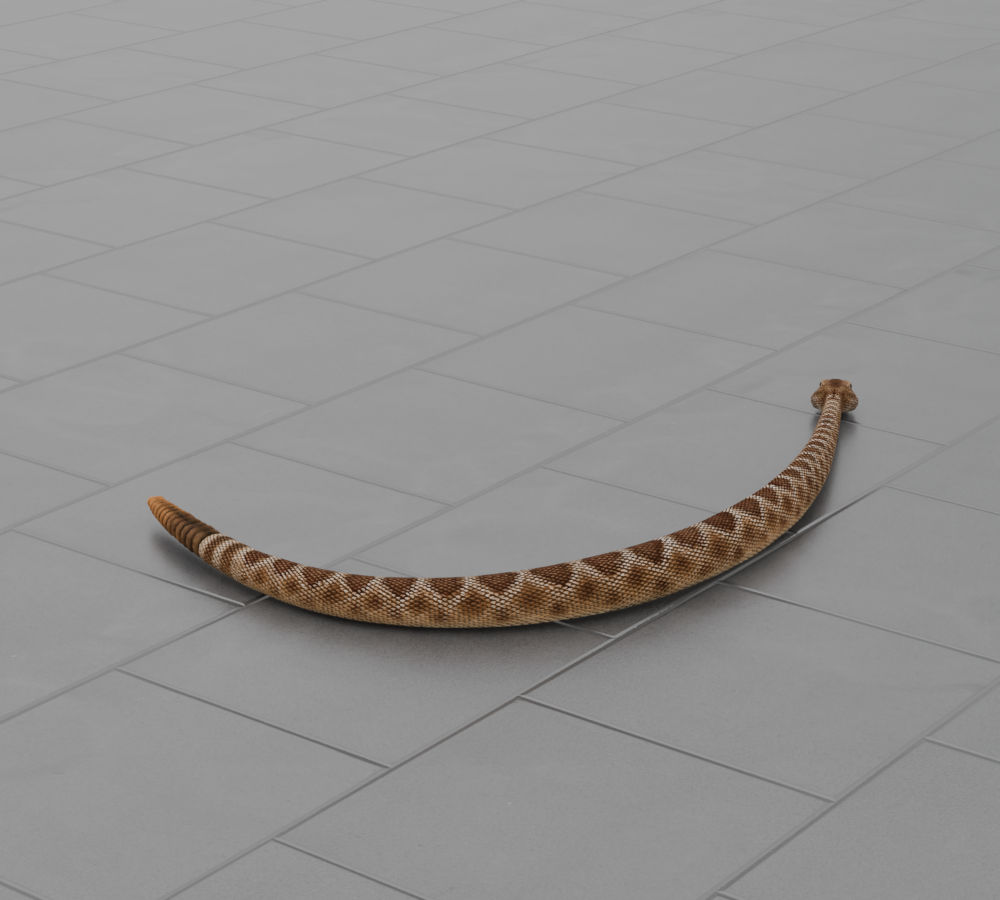}\hspace{2pt}%
\includegraphics[width=0.16\linewidth]{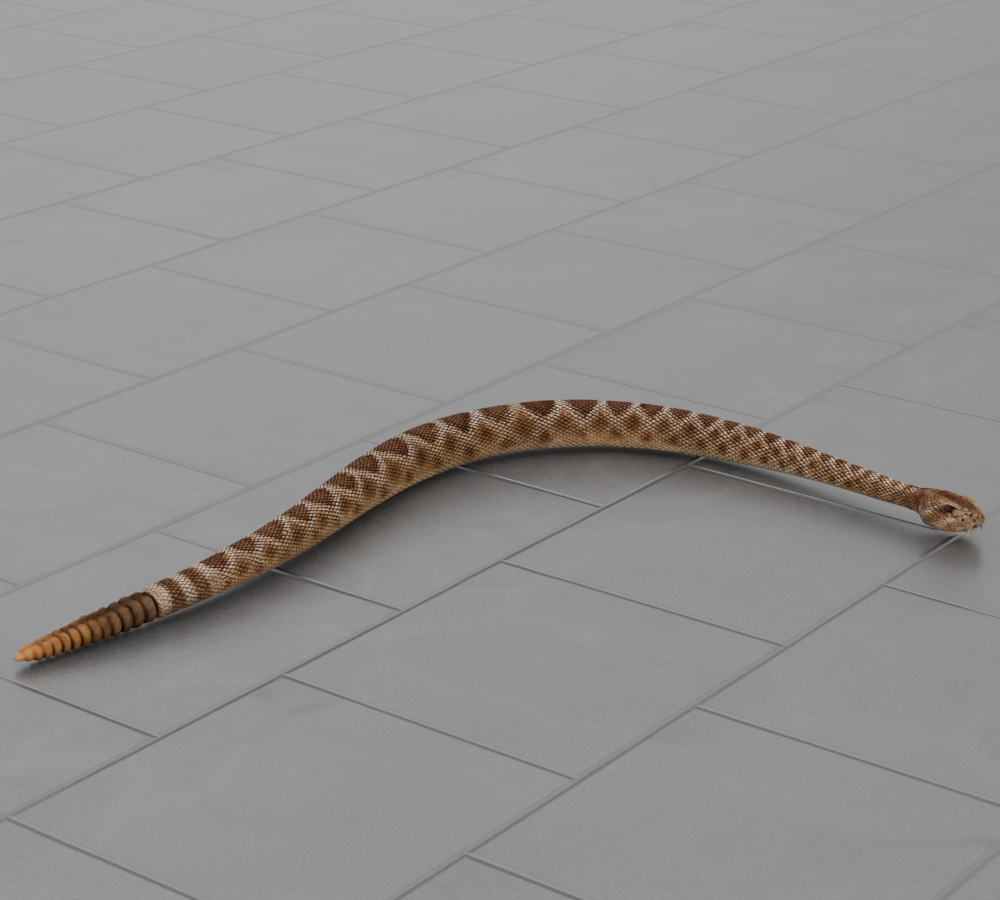}\hspace{2pt}%
\includegraphics[width=0.16\linewidth]{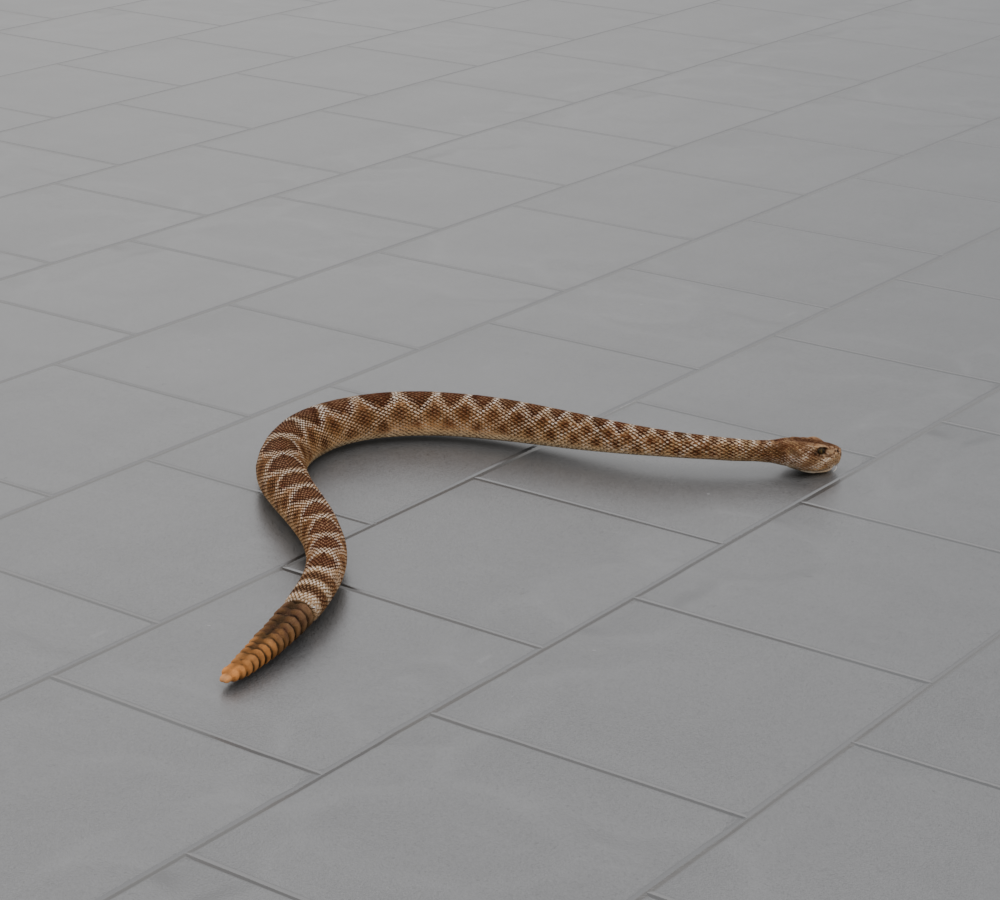}\hspace{2pt}%
\includegraphics[width=0.16\linewidth]{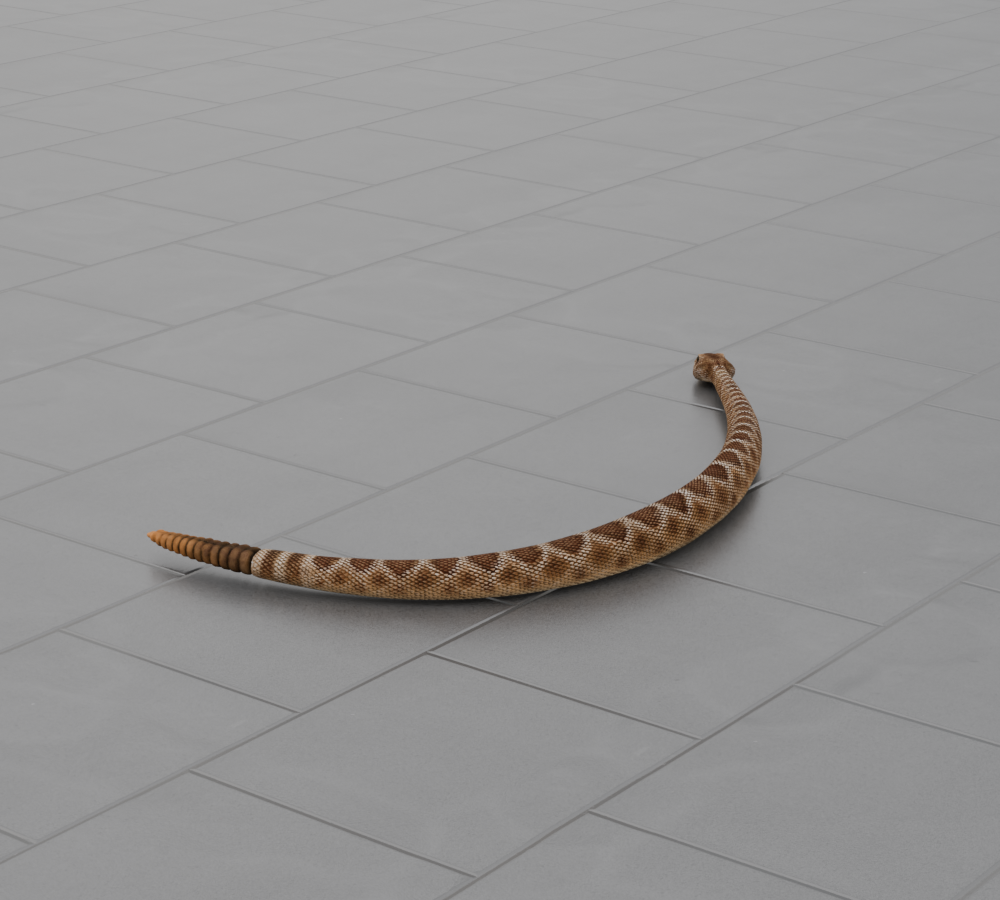}\hspace{2pt}%
\includegraphics[width=0.16\linewidth]{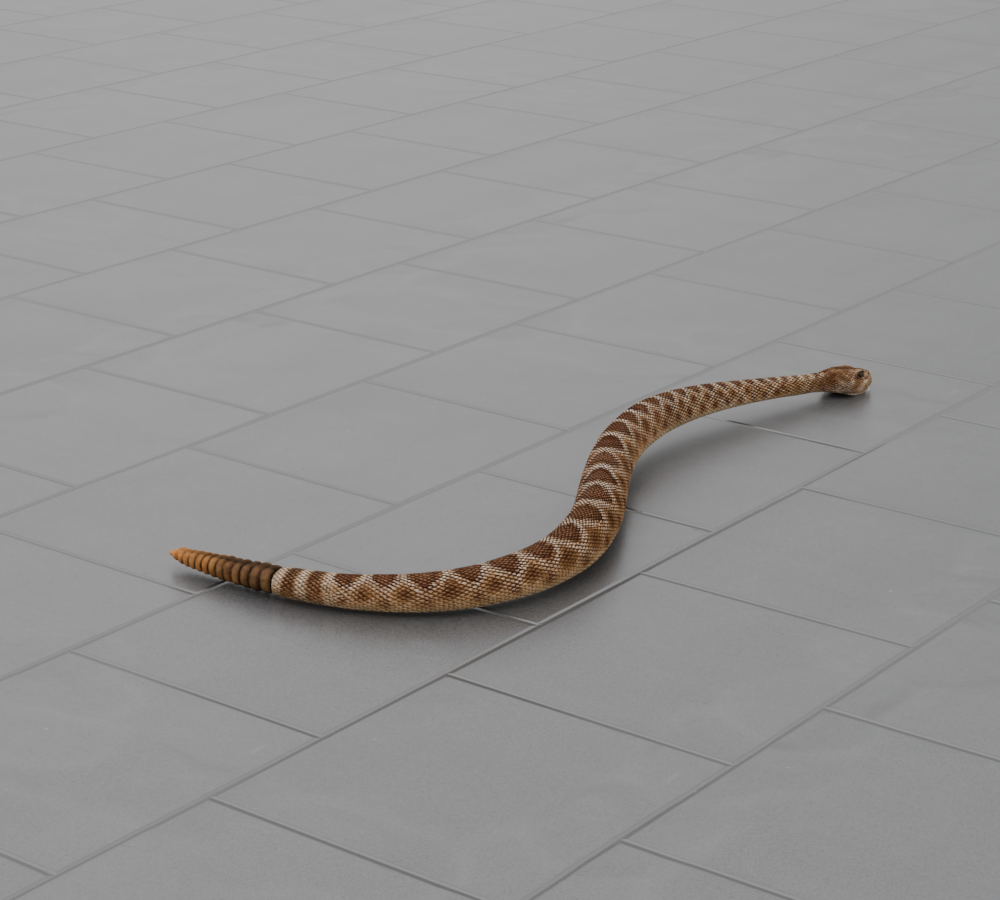}
\end{minipage}

\noindent
\begin{minipage}[t]{0.14\textwidth}
\vspace{10pt}
\raggedright
\textbf{Case 2:}\\
{$\max(A_{jk}) = 2$}\\
{$\max(B_{jk}) = 2$}\\
{$\max(C_{jk}) = 0.2$}\\
{$\max(D_{jk}) = 0.2$}\\
\end{minipage}%
\hfill
\begin{minipage}[t]{0.85\textwidth}
\vspace{2pt}
\centering
\includegraphics[width=0.16\linewidth]{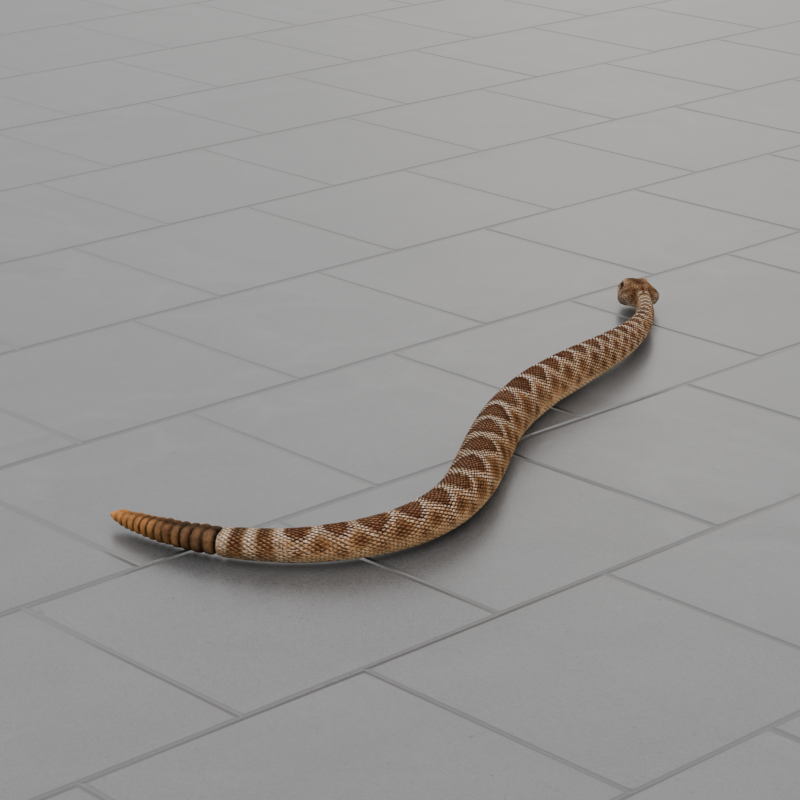}\hspace{2pt}%
\includegraphics[width=0.16\linewidth]{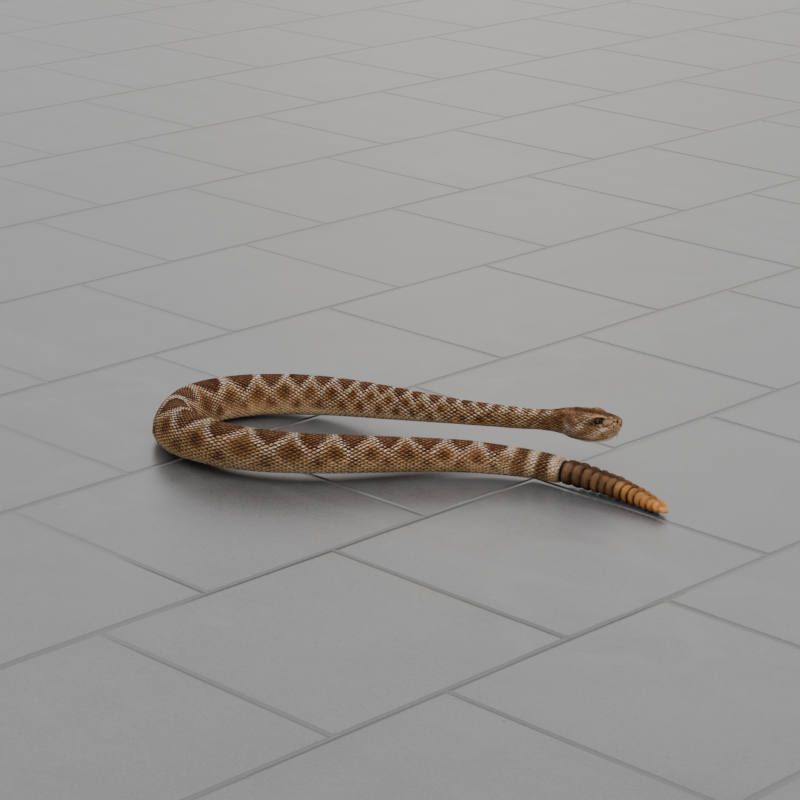}\hspace{2pt}%
\includegraphics[width=0.16\linewidth]{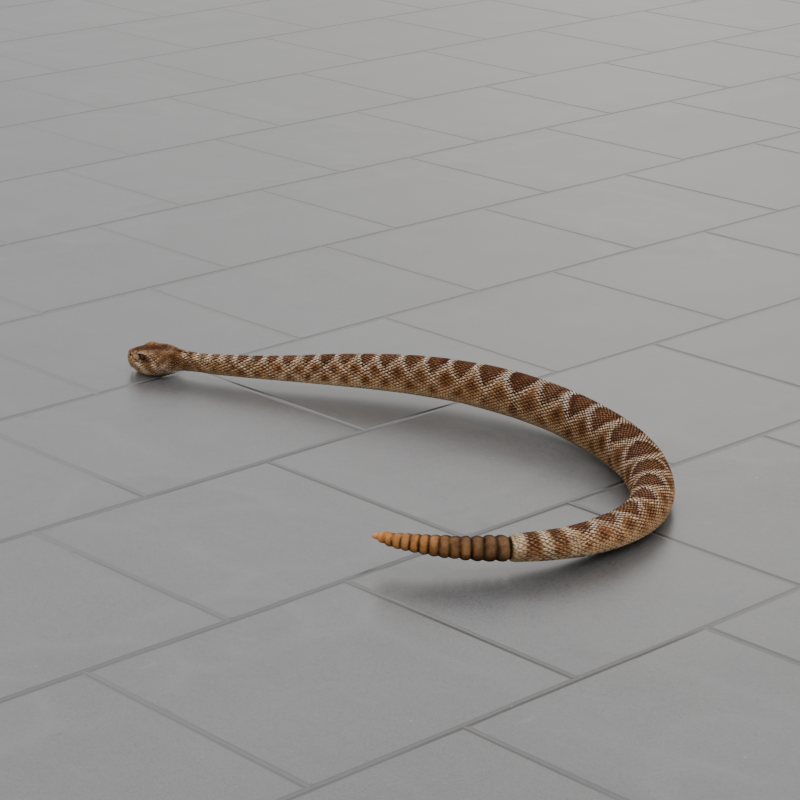}\hspace{2pt}%
\includegraphics[width=0.16\linewidth]{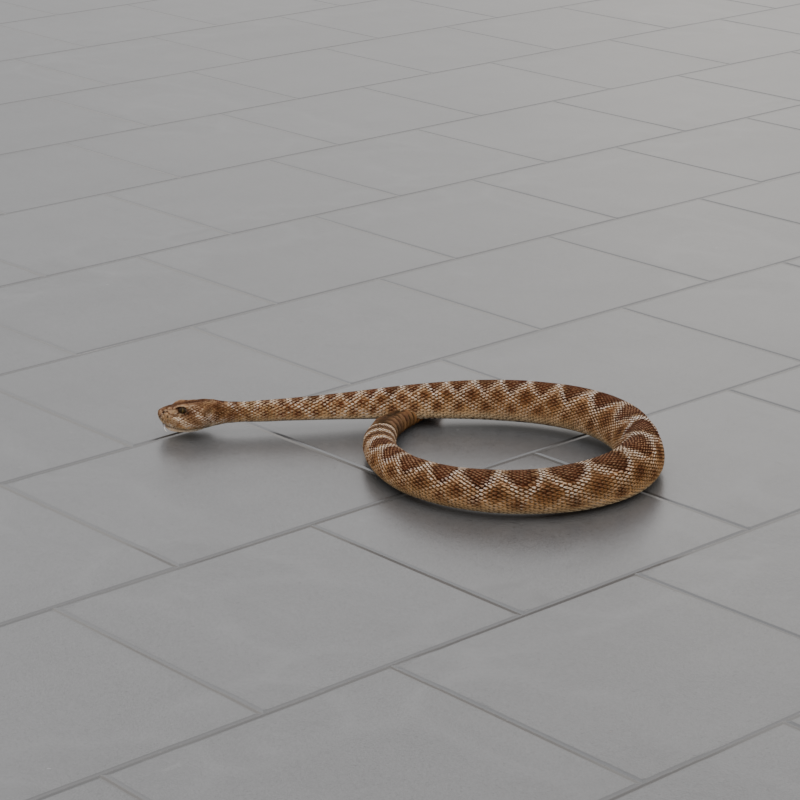}\hspace{2pt}%
\includegraphics[width=0.16\linewidth]{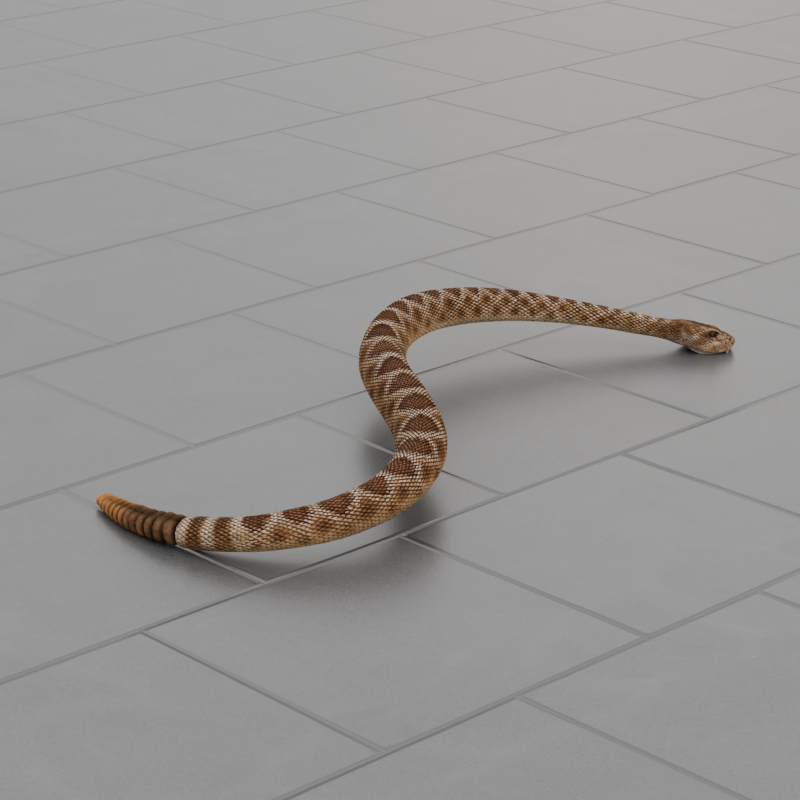}\hspace{2pt}%
\includegraphics[width=0.16\linewidth]{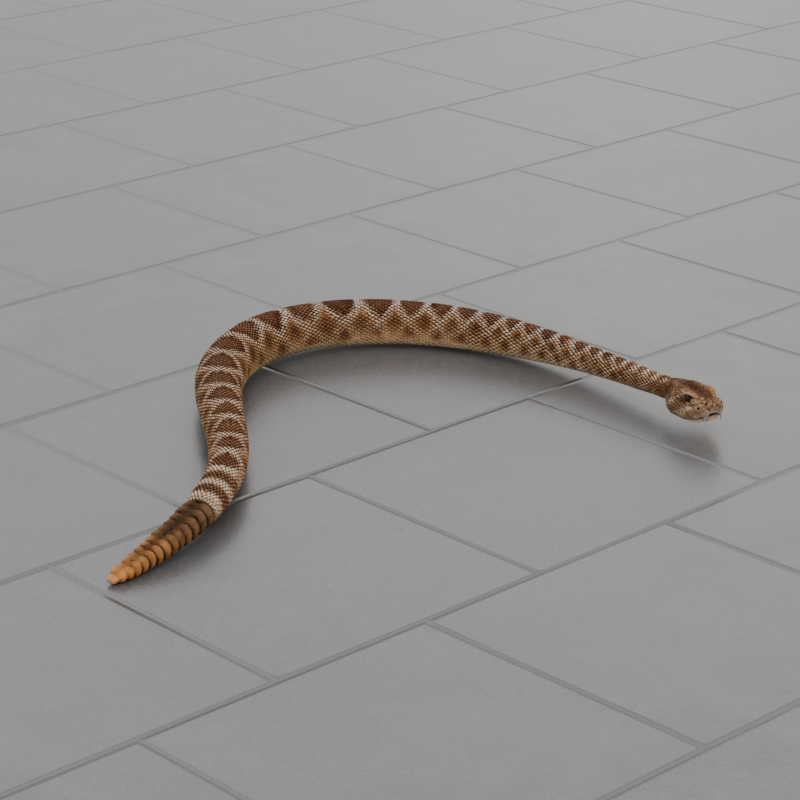}
\end{minipage}

\caption{Temporal evolution of motion, corresponding to an FDSA-optimized gait with friction ratios $\mu_n/\mu_f = 1$ and $\mu_b/\mu_f = 5$, a non-uniform mass distribution, and no bending energy term in the cost function. 
The non-uniform mass is generated from a quadratic profile $f(x) = x^2$, sampled over the interval $x \in [1,\,1.2]$, and subsequently normalized by interpolating the resulting values to lie within $[1,\,1.05]$, ensuring variations in mass along the body.}
\label{fig:Img4}
\end{figure*}


\subsection{Simultaneous Perturbation Stochastic Approximation}\label{subsec:SPSA}
In this faster method, a random perturbation vector $\displaystyle \delta_{\text{fix}} = \pm 1$ is used to estimate all gradient components at once
\begin{equation}
    \frac{\partial \mathcal{L}_T}{\partial \boldsymbol{\theta}} \approx \frac{\mathcal{L}_T(\boldsymbol{\theta}^{+}) - \mathcal{L}_T(\boldsymbol{\theta}^{-})}{2\mathbf{c}\delta_{\text{fix}}}.
\end{equation}
where $\displaystyle \boldsymbol{\theta}^{+} = \boldsymbol{\theta} + \delta_{\text{fix}}\mathbf{c}$ and  $\displaystyle \boldsymbol{\theta}^{-} = \boldsymbol{\theta} - \delta_{\text{fix}}\mathbf{c}$. Note that as $\displaystyle \delta_{\text{fix}}\mathbf{c}$ is a random perturbation vector, all the components of the cost function $\displaystyle \mathcal{L}_T(\boldsymbol{\theta})$ are perturbed equally in all directions. Unlike FDSA, which requires $\displaystyle 2p$ evaluations of $\displaystyle \mathcal{L}_T$ in each step, SPSA is computationally less expensive because it requires only two simulations per iteration.

The complete gradient vector $\nabla \mathcal{L}_T = \left[\frac{\partial \mathcal{L}_T}{\partial \theta_1}, \frac{\partial \mathcal{L}_T}{\partial \theta_2}, \ldots \frac{\partial \mathcal{L}_T}{\partial \theta_n}\right]^\top$ is normalized to prevent excessively large updates. The normalization is expressed as
\begin{equation}
    \nabla \hat{\mathcal{L}}_T = \frac{\nabla \mathcal{L}_T}{\|\nabla \mathcal{L}_T\| + \delta^2}
\end{equation}
where $\displaystyle \delta > 0$ is introduced to avoid division by zero when the gradient norm becomes small. This normalization ensures that the optimization proceeds according to the direction of steepest descent, while the learning rate independently governs the magnitude of each step.


\begin{figure*}[tbp]
\centering

\vspace{5pt}

\noindent
\begin{minipage}[t]{0.14\textwidth}
\vspace{10pt}
\raggedright
\textbf{Case 1:}\\
{$\max(A_{jk}) = 1$}\\
{$\max(B_{jk}) = 1$}\\
{$\max(C_{jk}) = 0.1$}\\
{$\max(D_{jk}) = 0.1$}\\
\end{minipage}%
\hfill
\begin{minipage}[t]{0.85\textwidth}
\vspace{2pt}
\centering
\includegraphics[width=0.16\linewidth]{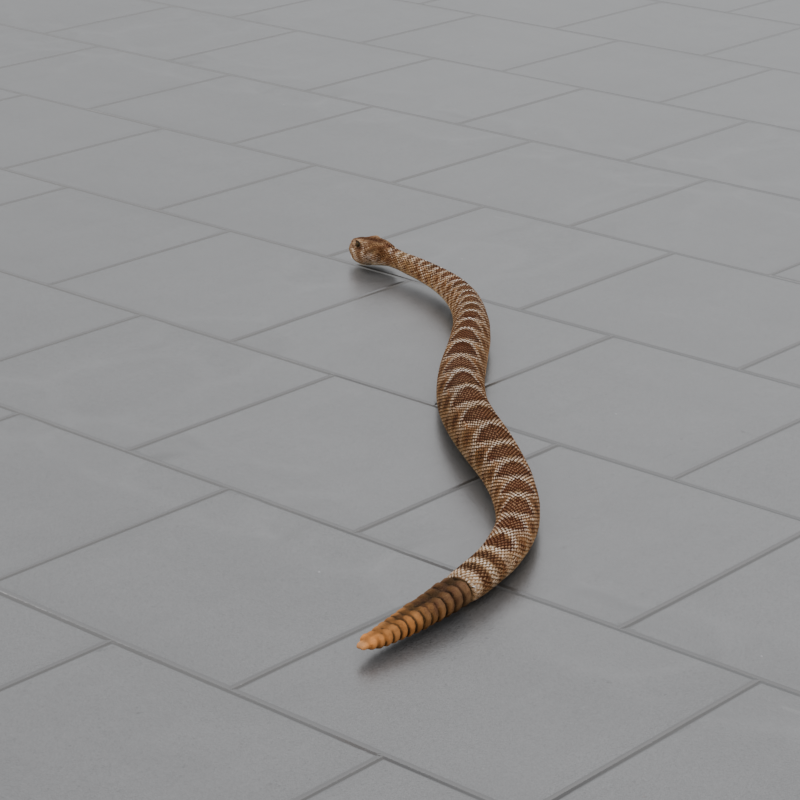}\hspace{2pt}%
\includegraphics[width=0.16\linewidth]{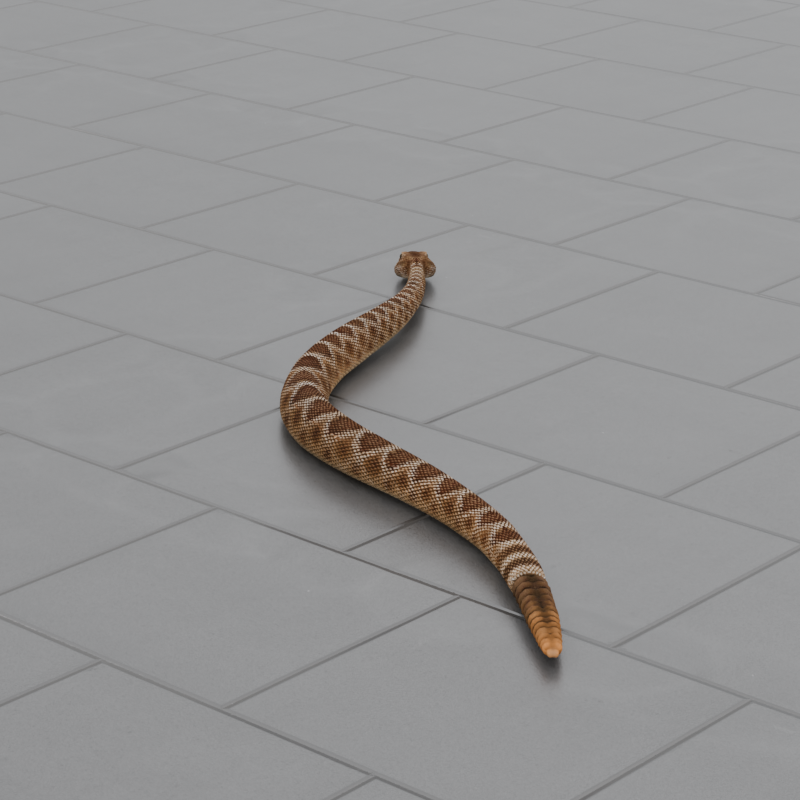}\hspace{2pt}%
\includegraphics[width=0.16\linewidth]{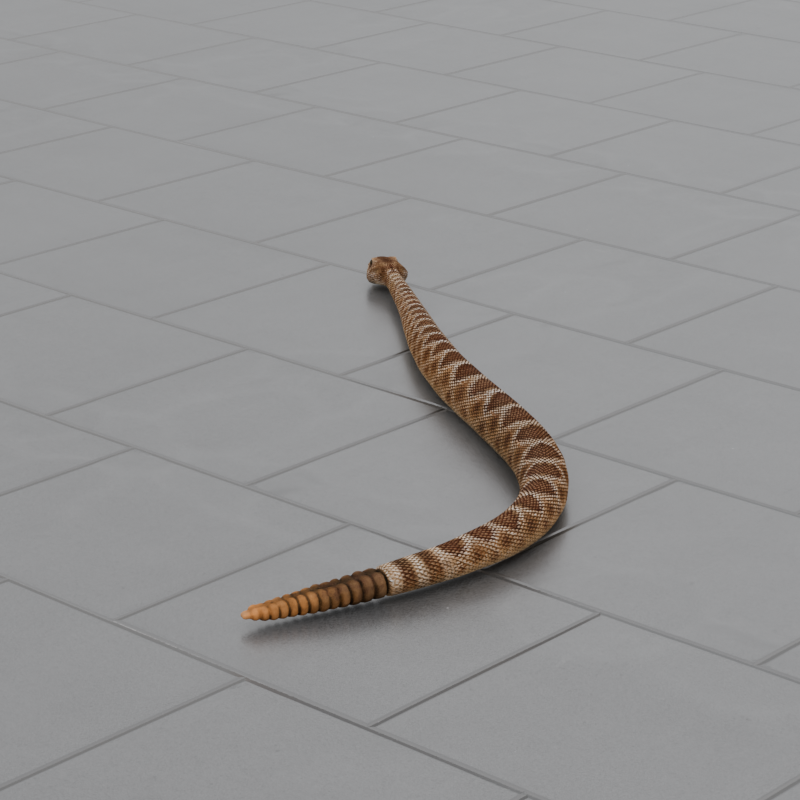}\hspace{2pt}%
\includegraphics[width=0.16\linewidth]{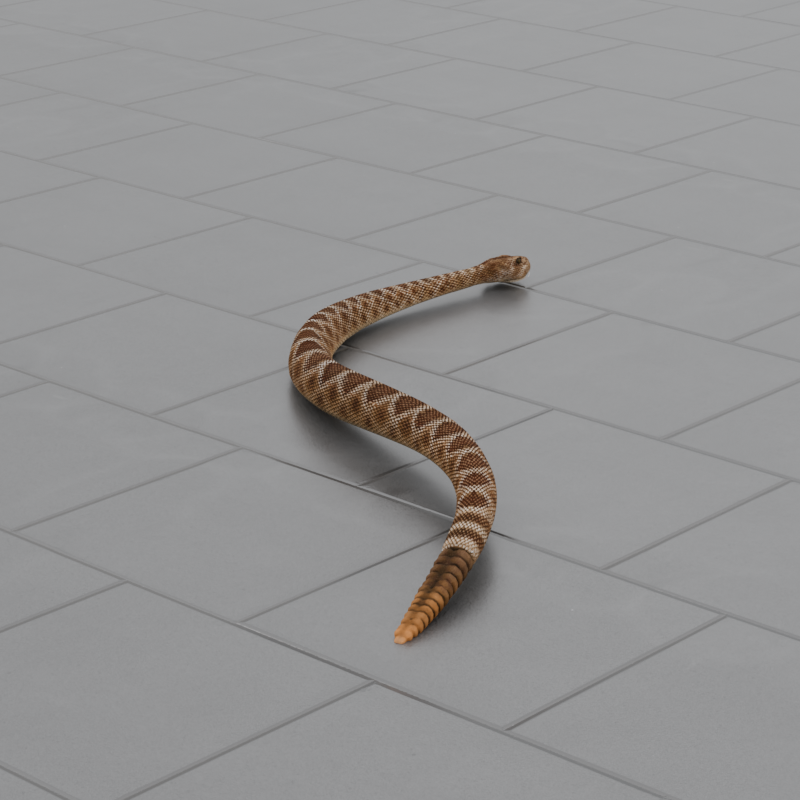}\hspace{2pt}%
\includegraphics[width=0.16\linewidth]{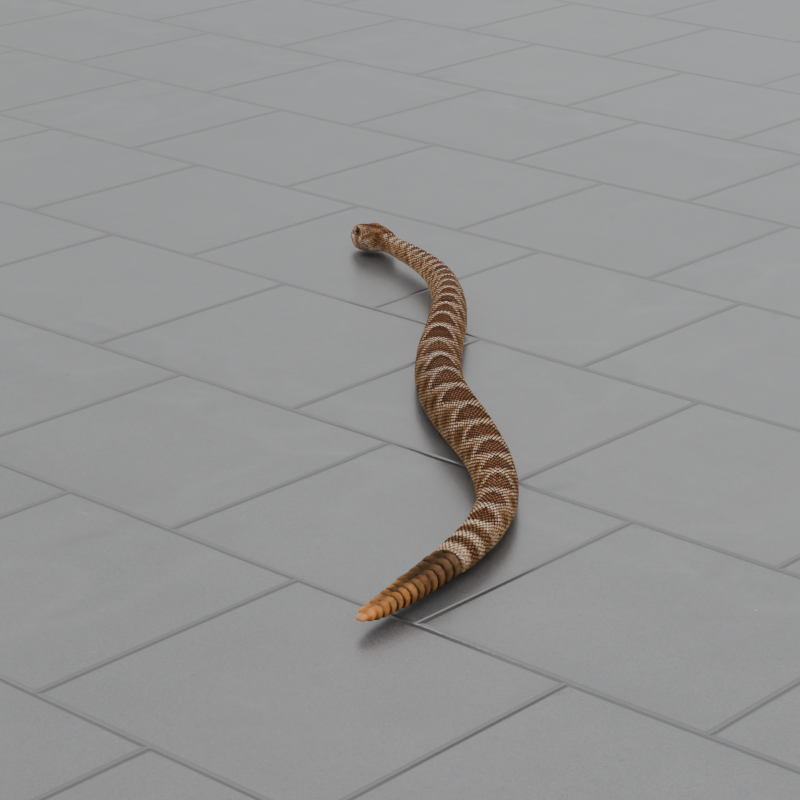}\hspace{2pt}%
\includegraphics[width=0.16\linewidth]{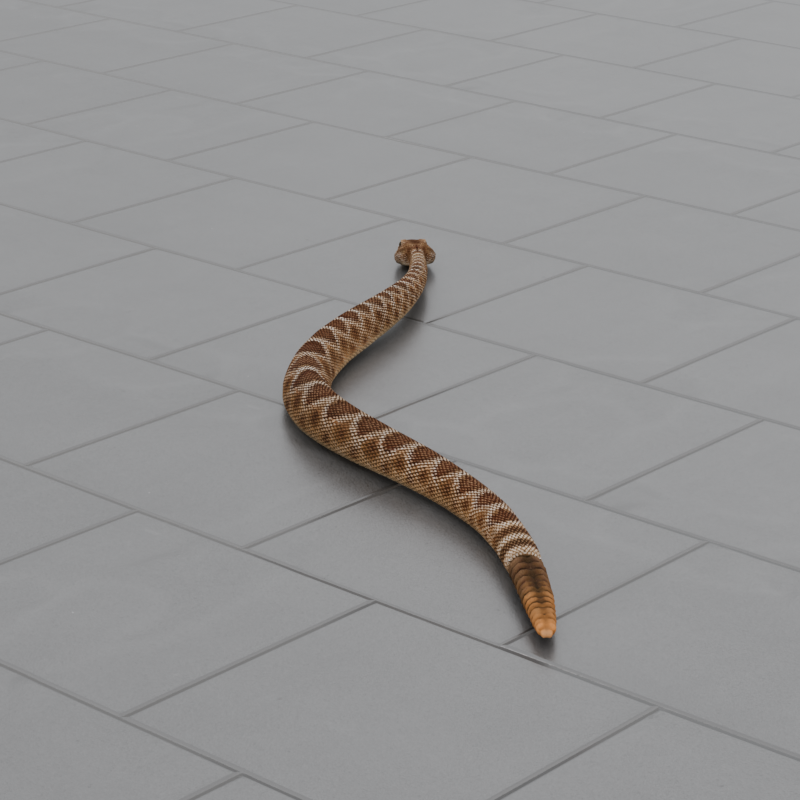}
\end{minipage}

\noindent
\begin{minipage}[t]{0.14\textwidth}
\vspace{10pt}
\raggedright
\textbf{Case 2:}\\
{$\max(A_{jk}) = 2$}\\
{$\max(B_{jk}) = 2$}\\
{$\max(C_{jk}) = 0.2$}\\
{$\max(D_{jk}) = 0.2$}\\
\end{minipage}%
\hfill
\begin{minipage}[t]{0.85\textwidth}
\vspace{2pt}
\centering
\includegraphics[width=0.16\linewidth]{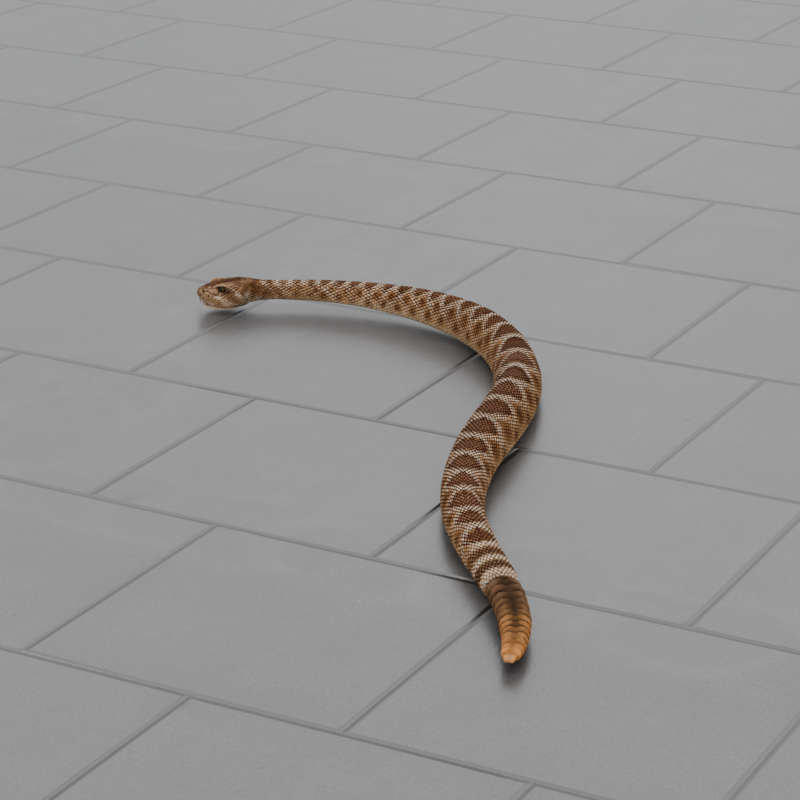}\hspace{2pt}%
\includegraphics[width=0.16\linewidth]{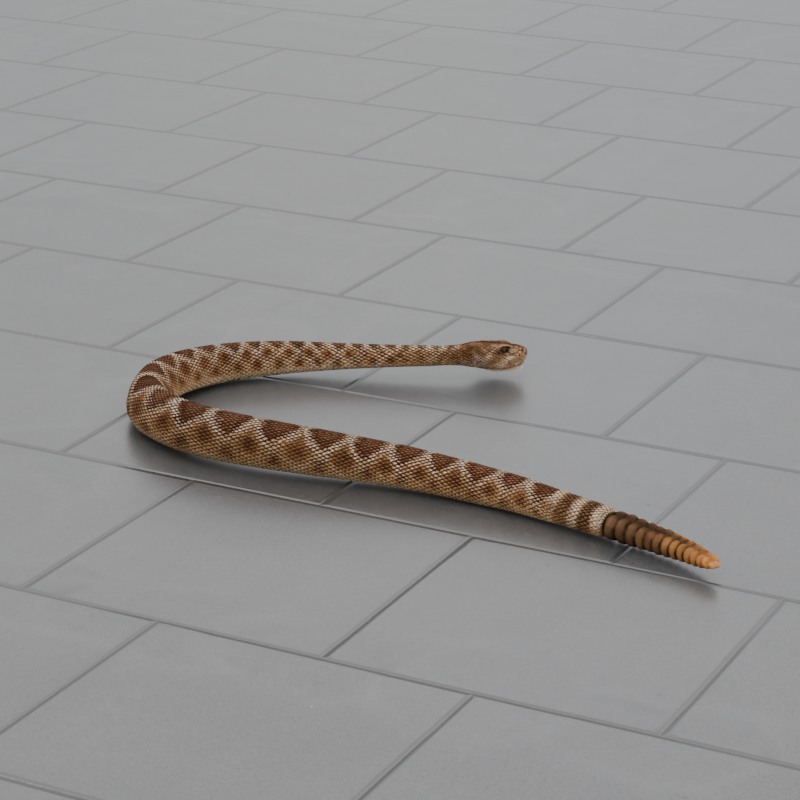}\hspace{2pt}%
\includegraphics[width=0.16\linewidth]{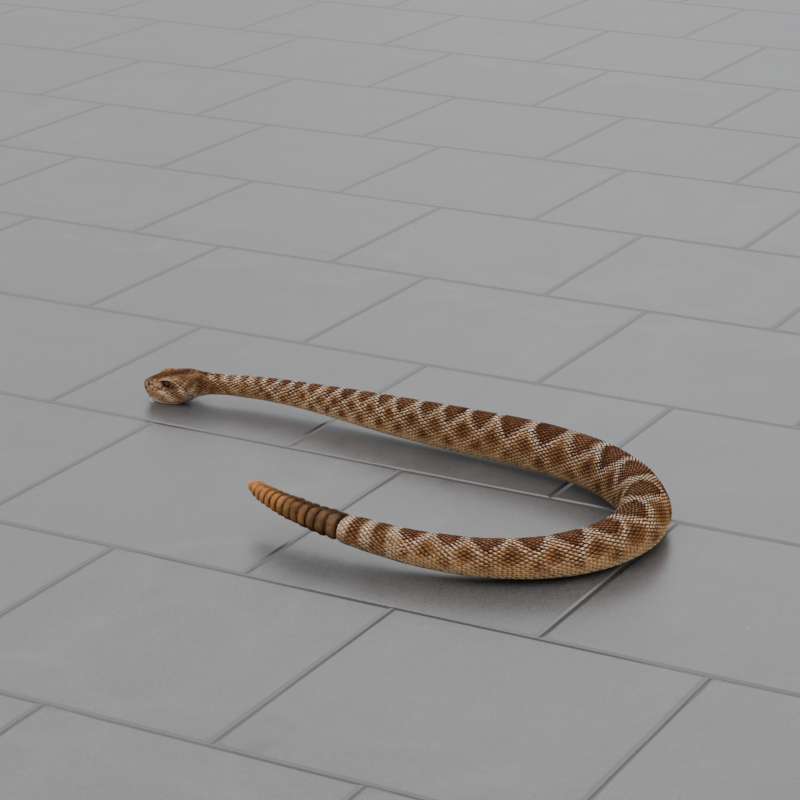}\hspace{2pt}%
\includegraphics[width=0.16\linewidth]{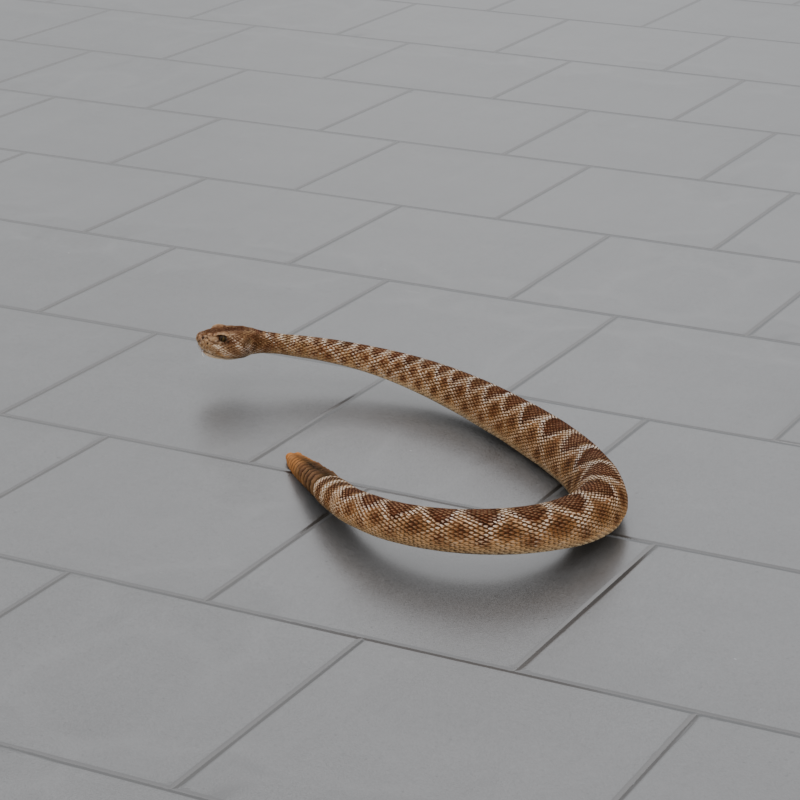}\hspace{2pt}%
\includegraphics[width=0.16\linewidth]{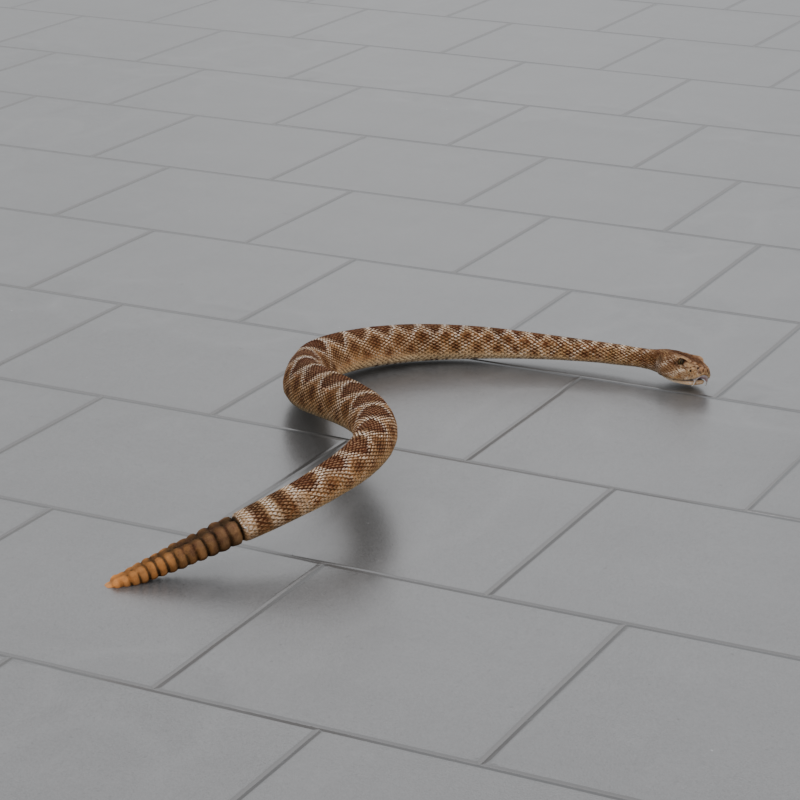}\hspace{2pt}%
\includegraphics[width=0.16\linewidth]{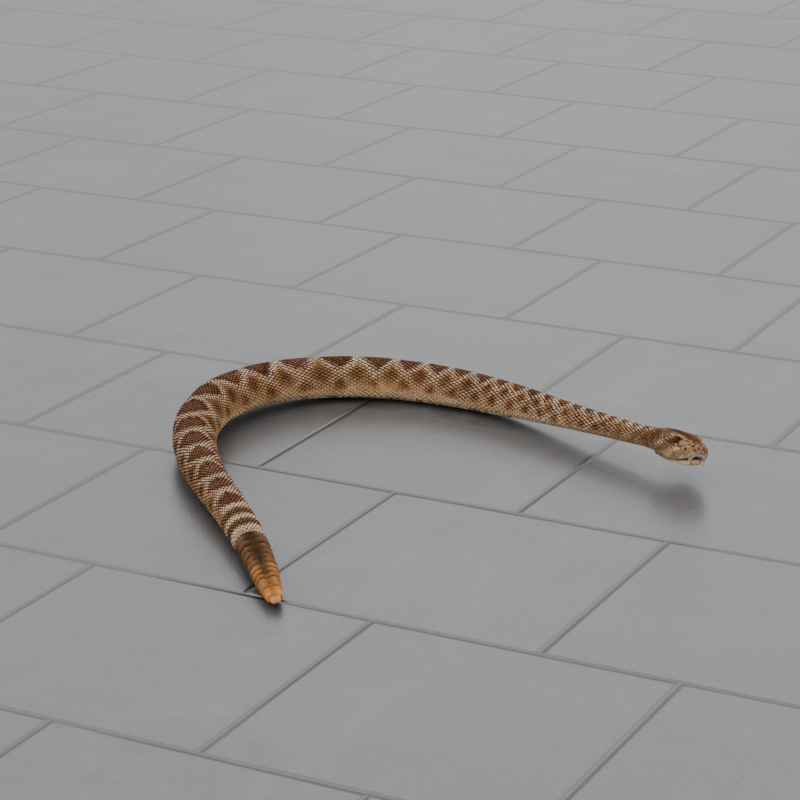}
\end{minipage}

\caption{Temporal evolution of motion, corresponding to an FDSA-optimized gait with friction ratios $\mu_n/\mu_f = 1$ and $\mu_b/\mu_f = 5$ with a non-uniform mass distribution, and the bending energy term included in the cost function. 
The non-uniform mass is constructed from a quadratic profile $f(x) = x^2$, sampled over $x \in [1,\,1.2]$, and rescaled to lie within $[1,\,1.05]$, resulting in variations in mass along the body.}
\label{fig:Img5}
\end{figure*}


\begin{figure*}[tbp]
\centering

\vspace{0pt}

\noindent
\begin{minipage}[t]{0.14\textwidth}
\vspace{17pt}
\raggedright
\textbf{Case 1:}\\
{Without bending energy term in cost function}
\end{minipage}%
\hfill
\begin{minipage}[t]{0.85\textwidth}
\vspace{2pt}
\centering
\includegraphics[width=0.16\linewidth]{Figures/k1.png}\hspace{2pt}%
\includegraphics[width=0.16\linewidth]{Figures/k2.png}\hspace{2pt}%
\includegraphics[width=0.16\linewidth]{Figures/k3.png}\hspace{2pt}%
\includegraphics[width=0.16\linewidth]{Figures/k4.png}\hspace{2pt}%
\includegraphics[width=0.16\linewidth]{Figures/k5.png}\hspace{2pt}%
\includegraphics[width=0.16\linewidth]{Figures/k6.png}
\end{minipage}

\noindent
\begin{minipage}[t]{0.14\textwidth}
\vspace{17pt}
\raggedright
\textbf{Case 2:}\\
{Without torsion energy term in cost function}
\end{minipage}%
\hfill
\begin{minipage}[t]{0.85\textwidth}
\vspace{2pt}
\centering
\includegraphics[width=0.16\linewidth]{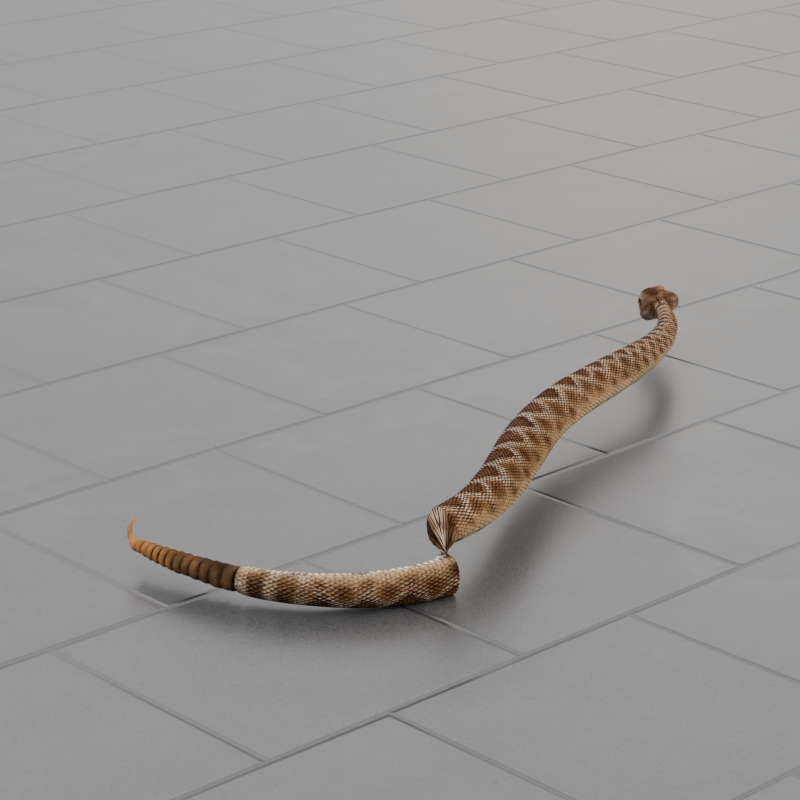}\hspace{2pt}%
\includegraphics[width=0.16\linewidth]{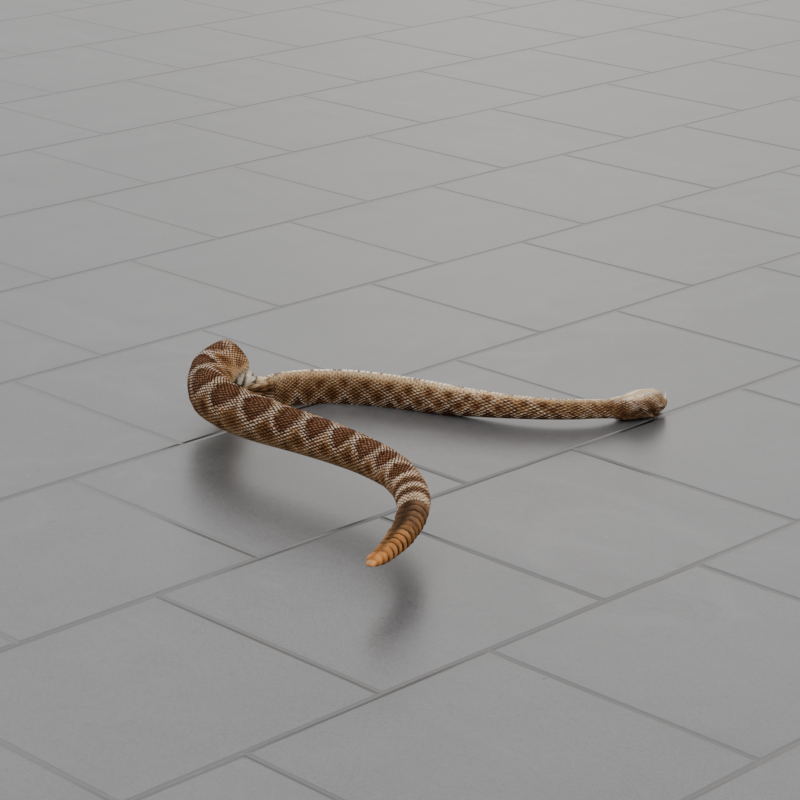}\hspace{2pt}%
\includegraphics[width=0.16\linewidth]{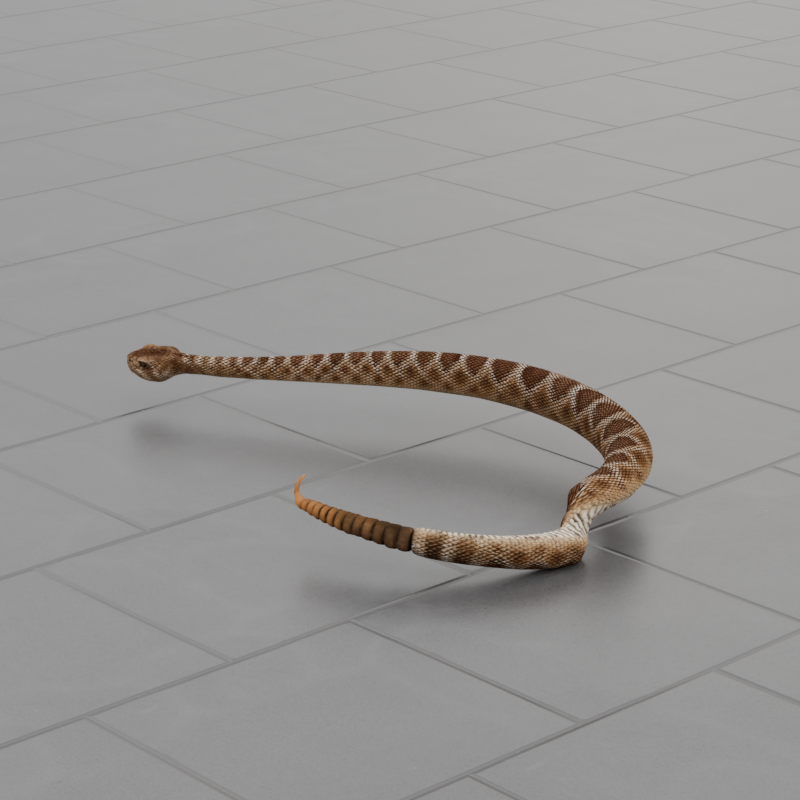}\hspace{2pt}%
\includegraphics[width=0.16\linewidth]{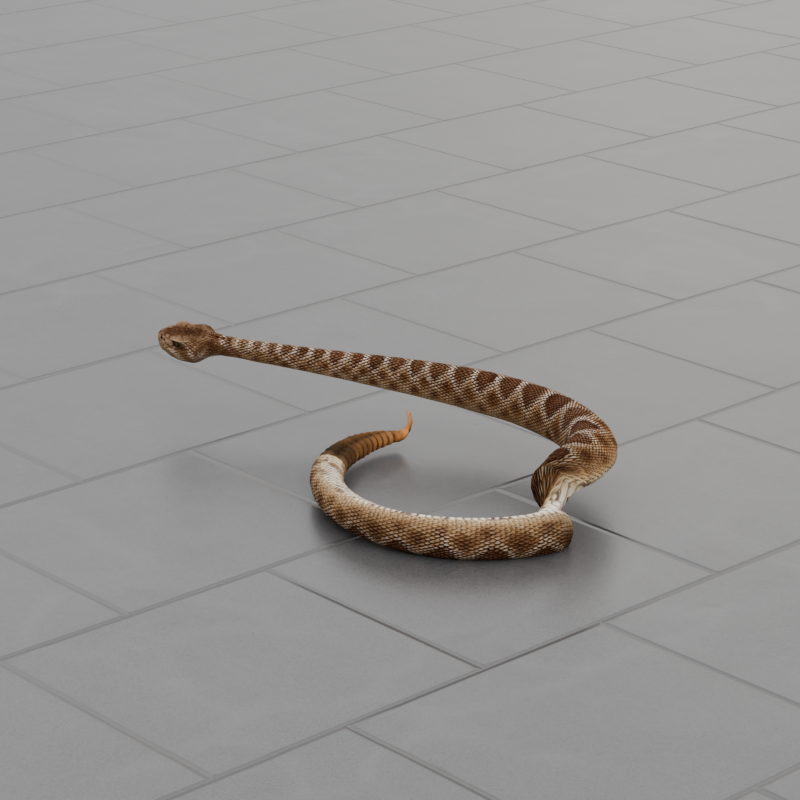}\hspace{2pt}%
\includegraphics[width=0.16\linewidth]{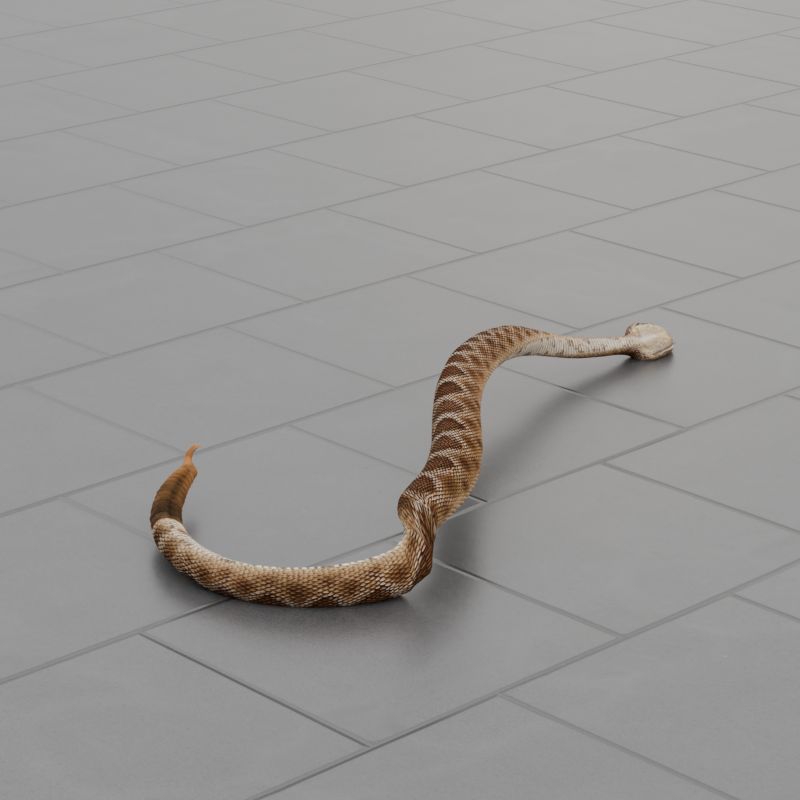}\hspace{2pt}%
\includegraphics[width=0.16\linewidth]{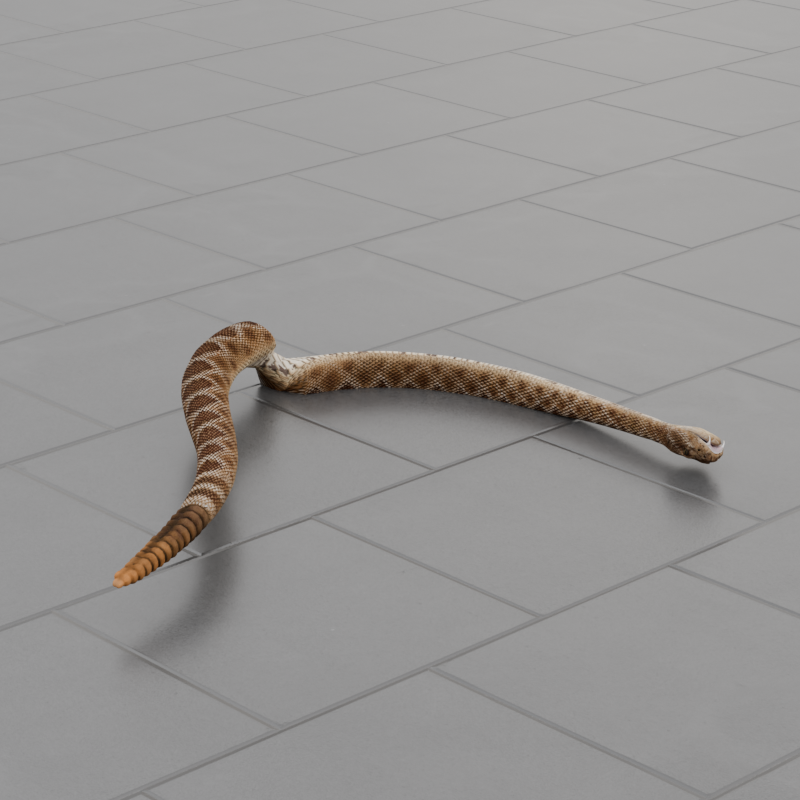}
\end{minipage}

\noindent
\begin{minipage}[t]{0.14\textwidth}
\vspace{17pt}
\raggedright
\textbf{Case 3:}\\
{With bending and torsion energy terms in cost function}
\end{minipage}%
\hfill
\begin{minipage}[t]{0.85\textwidth}
\vspace{2pt}
\centering
\includegraphics[width=0.16\linewidth]{Figures/m1.png}\hspace{2pt}%
\includegraphics[width=0.16\linewidth]{Figures/m2.png}\hspace{2pt}%
\includegraphics[width=0.16\linewidth]{Figures/m3.png}\hspace{2pt}%
\includegraphics[width=0.16\linewidth]{Figures/m4.png}\hspace{2pt}%
\includegraphics[width=0.16\linewidth]{Figures/m5.png}\hspace{2pt}%
\includegraphics[width=0.16\linewidth]{Figures/m6.png}
\end{minipage}

\caption{Temporal evolution of motion, corresponding to $\mu_n/\mu_f = 1,\ \mu_b/\mu_f = 5$, $\max(A_{jk}) = \max(B_{jk}) = 2$, $\max(C_{jk}) = \max(D_{jk}) = 0.2$, with a non-uniform mass distribution, using the FDSA optimizer. A clear self-intersection (Top Row) and a twist (Middle Row) are observed when the bending and torsion energy terms are not included, respectively; this artifact vanishes when the bending and torsion energy terms are incorporated (Bottom Row).}
\label{fig:Img6}
\end{figure*}


\subsection{Gradient Update Rule}\label{subsec:Gradient Update}
To update the parameters iteratively, a momentum-based formulation of gradient descent is adopted. This approach enhances convergence by accumulating information from previous steps, allowing the search to maintain inertia in beneficial directions while damping oscillations in regions where the cost landscape is irregular. The update equations are written as
\begin{equation}
\begin{aligned}
    \boldsymbol{v}^{k+1} &= \chi \boldsymbol{v}^{k} - \xi^k \nabla \widehat{\mathcal{L}_T}^{k}, \\
    \boldsymbol{\theta}^{k+1} &= \Pi\left(\boldsymbol{\theta}^{k} + \boldsymbol{v}^{k+1}\right)
\end{aligned}
\end{equation}
where $\displaystyle \boldsymbol{v}^{k}$ represents the accumulated velocity vector containing the exponentially weighted history of past gradients using $\displaystyle \chi \in [0,1)$, which is a the momentum coefficient, and $\displaystyle \xi^{k}$ is the learning rate at iteration $\displaystyle k$. The first equation combines the influence of the most recent gradient with the directional information from previous steps, and the second applies the resulting velocity to the parameters, ensuring that the projected vector remains within the feasible region defined by the amplitude limits. The inclusion of momentum accelerates convergence in long, narrow valleys of the cost surface and prevents oscillatory motion near minima.

The learning rate determines the distance moved in parameter space at each iteration. Because the curvature of the cost surface varies across regions, an adaptive strategy is employed to adjust the learning rate dynamically in response to the success or failure of the most recent step. The adaptive rule is given by

\begin{equation}
\xi^{k+1} =
\begin{cases}
    1.05\xi^{k}, & \mathcal{L}_T\left(\boldsymbol{\theta}^{k+1}\right) < \mathcal{L}_T\left(\boldsymbol{\theta}^{k}\right) \\
    0.5\xi^{k}, & \mathcal{L}_T\left(\boldsymbol{\theta}^{k+1}\right) \geq \mathcal{L}_T\left(\boldsymbol{\theta}^{k}\right)
\end{cases}
\end{equation}
where the learning rate is slightly increased when the cost decreases, allowing the search to progress more aggressively in promising regions, and is halved when the cost increases, providing a mechanism for stabilization. This dynamic adjustment enables the optimizer to self-regulate its exploration step without prior knowledge of the energy landscape. When several consecutive iterations fail to improve the cost, the velocity vector $\displaystyle \boldsymbol{v}$ is reset to zero to eliminate accumulated momentum that might otherwise keep the search trapped in a nonproductive region.

The iterative process continues until convergence is achieved according to one or more defined criteria. The first criterion is the stagnation of cost improvement over a specified number of past iterations, expressed as
\begin{equation}
    \|\mathcal{L}_T^{k} - \mathcal{L}_T^{k-p}\| < \epsilon_{thres}
\end{equation}
where $\displaystyle p$ denotes the last $\displaystyle p$-step of the observation window and $\epsilon_{thres}$ is a small threshold value indicating a negligible change in cost. This condition ensures termination once the optimization has reached a steady state. Convergence is also achieved if the learning rate falls below a minimum permissible value, indicating that further adjustments would have a negligible effect, or if the total number of iterations exceeds the prescribed maximum.

To safeguard against non-physical or unstable parameter sets that may arise during the process, a penalization strategy is incorporated. Whenever the cost evaluation results in a non-finite value or an unrealistic mechanical state, a large artificial penalty is applied according to
\begin{equation}
    \mathcal{L}_T = 10^{10}\left(1 + \mathcal{U}\right),
\end{equation}
where $\displaystyle \mathcal{U}$ is a small uniformly distributed random number used to prevent repetitive behavior. This mechanism automatically discourages the search from revisiting invalid regions of the parameter space and guides it toward stable, physically meaningful solutions.

The described optimization framework integrates a derivative-free gradient estimation with adaptive control of step size and momentum-based acceleration. Together, these components enable the search to efficiently explore a high-dimensional, non-linear cost surface, ensuring convergence to physically plausible and energetically optimal solutions. The constraints on curvature and torsion guarantee that all resulting body configurations remain smooth and mechanically feasible, while the rotational penalty maintains directional stability throughout the motion. This unified process forms the theoretical core of the present study, enabling the autonomous discovery of efficient, body-deformation patterns that minimize the mechanical cost of limbless body locomotion.

\subsection{Time-stepping Soft-Body Movement}\label{sec:extrinsic} 
Once the $\displaystyle \kappa(s,t)$ and $\displaystyle \tau(s,t)$ are estimated, we need to move the soft-body forward in the next time frame. Linear and angular momentum must be preserved when updating the extrinsic parameter space. As explained in the section~\ref{subsec:fs} before, six extrinsic parameters control the locomotion of the soft body --- three initial tail position and three initial tail orientation denoted by $\displaystyle \boldsymbol{\psi}$ from Equation~\eqref{eq:RigidBodyVar}. In the following section, we outline the algorithm for estimating the extrinsic parameters. 

\subsubsection{Extrinsic Parameters Estimation}\label{subsec: Calculate_RBVar}
At the beginning of each time step, an initial estimate at $\displaystyle t=0$ for $\displaystyle \boldsymbol{\psi}^{0}$ is assigned. The iterative update rule of $\displaystyle \boldsymbol{\psi}$ follows a leapfrog integration method
\begin{equation}
    \boldsymbol{\psi}^{t+1} = 2\boldsymbol{\psi}^{t} - \boldsymbol{\psi}^{t-1}
\end{equation}
providing a smooth continuation of motion from the recent history of the system. 

For a given estimate of $\displaystyle \boldsymbol{\psi}$, the residual vector $\displaystyle \mathcal{R}(\boldsymbol{\psi})$ is evaluated. This vector comprises six components, corresponding to the force and torque balance conditions from Equations~\eqref{eq:force_balance} and ~\eqref{eq:torque_balance}, respectively. Together, they represent the degree of mechanical imbalance in the system. The magnitude of this residual, $\|\mathcal{R}(\boldsymbol{\psi})\|$, quantifies how far the current configuration is from equilibrium. If this imbalance exceeds a specified tolerance, a Newton–Raphson iteration is performed to minimize it. Let the iteration steps be defined as $\displaystyle m \in \mathbb{Z}^+$ in our experiment.

Within each iteration, the Jacobian, $\displaystyle \boldsymbol{J}_m$, of the vector, $\displaystyle \boldsymbol{\psi}^t_m$, is constructed to describe how the components of $\displaystyle \mathcal{R}\left(\boldsymbol{\psi}^t_m\right)$ vary with small perturbations in the six parameters of $\displaystyle \boldsymbol{\psi}$. Because the system is highly non-linear and it is difficult to obtain an analytical expression for the Jacobian, it is evaluated numerically using finite differences. Each column of the Jacobian matrix is obtained by perturbing one element of the parameter vector $\displaystyle \boldsymbol{\psi}$ at a time, keeping all others fixed.
\begin{equation}
    \boldsymbol{J} = \frac{\mathcal{R}\left(\boldsymbol{\psi}^t_m + \boldsymbol{\delta} \otimes \hat{\boldsymbol{\psi}}^t_m \right) - \mathcal{R}\left(\boldsymbol{\psi}^t_m \right)}{\boldsymbol{\delta} \otimes \hat{\boldsymbol{\psi}}^t_m},
\end{equation}
where $\displaystyle \hat{\boldsymbol{\psi}}^t_m$ denotes the normalized vector and $\boldsymbol{\delta} = 10^{-8}\max\left(1, |\boldsymbol{\psi}^t_m|_i|\right)$ to maintain appropriate scaling and numerical stability.

After assembling the Jacobian matrix, the Newton correction $\Delta V$ is obtained by solving the linearized system
\begin{equation}
    \left(\boldsymbol{\delta} \otimes \boldsymbol{\psi}^t_m\right)\boldsymbol{J}_m = -\mathcal{R}\left(\boldsymbol{\psi}^t_m\right)
\end{equation}
which yields the direction and magnitude of the parameter update required to reduce the residual. The parameters are then updated from the Newton-Raphson iteration according to 
\begin{equation}
    \boldsymbol{\psi}^t_{m} = \boldsymbol{\psi}^t_{m-1} + \kappa_D \left(\boldsymbol{\delta} \otimes\boldsymbol{\psi} \right)
\end{equation}
where $\displaystyle \kappa_D \in (0,1]$ is a damping factor used to control the step size. To ensure stable and monotonic convergence, $\displaystyle \kappa_D$ is determined through a backtracking line search. Successive step sizes of the form are iterated until the updated configuration yields a smaller residual norm
\begin{equation}
    \|\mathcal{R}\left(\boldsymbol{\psi}^t_{m}\right)\| < \|\mathcal{R}\left(\boldsymbol{\psi}^t_{m-1}\right)\|, \;\; \text{s.t.} \;\; \kappa_D = 2^{-m} 
\end{equation}
This adaptive adjustment prevents overshooting and guarantees that each iteration improves the solution.

After each update, the residual vector $\displaystyle \mathcal{R}\left(\boldsymbol{\psi}^t_m\right)$ and its norm are re-evaluated. The iteration continues until the residual magnitude falls below the prescribed tolerance, indicating that the total force and torque are balanced to the desired accuracy.


\begin{figure*}[tbp]
\centering

\vspace{5pt}

\noindent
\begin{minipage}[t]{0.14\textwidth}
\vspace{10pt}
\raggedright
\textbf{Case 1:}\\
{$\max(A_{jk}) = 1$}\\
{$\max(B_{jk}) = 1$}\\
{$\max(C_{jk}) = 0.1$}\\
{$\max(D_{jk}) = 0.1$}\\
\end{minipage}%
\hfill
\begin{minipage}[t]{0.85\textwidth}
\vspace{2pt}
\centering
\includegraphics[width=0.16\linewidth]{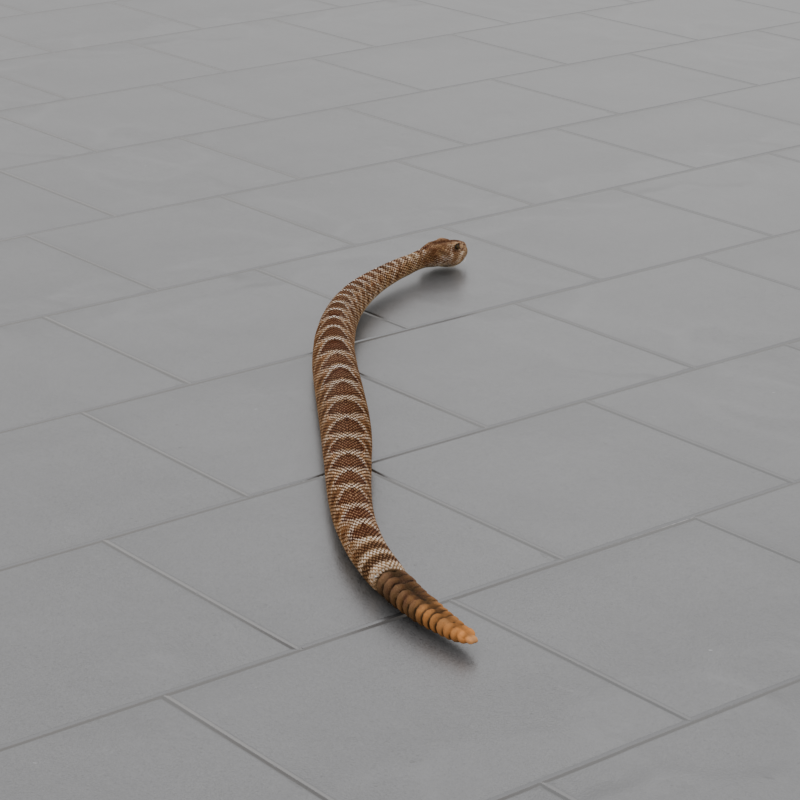}\hspace{2pt}%
\includegraphics[width=0.16\linewidth]{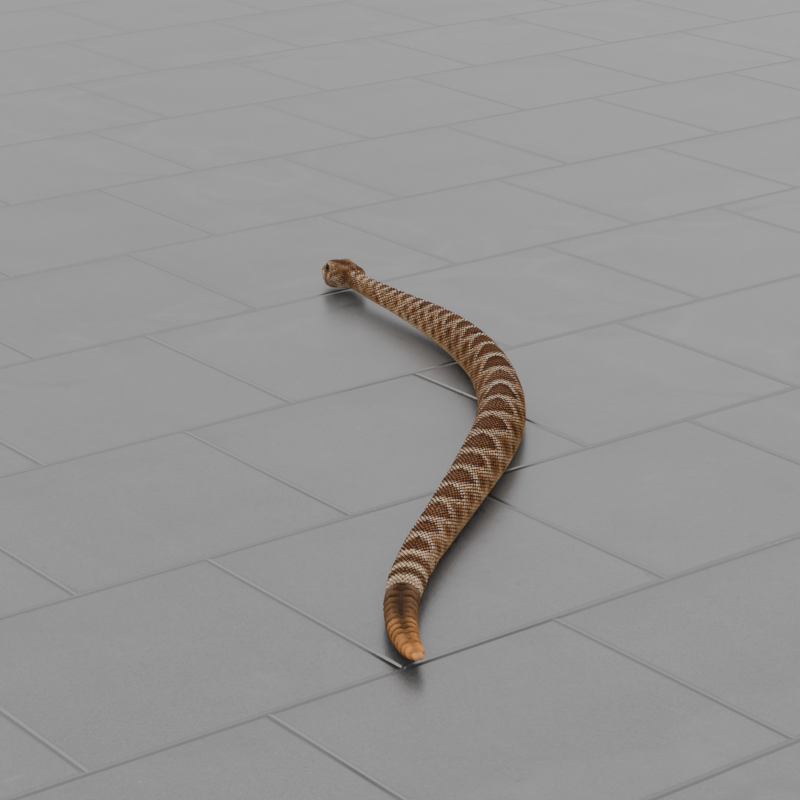}\hspace{2pt}%
\includegraphics[width=0.16\linewidth]{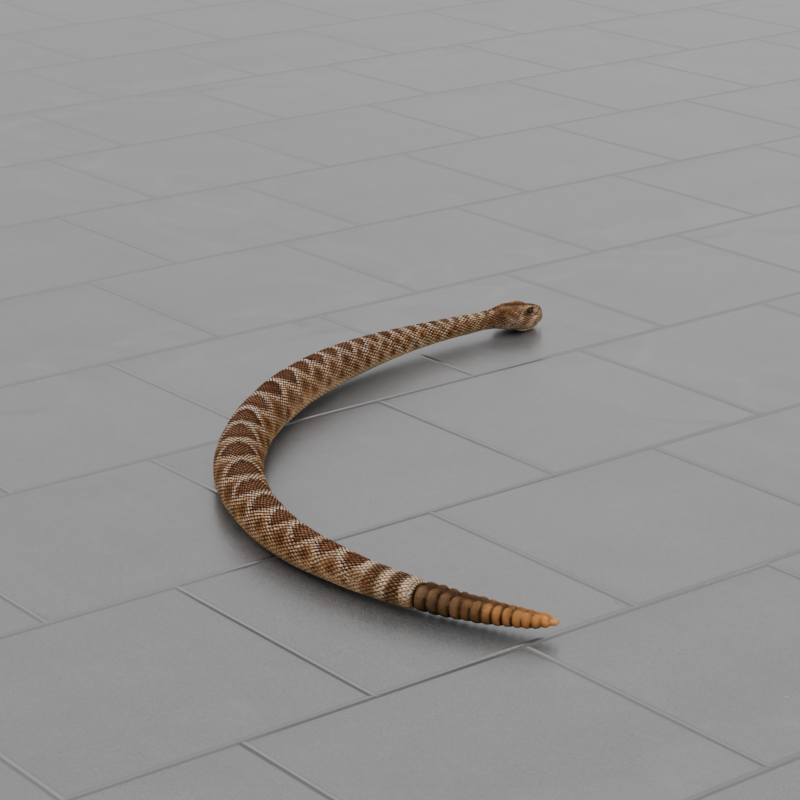}\hspace{2pt}%
\includegraphics[width=0.16\linewidth]{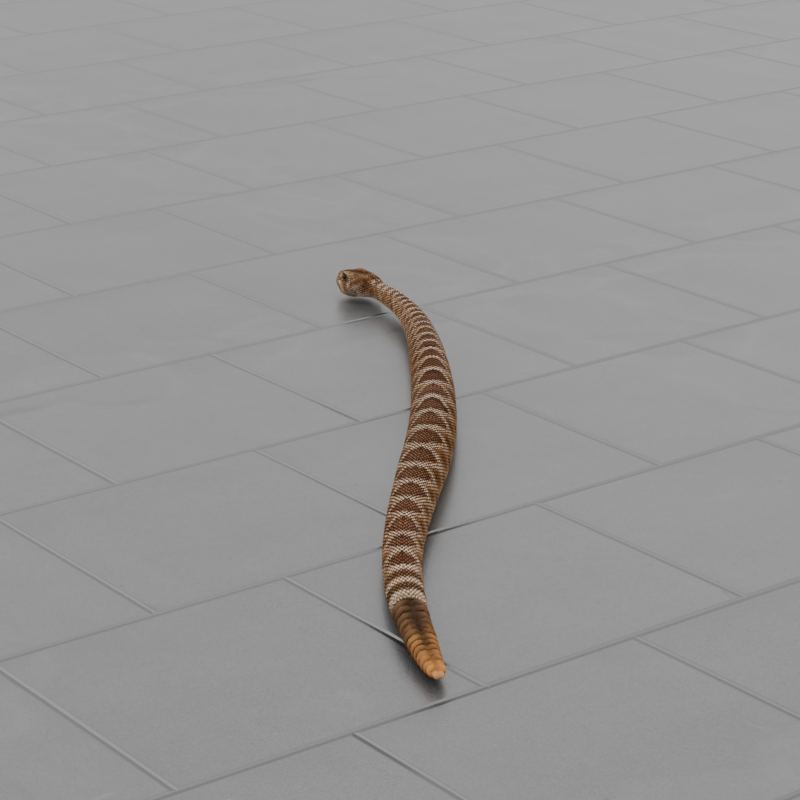}\hspace{2pt}%
\includegraphics[width=0.16\linewidth]{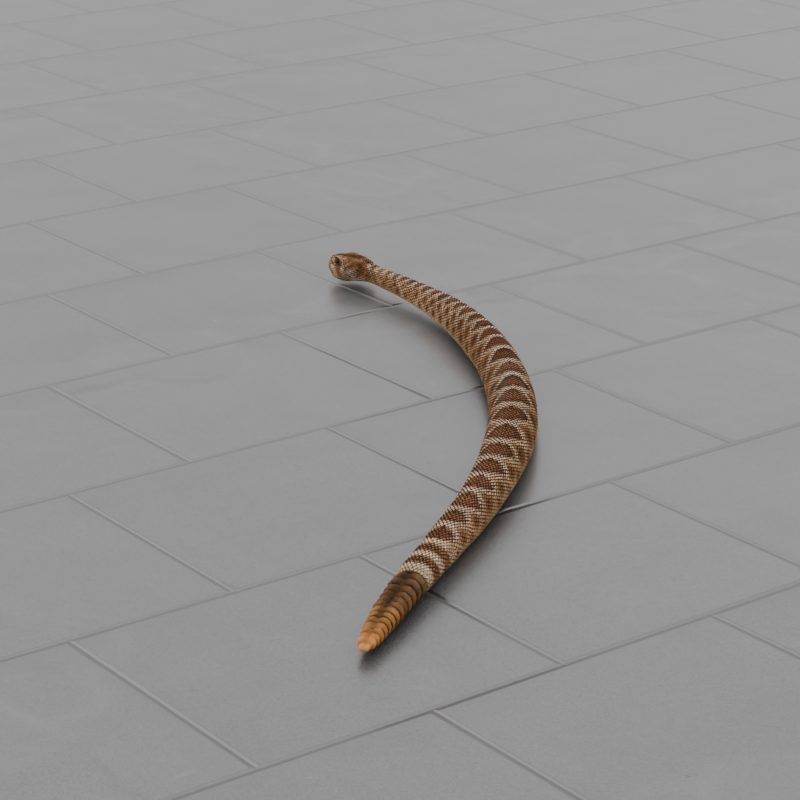}\hspace{2pt}%
\includegraphics[width=0.16\linewidth]{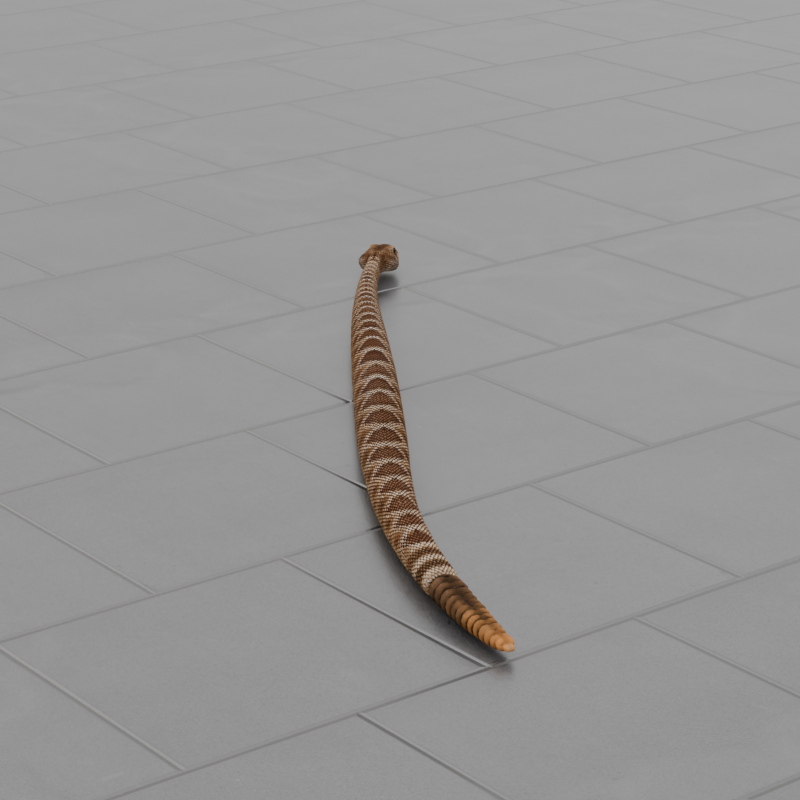}
\end{minipage}

\noindent
\begin{minipage}[t]{0.14\textwidth}
\vspace{10pt}
\raggedright
\textbf{Case 2:}\\
{$\max(A_{jk}) = 2$}\\
{$\max(B_{jk}) = 2$}\\
{$\max(C_{jk}) = 0.1$}\\
{$\max(D_{jk}) = 0.1$}\\
\end{minipage}%
\hfill
\begin{minipage}[t]{0.85\textwidth}
\vspace{2pt}
\centering
\includegraphics[width=0.16\linewidth]{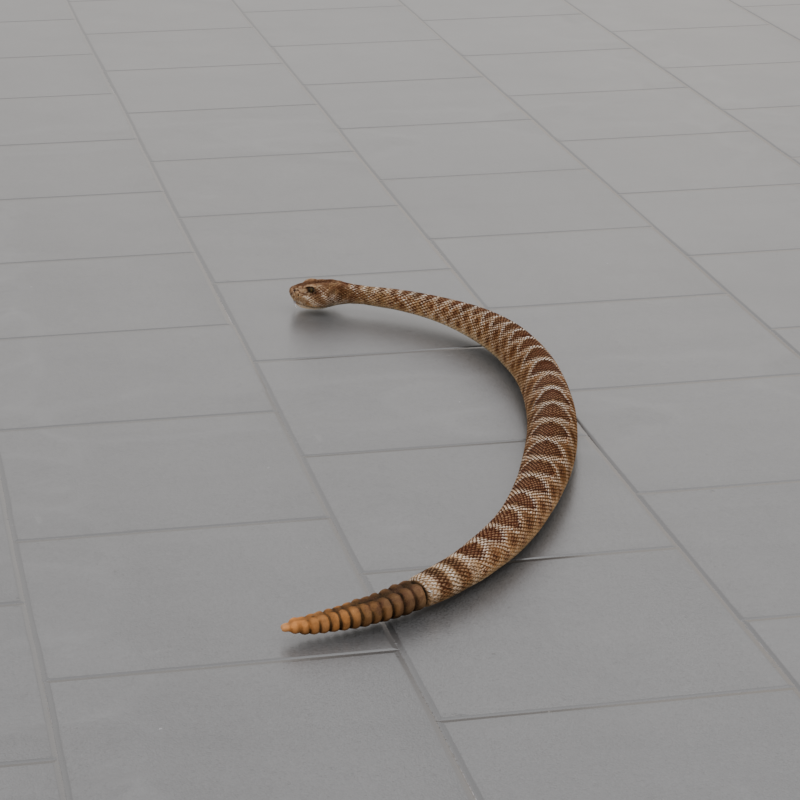}\hspace{2pt}%
\includegraphics[width=0.16\linewidth]{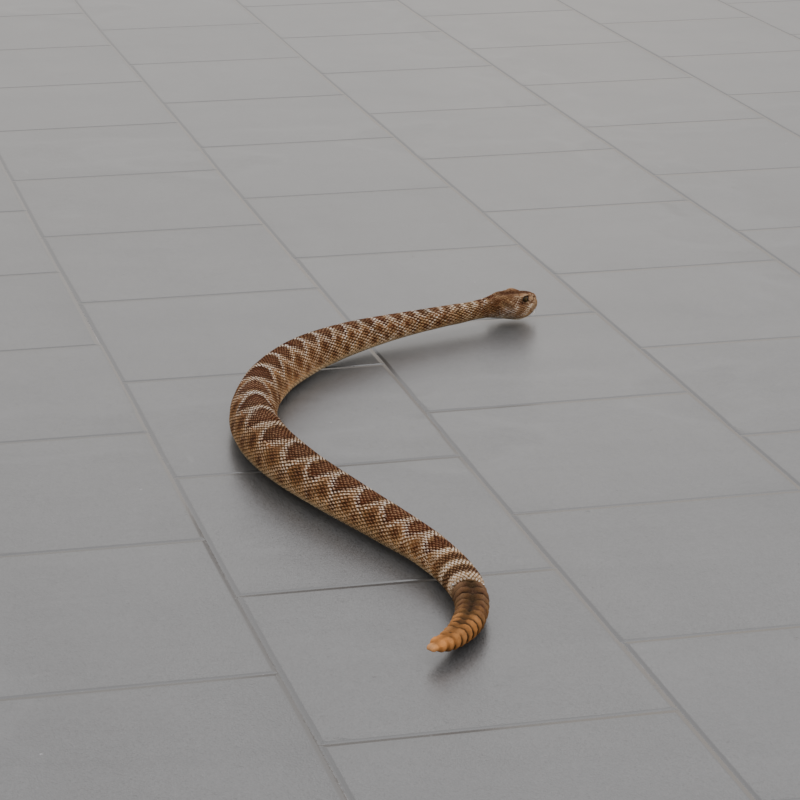}\hspace{2pt}%
\includegraphics[width=0.16\linewidth]{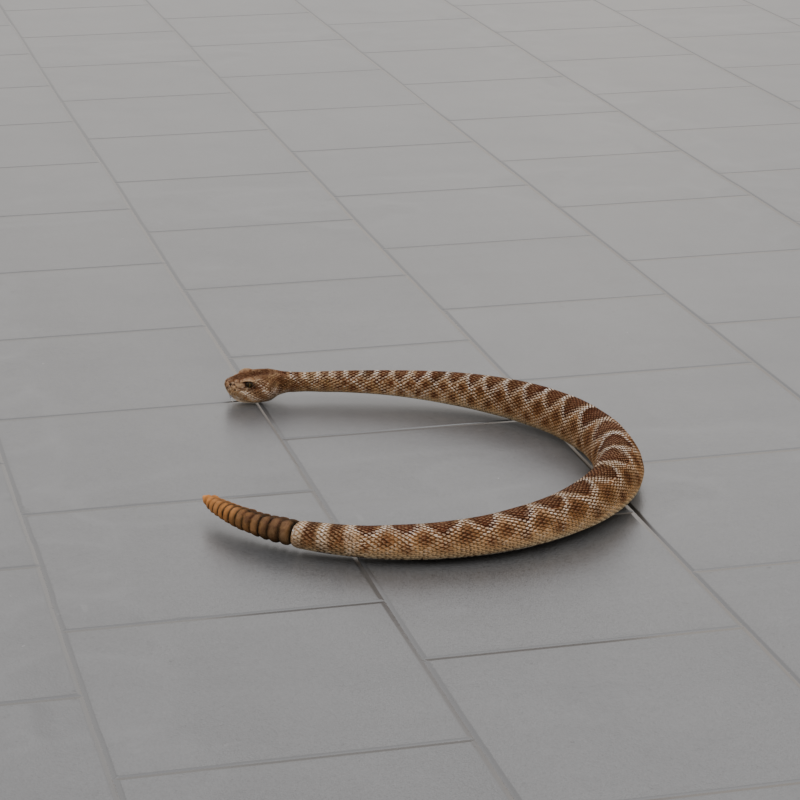}\hspace{2pt}%
\includegraphics[width=0.16\linewidth]{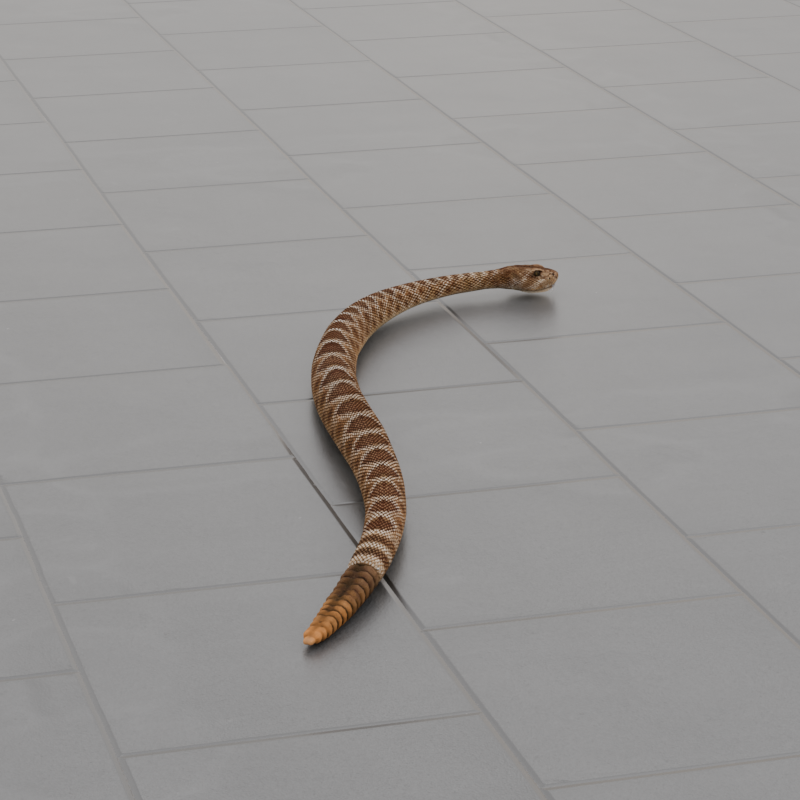}\hspace{2pt}%
\includegraphics[width=0.16\linewidth]{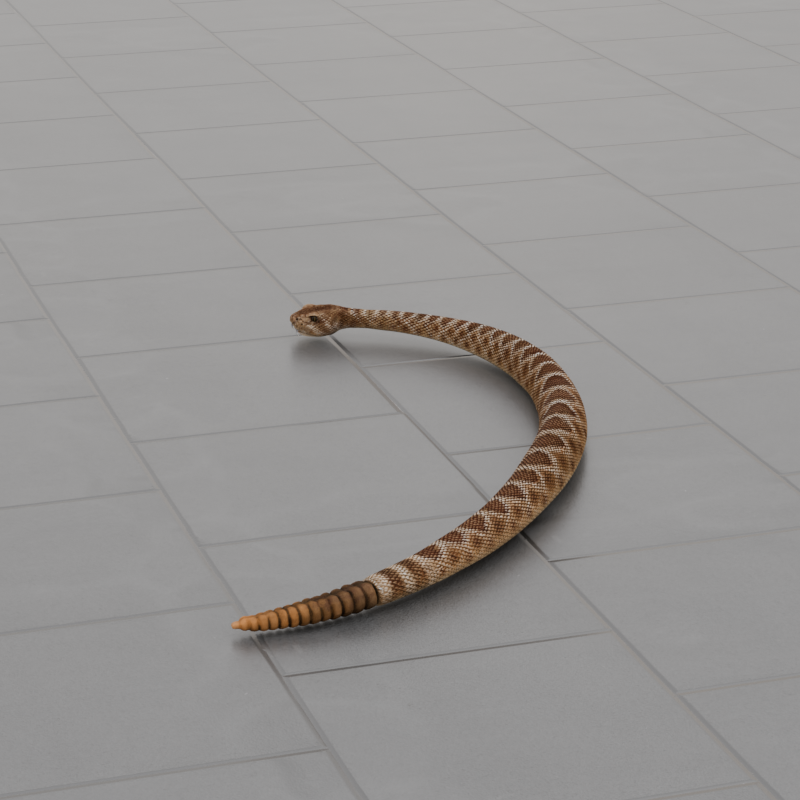}\hspace{2pt}%
\includegraphics[width=0.16\linewidth]{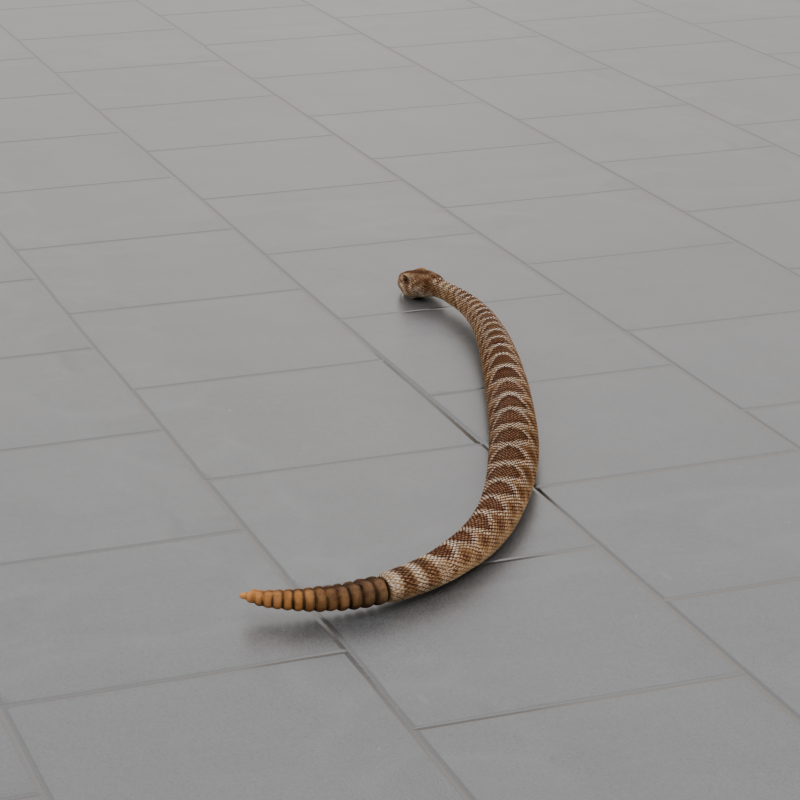}
\end{minipage}

\noindent
\begin{minipage}[t]{0.14\textwidth}
\vspace{10pt}
\raggedright
\textbf{Case 3:}\\
{$\max(A_{jk}) = 3$}\\
{$\max(B_{jk}) = 3$}\\
{$\max(C_{jk}) = 0.3$}\\
{$\max(D_{jk}) = 0.3$}\\
\end{minipage}%
\hfill
\begin{minipage}[t]{0.85\textwidth}
\vspace{2pt}
\centering
\includegraphics[width=0.16\linewidth]{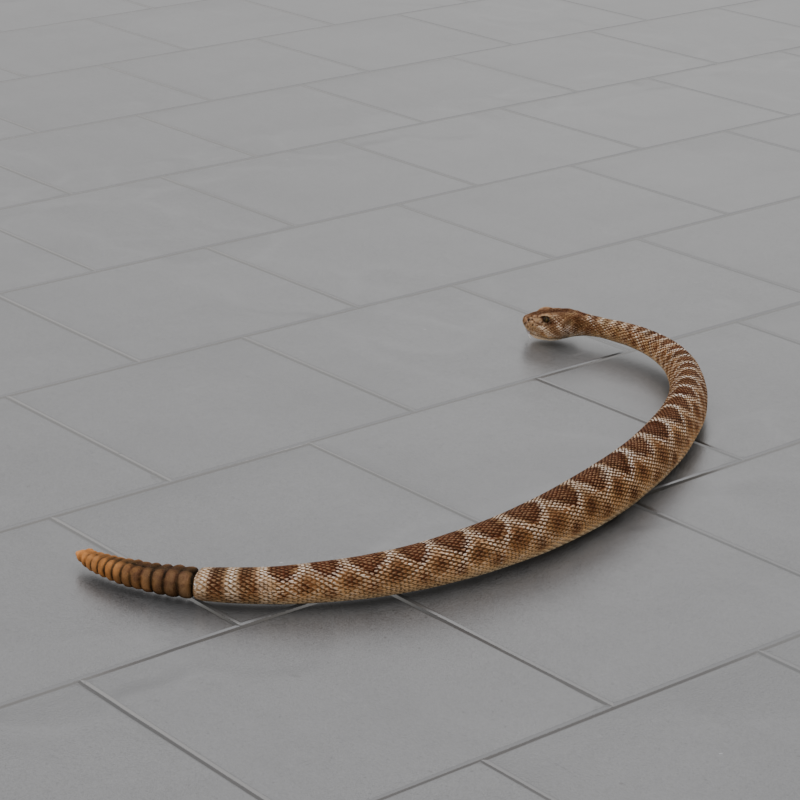}\hspace{2pt}%
\includegraphics[width=0.16\linewidth]{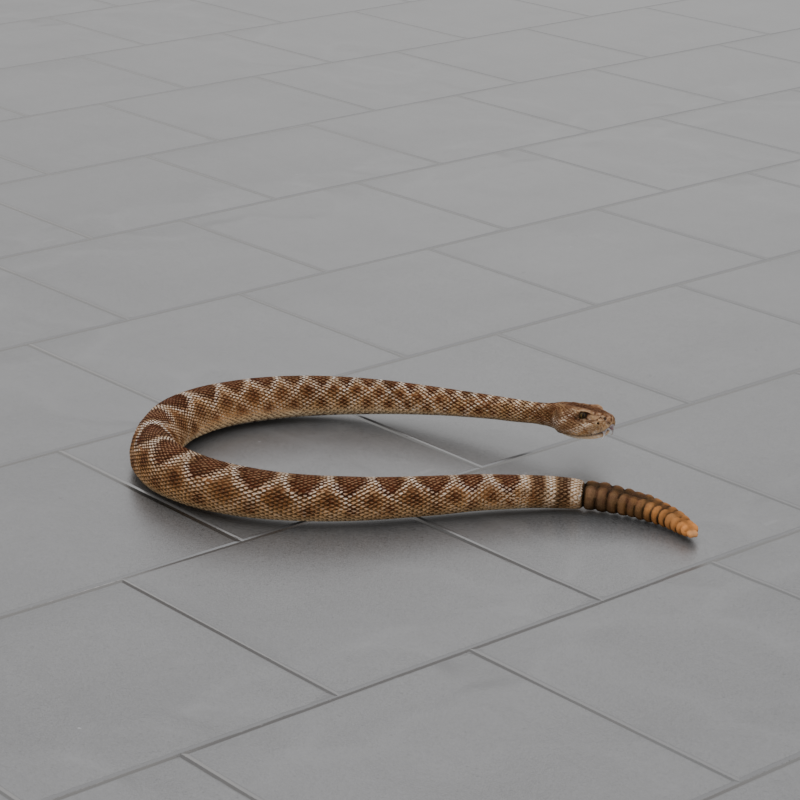}\hspace{2pt}%
\includegraphics[width=0.16\linewidth]{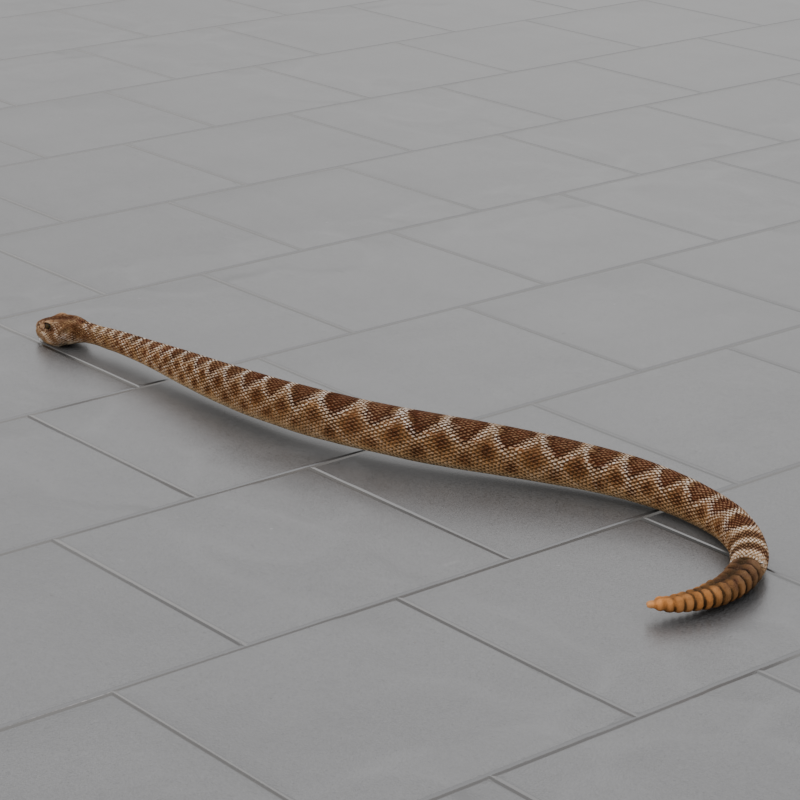}\hspace{2pt}%
\includegraphics[width=0.16\linewidth]{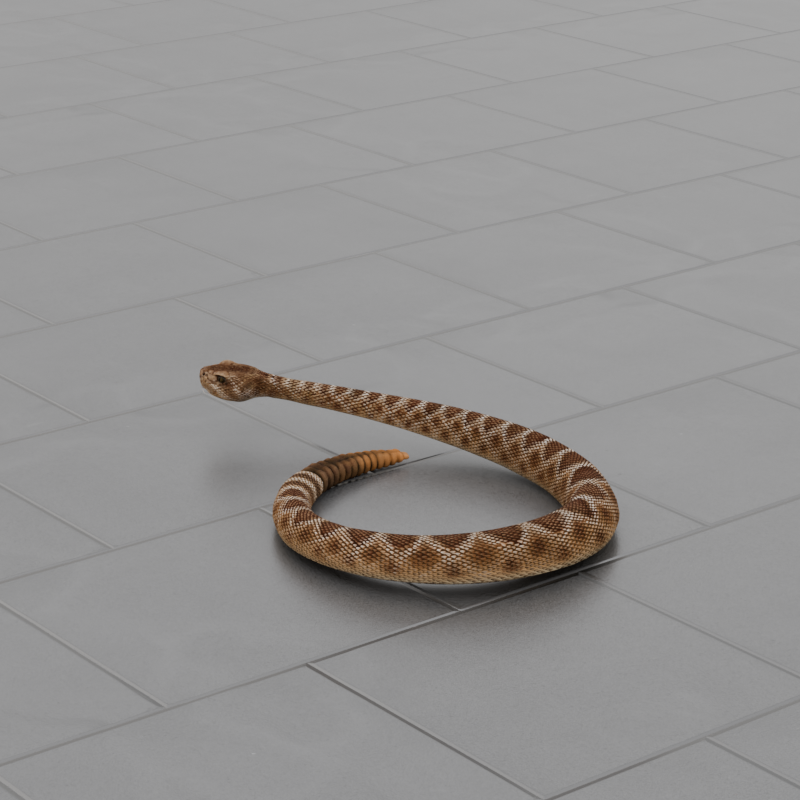}\hspace{2pt}%
\includegraphics[width=0.16\linewidth]{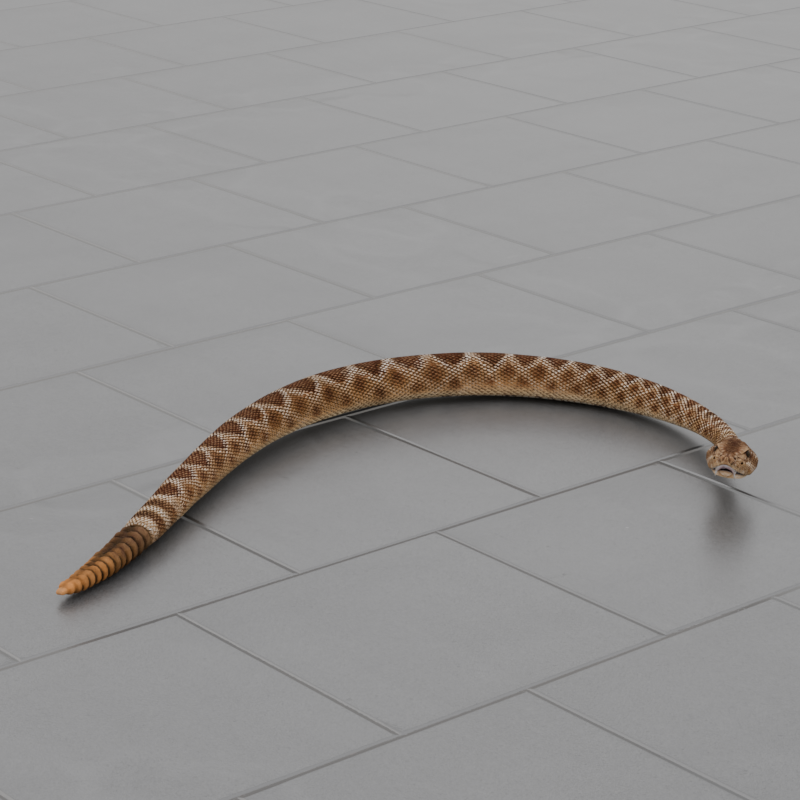}\hspace{2pt}%
\includegraphics[width=0.16\linewidth]{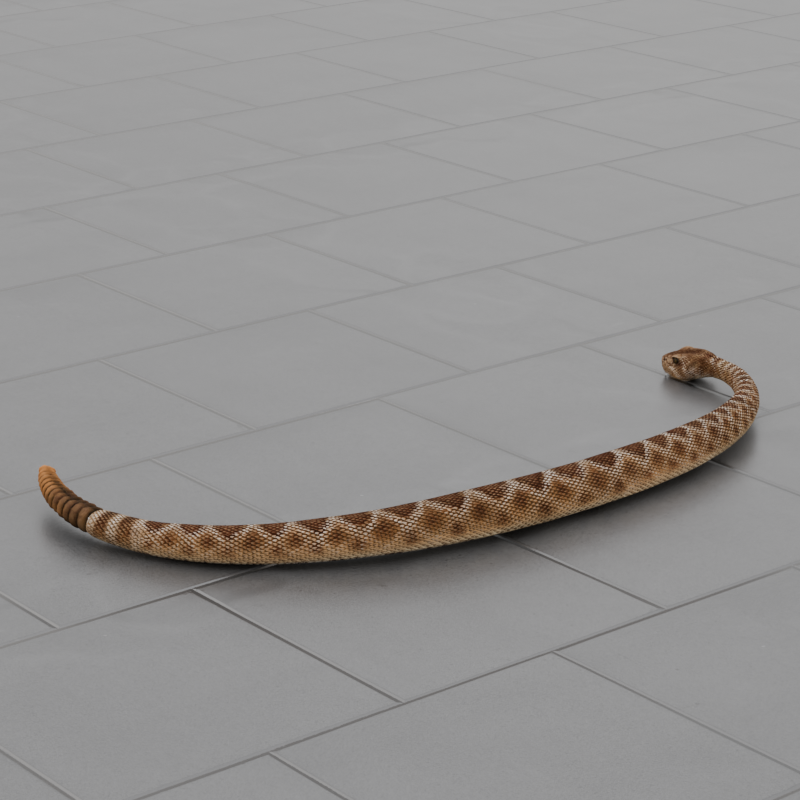}
\end{minipage}

\caption{Temporal evolution of motion, corresponding to an SPSA-optimized gait with friction ratios $\mu_n/\mu_f = 1$ and $\mu_b/\mu_f = 5$, and no bending energy term in the cost function. Cases 1 and 2 correspond to a uniform mass distribution with different coefficient bounds, while Case 3 corresponds to a non-uniform mass distribution.}
\label{fig:Img7}
\end{figure*}

\begin{figure*}[tbp]
\centering

\vspace{5pt}

\noindent
\begin{minipage}[t]{0.14\textwidth}
\vspace{10pt}
\raggedright
\textbf{Parameters:}\\
{$\max(A_{jk}) = 2$}\\
{$\max(B_{jk}) = 2$}\\
{$\max(C_{jk}) = 0.1$}\\
{$\max(D_{jk}) = 0.1$}\\
\end{minipage}%
\hfill
\begin{minipage}[t]{0.85\textwidth}
\vspace{2pt}
\centering
\includegraphics[width=0.16\linewidth]{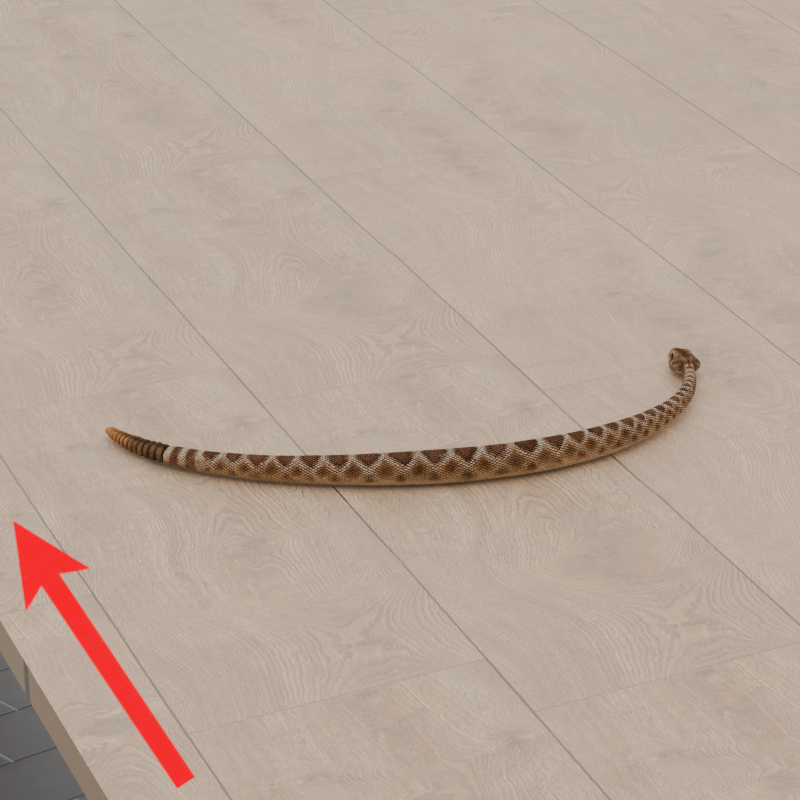}\hspace{2pt}%
\includegraphics[width=0.16\linewidth]{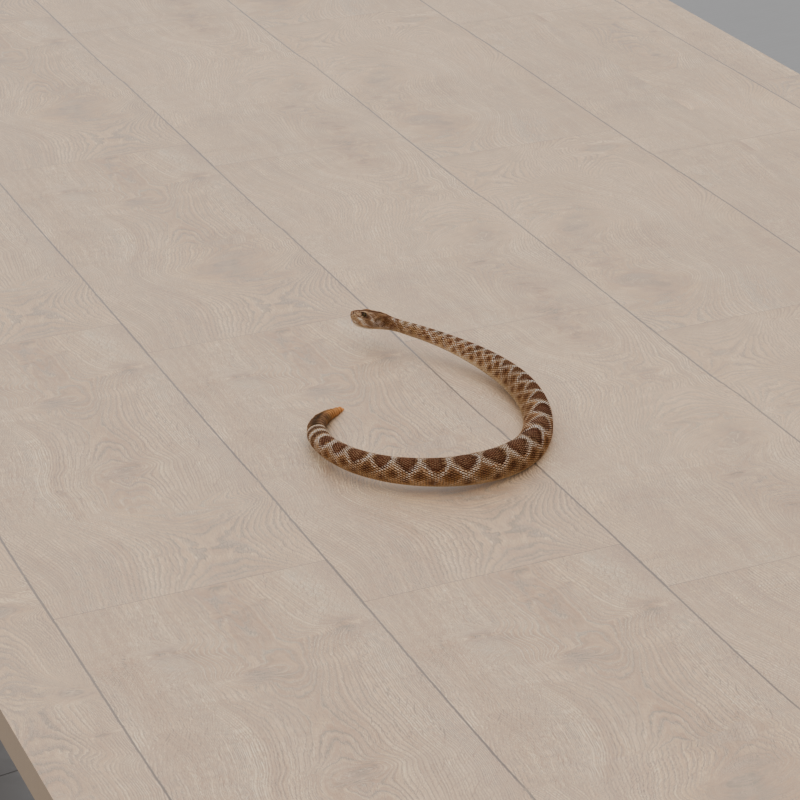}\hspace{2pt}%
\includegraphics[width=0.16\linewidth]{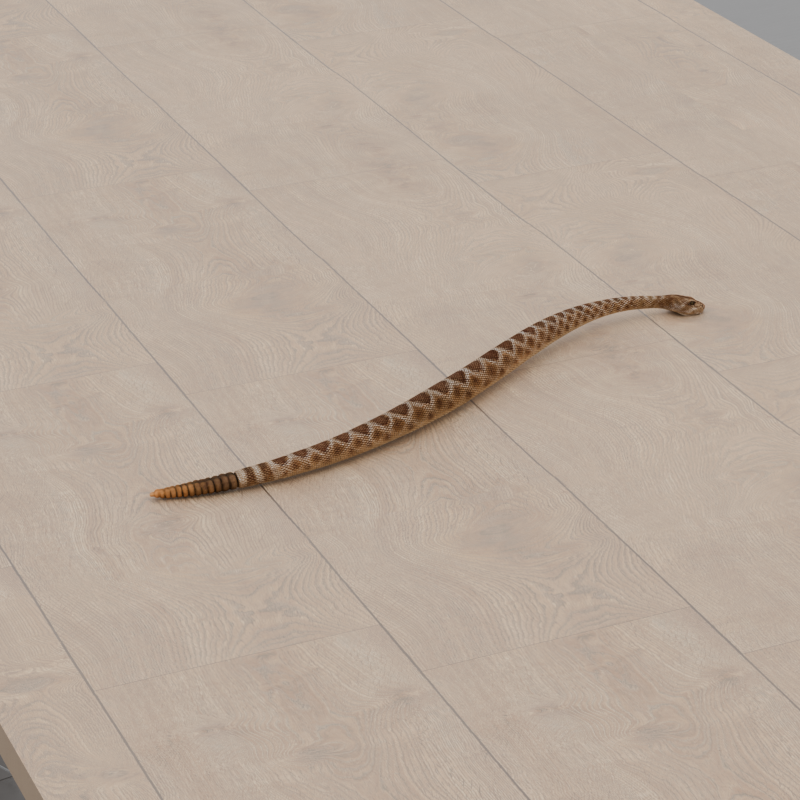}\hspace{2pt}%
\includegraphics[width=0.16\linewidth]{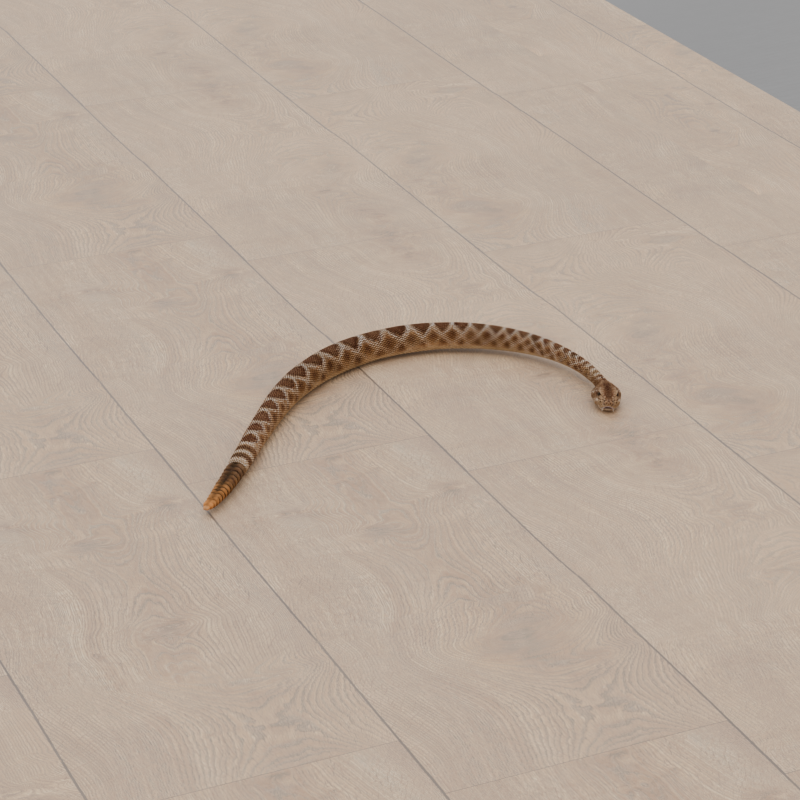}\hspace{2pt}%
\includegraphics[width=0.16\linewidth]{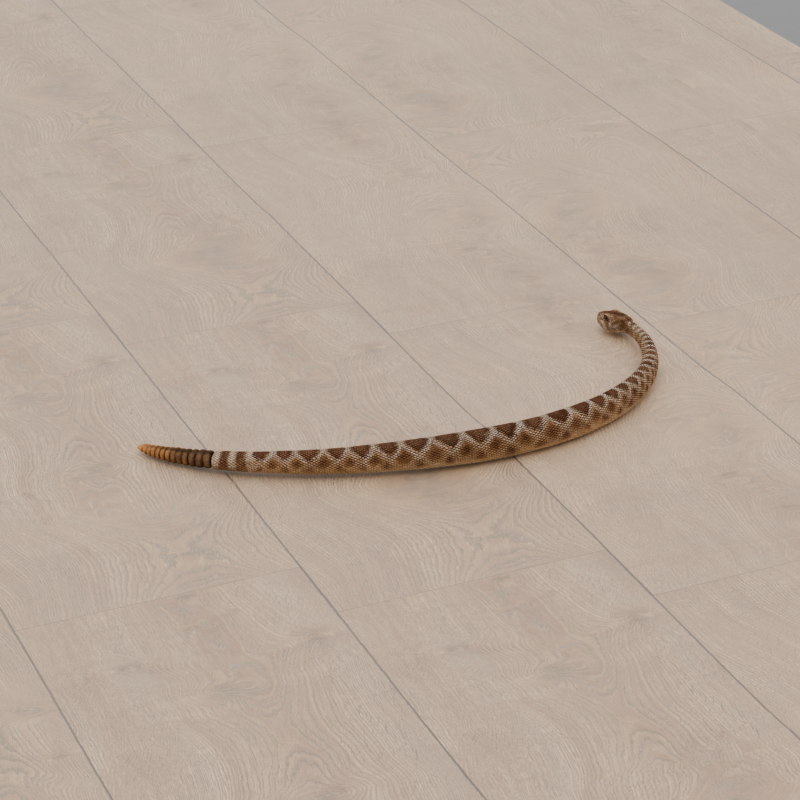}\hspace{2pt}%
\includegraphics[width=0.16\linewidth]{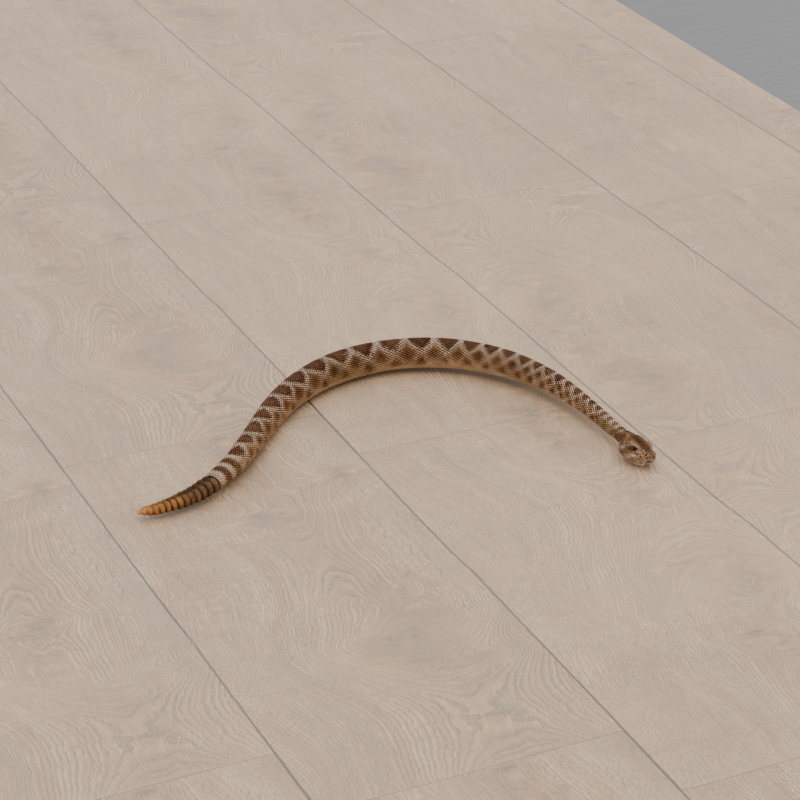}
\end{minipage}

\caption{Temporal evolution of motion for \emph{sidewinding} motion. The solution is obtained using FDSA under anisotropic friction characterized by $\mu_n/\mu_f = 1$ and $\mu_b/\mu_f = 5$, with a uniform mass distribution and no bending energy regularization. The \emph{sidewinding} motion is generated on a $\boldsymbol{10^\circ}$ inclined plane, demonstrating the influence of gravitational force and surface inclination on the locomotion dynamics. The first image includes a red arrow indicating the direction of inclination.}
\label{fig:Img7-1}
\end{figure*}


\subsubsection{Numerical Integration of the Frenet-Serret System}\label{subsec:numint}
Finally, we need to calculate the position and orientation of the whole body. The spinal trajectory in the body-fixed coordinate frame, denoted by $\displaystyle \boldsymbol{\gamma}(s,t)$, satisfies
\begin{equation}
    \frac{\partial \boldsymbol{\gamma}}{\partial s} = \boldsymbol{T}(s,t)
    \label{eq:num_x}
\end{equation}
with the initial conditions at the tail ($s = 0$) defined by Equation~\eqref{eq:init-pose}.

To compute the evolution of these quantities along the body, the fourth-order Runge-Kutta (RK4) algorithm~\cite{gill1981practical} is used. The arc length domain is discretized into $\displaystyle n$ uniform segments with step size $\displaystyle \delta s = 1/n$. At each discretization step $\displaystyle i$, the intermediate evaluations are computed using Equation~\eqref{eq:fs} and Equation~\eqref{eq:GlobalPosition}. Thus, we get the final complete body configuration. Algorithm~\ref{alg:snake_optimization} details the complete flow of our proposed method.

\section{Results}\label{sec:result}

The proposed framework demonstrates that efficient three dimensional locomotion can be systematically achieved by directly controlling curvature and torsion along the body curve. By formulating locomotion as an optimization problem over these geometric quantities, the approach leverages differential geometry to represent body deformation, a physically consistent force model to capture environmental interactions, and stochastic optimization techniques to explore the control space. As a result, realistic locomotion patterns are not prescribed a priori, unlike~\cite{gross2023shapechange}~\cite{soliman2024goflow} but instead emerge naturally from the interplay between body deformation and frictional forces. The optimization process consistently converges to smooth, stable, and robust locomotion strategies. This indicates that the chosen parameterization and cost formulation are well-suited for capturing the essential dynamics of limbless locomotion. Also, the inclusion of torsion enables the body to exploit out-of-plane deformation. This additional degree of freedom allows the system to modulate ground contact through lifting and twisting motions, thereby reducing frictional resistance and improving locomotion efficiency, particularly in environments with anisotropic friction. From an energetic perspective, the framework explicitly balances frictional work against deformation-related costs. The optimizer seeks configurations that minimize the total energy expenditure required to achieve displacement, effectively identifying energetically efficient locomotion strategies. The inclusion of bending energy further regularizes the solution space, preventing physically unrealistic deformations and ensuring that the resulting motions remain smooth and geometrically consistent. An important feature of the proposed approach is its flexibility. Since the locomotion is governed by a compact set of geometric control parameters, the framework can adapt to different physical settings, including variations in frictional conditions, mass distribution, and regularization terms. This makes it a general and extensible tool for studying three-dimensional soft-body locomotion.

All experiments were conducted on a system equipped with an Intel Core i7-14700K processor (20 cores) and 64 GB of DDR5 RAM. For visual validation, the resulting locomotion patterns were rendered using Blender~\cite{blender}. The simulated curve was exported and embedded into high-quality 3D meshes, allowing realistic visualization of the deforming body over time. Figure~\ref{fig:teaser} shows the locomotion of a snake on solid ground and an eel in fluid media.

\subsection{Visualization}

To validate the proposed framework on realistic geometries, high-quality three-dimensional meshes were incorporated into the simulation pipeline. This enables a physically meaningful and visually interpretable representation of the optimized locomotion patterns, bridging the gap between abstract curve animation and realistic body geometries. All the meshes used in this work were obtained from the \texttt{cgtrader.com} platform.

\subsubsection{Centerline Generation}
The locomotion model produces a time-dependent space curve representing the body centerline. This curve is reconstructed from the prescribed curvature $\kappa(s,t)$ and torsion $\tau(s,t)$ using the Frenet–Serret framework. The resulting curve is discretized into a sequence of points along the arc-length parameter $s \in [0,1]$ for each time instance.

\subsubsection{Embedding of the Curve into the Mesh}
The embedding process is performed in a two-stage pipeline involving curve generation in the simulation environment, and geometric deformation in \textit{Blender}~\cite{blender}. First, for each time step, the computed curve is exported as a discrete curve (e.g., in \emph{.obj} format). This curve represents the desired backbone of the deforming body.

\subsubsection{Design Overview} 
This embedding strategy ensures that the mesh's geometric deformation is fully consistent with the underlying mathematical model. By separating the physical simulation (curve computation) from the geometric rendering (mesh deformation), the framework maintains both computational efficiency and visual fidelity.
Moreover, this approach allows different meshes to be used interchangeably with the same curve data, enabling flexible visualization across multiple body geometries without modifying the core simulation.

\subsection{Optimized Locomotion Patterns}
\begin{figure*}[tbp]
\centering

\vspace{5pt}

\noindent
\begin{minipage}[t]{0.14\textwidth}
\vspace{50pt}
\raggedright
\textbf{Parameters:}\\
{$\max(A_{jk}) = 1$}\\
{$\max(B_{jk}) = 1$}\\
{$\max(C_{jk}) = 0.1$}\\
{$\max(D_{jk}) = 0.1$}\\
\end{minipage}%
\hfill
\begin{minipage}[t]{0.85\textwidth}
\vspace{2pt}
\centering
\includegraphics[width=0.32\linewidth]{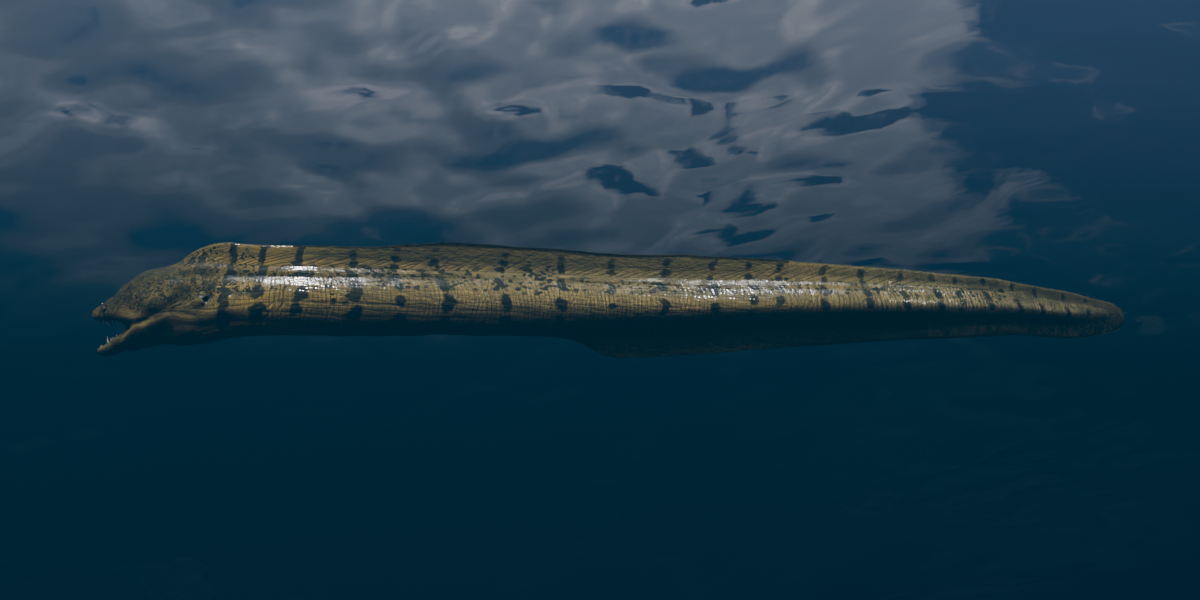}\hspace{2pt}%
\includegraphics[width=0.32\linewidth]{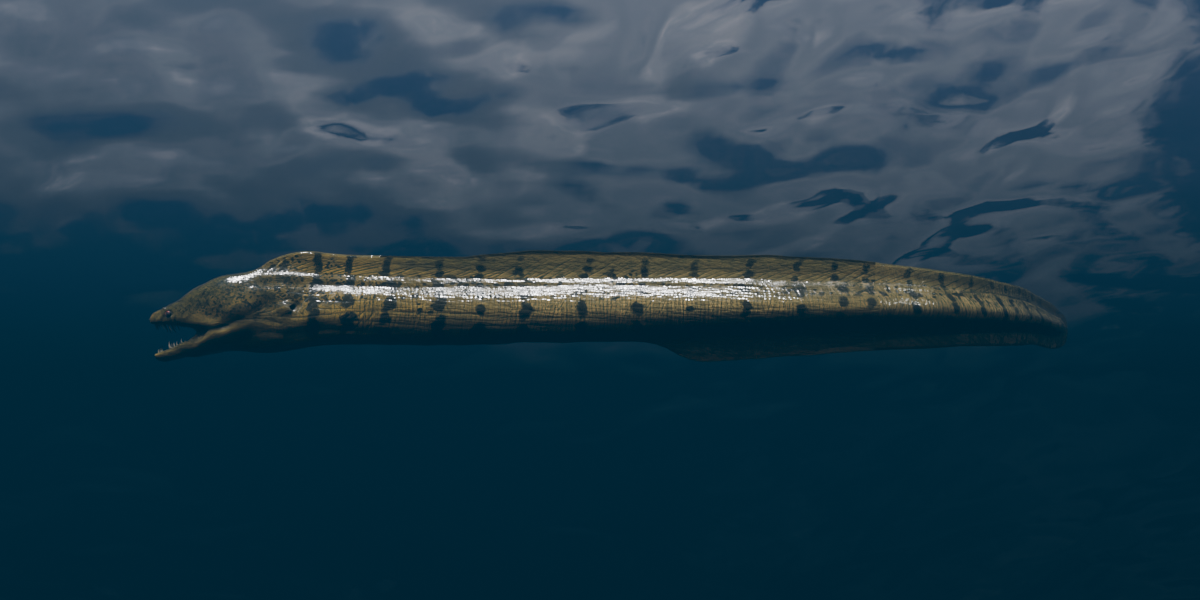}\hspace{2pt}%
\includegraphics[width=0.32\linewidth]{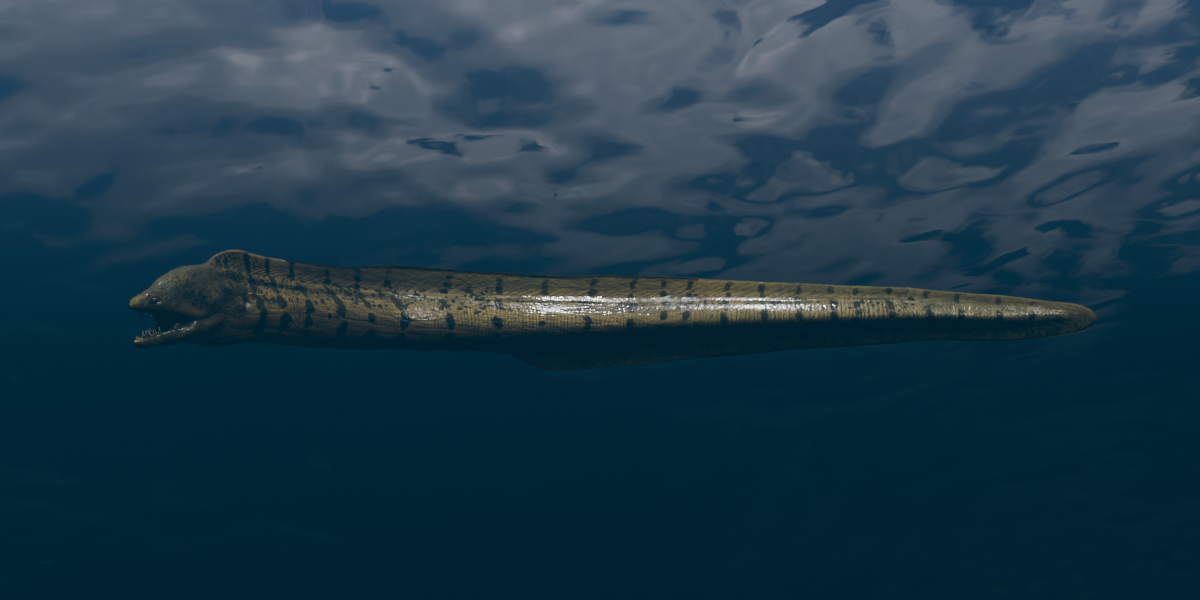}\hspace{2pt}%
\includegraphics[width=0.32\linewidth]{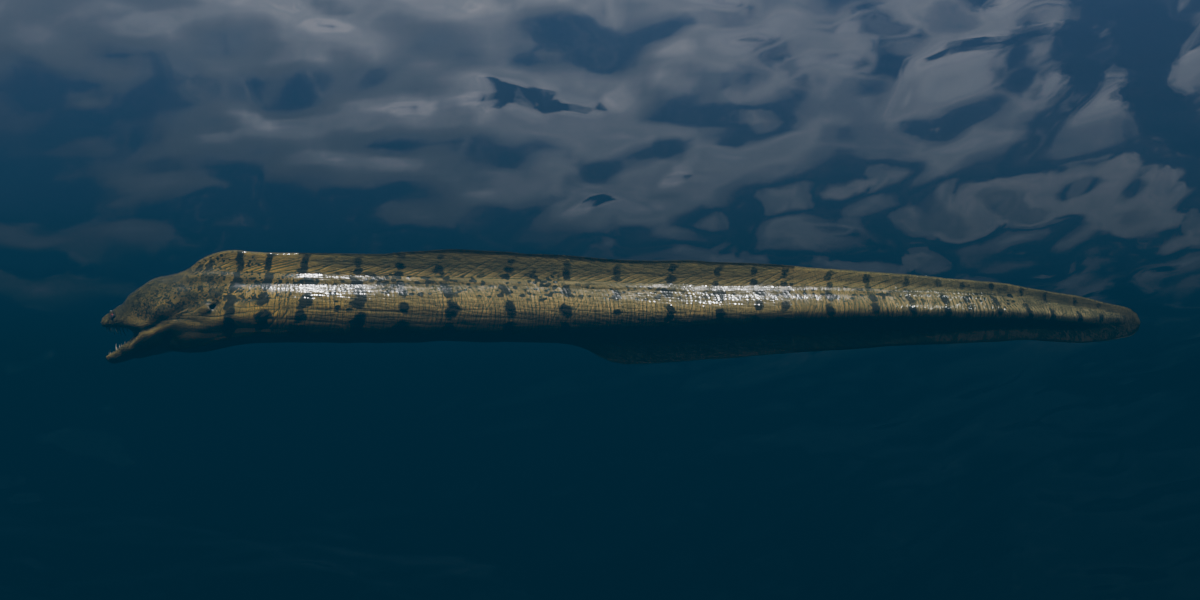}\hspace{2pt}%
\includegraphics[width=0.32\linewidth]{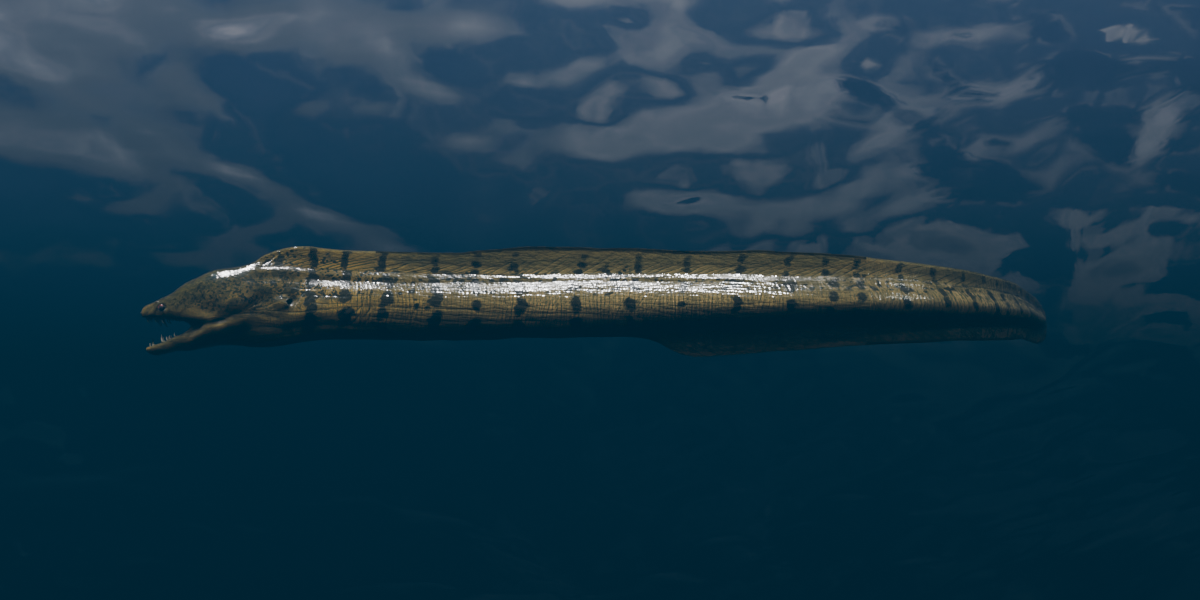}\hspace{2pt}%
\includegraphics[width=0.32\linewidth]{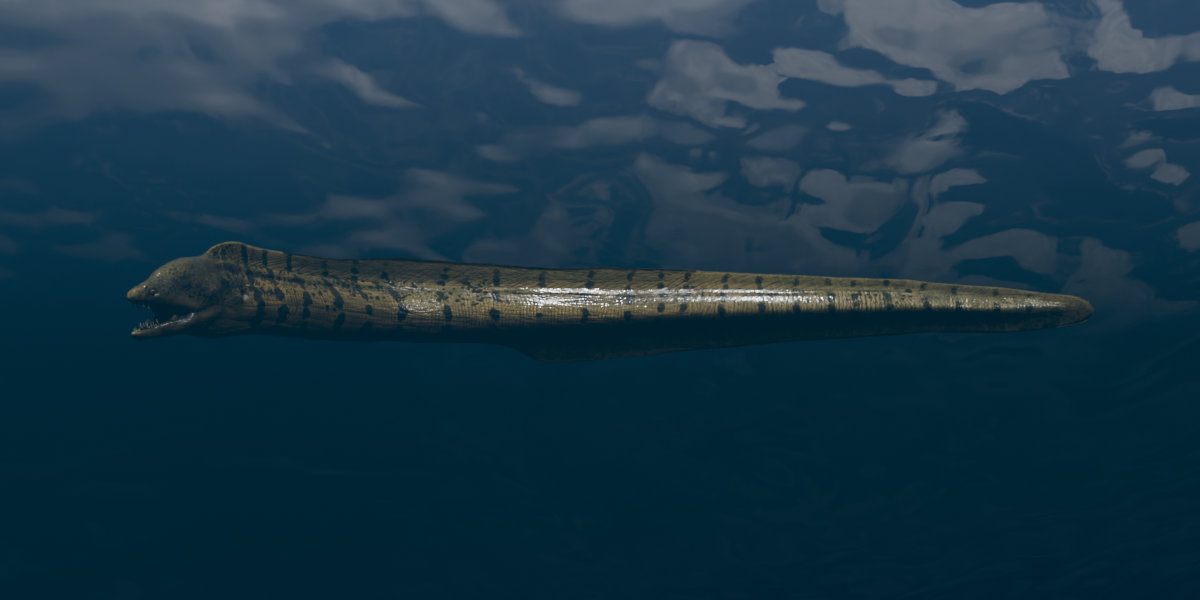}
\end{minipage}

\caption{Eel-like locomotion generated using the same framework. The motion exhibits wave-like undulatory patterns propagating along the body, approximating fluid-like swimming behavior. The simulation uses uniform mass distribution and high bending and torsional energy terms. The friction ratios used are $\mu_n/\mu_f = 1$ and $\mu_b/\mu_f = 5$.}
\label{fig:Eel}
\end{figure*}

To systematically investigate the factors governing optimized locomotion, we design a sequence of controlled experiments, presented in Figures~\ref{fig:Img1}--\ref{fig:Eel}. Each experiment isolates a specific physical or modeling parameter, allowing us to understand its role in shaping the optimization landscape and the resulting locomotion strategy.

\subsubsection{Influence of Friction Anisotropy (Figure~\ref{fig:Img1})}
The primary objective of this experiment is to examine how different frictional regimes affect the feasibility and effectiveness of locomotion. Frictional anisotropy is a fundamental requirement in limbless locomotion, as it determines whether net propulsion can be achieved. To study this, we vary the ratios $\mu_n/\mu_f$ and $\mu_b/\mu_f$, covering three distinct regimes --- near-isotropic friction, strong backward-friction dominance, and strong normal-friction dominance. The results show a clear dependence of the optimized motion on the frictional anisotropy. In \emph{Case 1}, where both $\mu_n/\mu_f$ and $\mu_b/\mu_f$ are close to unity, the motion is dominated by slipping. Since the friction is nearly isotropic, the body lacks sufficient directional resistance to anchor itself to the ground, and therefore, the generated deformation does not translate into effective forward motion. In \emph{Case 2}, with $\mu_n/\mu_f = 1$ and $\mu_b/\mu_f = 5$, the optimizer produces a more organized side-wise motion. The larger backward friction reduces reverse slipping and allows the body to generate lateral displacement through coordinated deformation. In \emph{Case 3}, where normal friction is significantly greater than forward friction, the optimized gait produces substantial forward motion. The high normal resistance prevents excessive lateral sliding, allowing the body undulations to be converted more effectively into forward propulsion. These observations indicate that frictional anisotropy strongly controls the direction and efficiency of locomotion, and that increasing normal friction relative to forward friction can significantly improve forward displacement.


\begin{figure*}[tbp]
\centering

\vspace{5pt}

\noindent
\vspace{8pt}
\begin{minipage}[t]{0.995\textwidth}
\vspace{2pt}
\centering
\includegraphics[width=0.1625\linewidth]{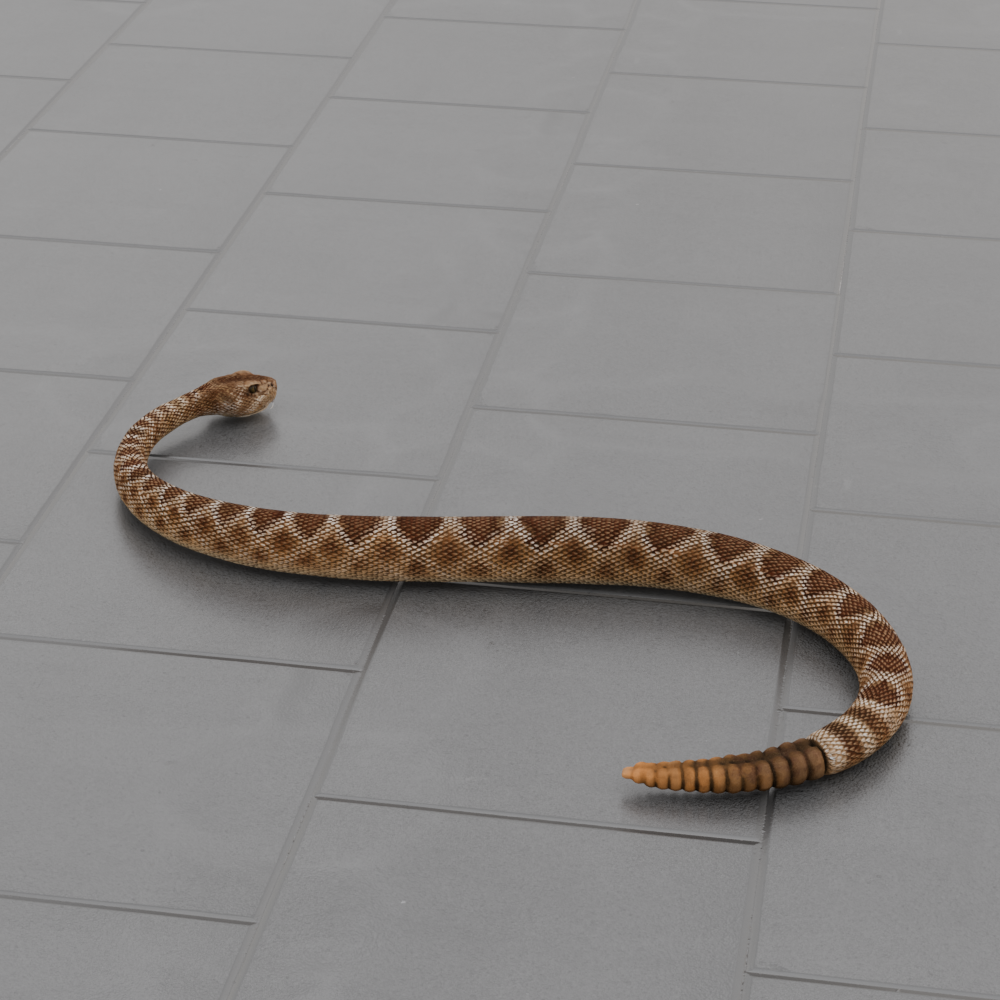}\hspace{2pt}%
\includegraphics[width=0.1625\linewidth]{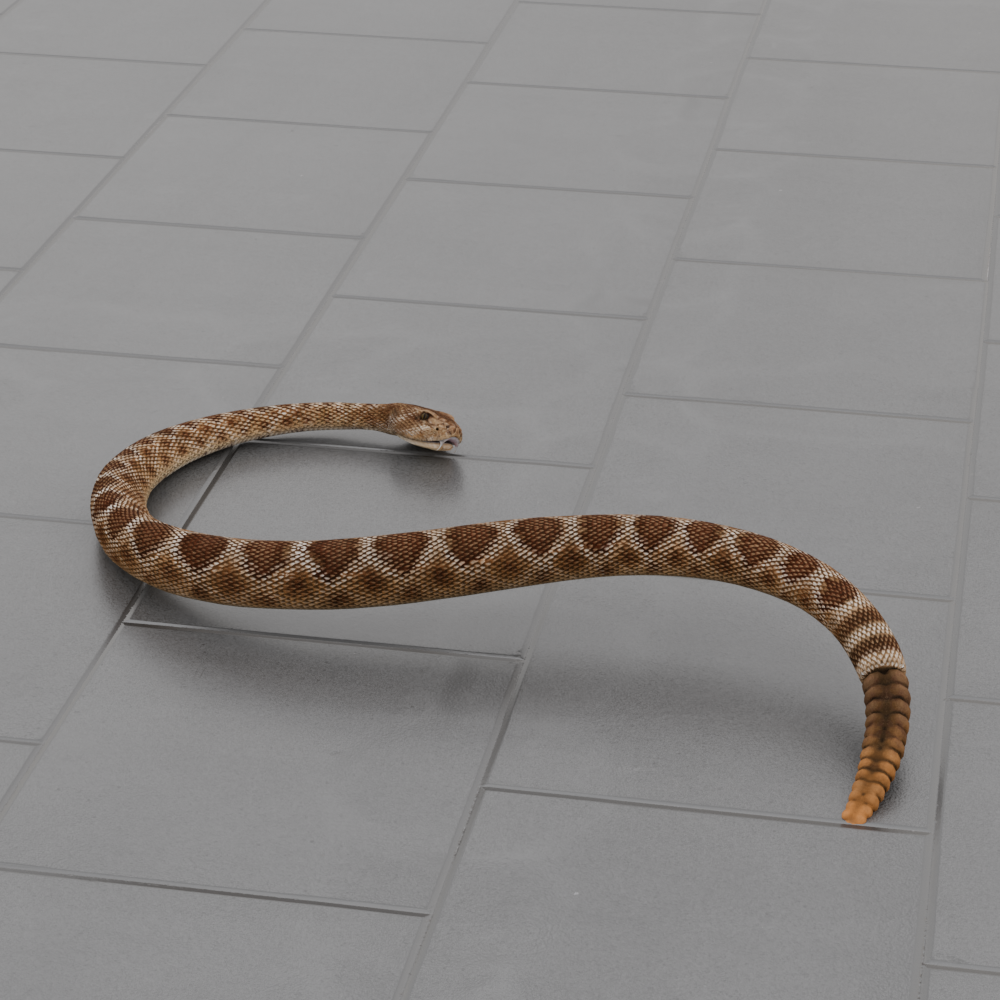}\hspace{2pt}%
\includegraphics[width=0.1625\linewidth]{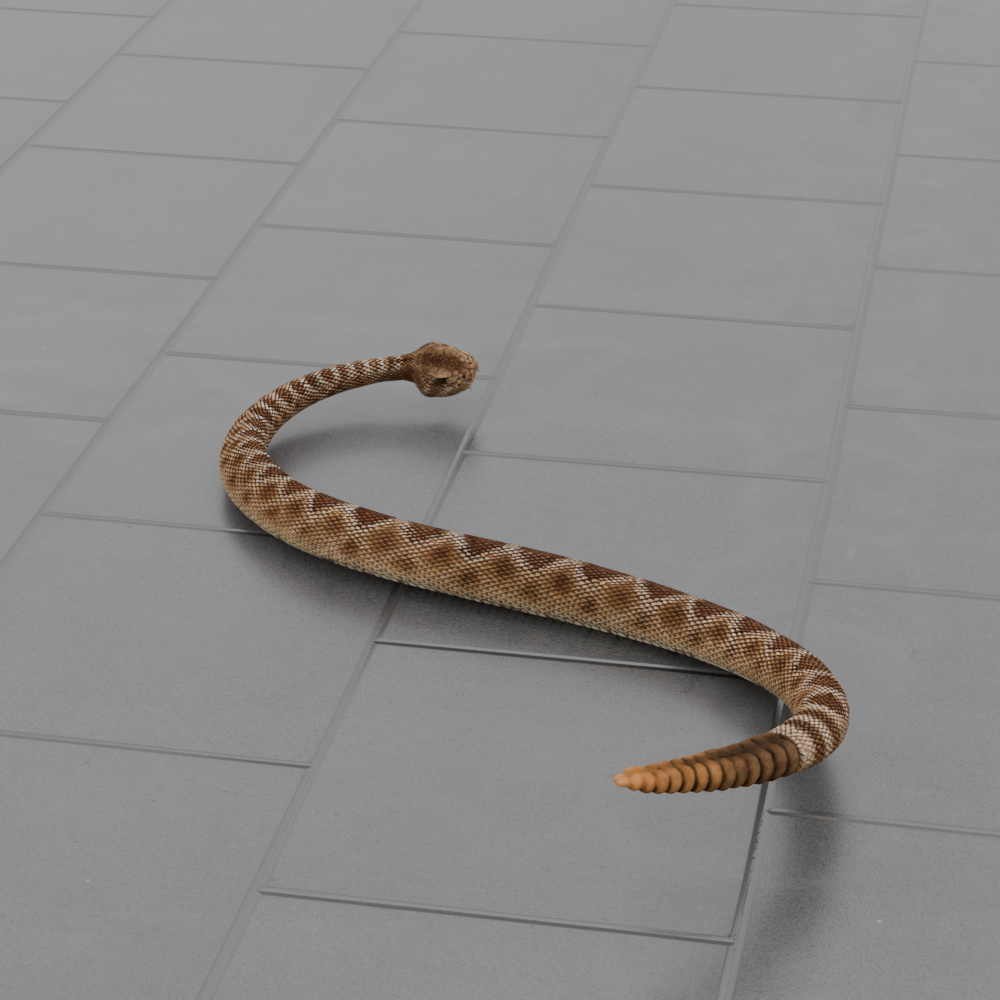}\hspace{2pt}%
\includegraphics[width=0.1625\linewidth]{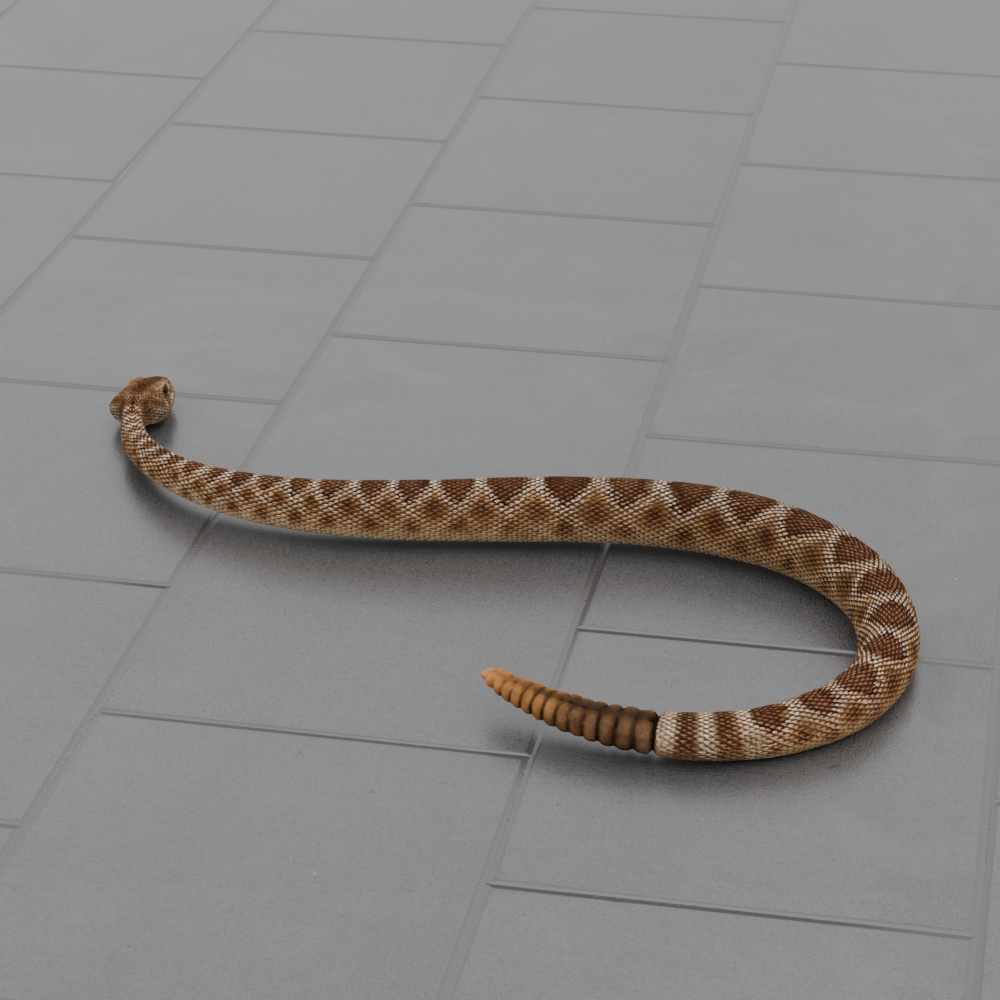}\hspace{2pt}%
\includegraphics[width=0.1625\linewidth]{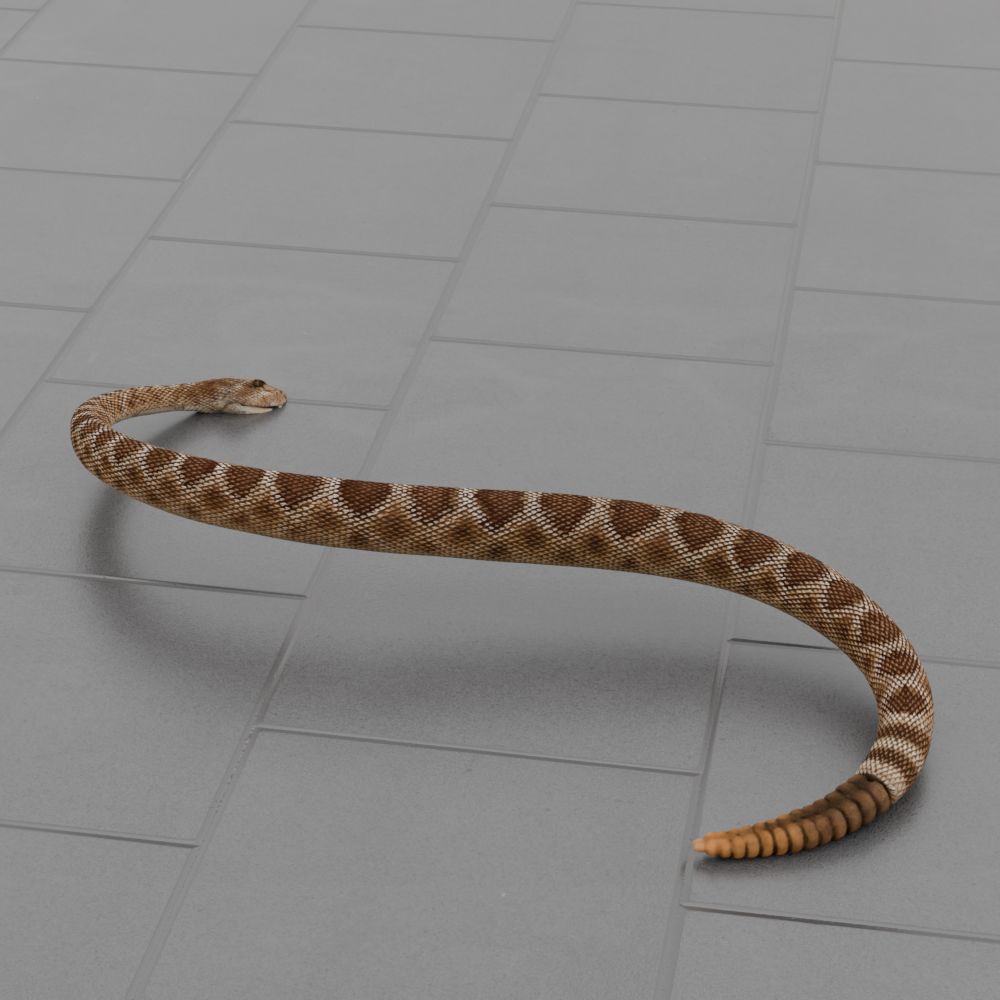}\hspace{2pt}%
\includegraphics[width=0.1625\linewidth]{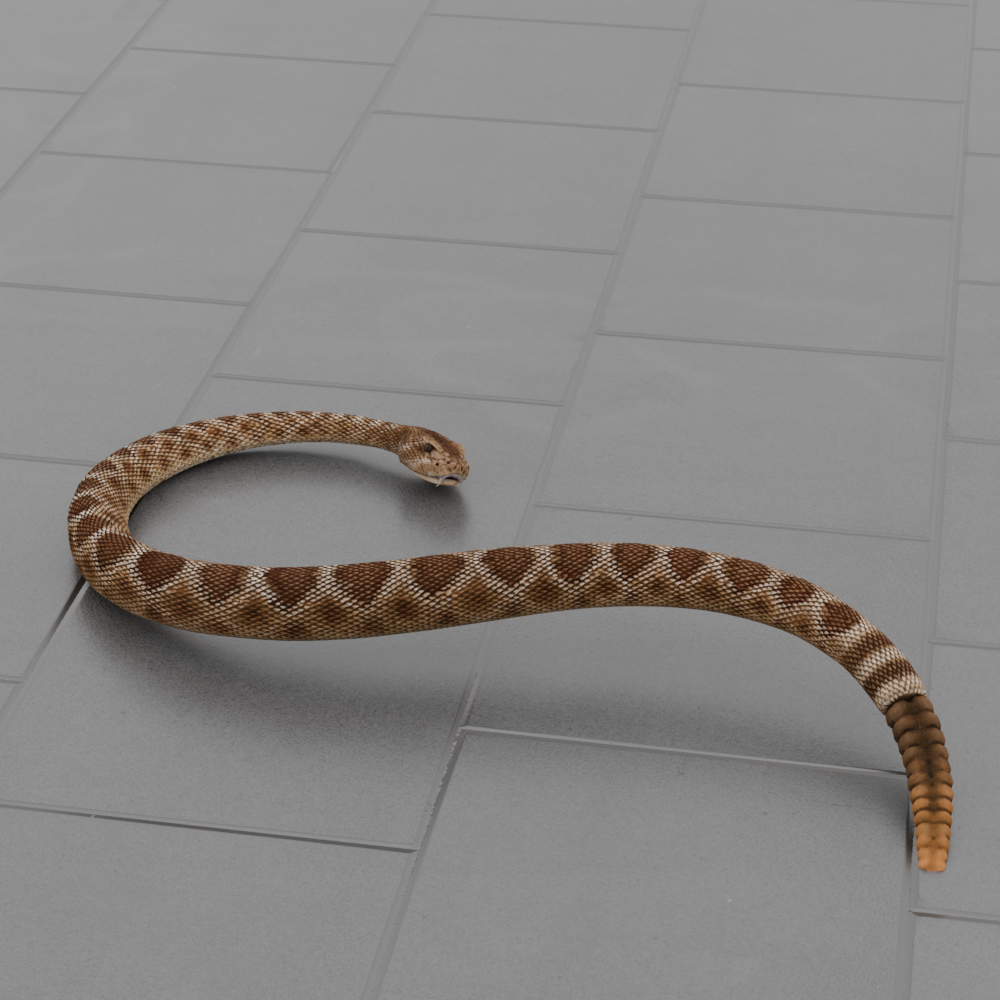}
\end{minipage}

\caption{Temporal evolution of motion using a Figures algorithm. The motion exhibits irregular, jittery body deformation, with non-uniform wave propagation along the body and weaker coordination between segments, resulting in a less stable, less consistent locomotion pattern. The friction ratios used are $\mu_n/\mu_f = 1$ and $\mu_b/\mu_f = 5$. And there is no explicit bounds on $A_{ij}, B_{ij}, C_{ij}, D_{ij}$.}
\label{fig:Genetic}
\end{figure*}


\subsubsection{Effect of Control Authority via Coefficient Bounds (Figures~\ref{fig:Img2} and \ref{fig:Img3})}
The goal of this experiment is to analyze how the allowable range of curvature and torsion coefficients influences the optimization process. The coefficient bounds directly control the expressiveness of the body deformation, effectively determining the size of the admissible control space. By progressively increasing $A_{\max}, B_{\max}, C_{\max}, D_{\max}$, we test whether expanding the control space enables the optimizer to access more efficient or diverse solutions. At the same time, we assess whether overly restrictive bounds limit the ability to achieve effective locomotion. For smaller bounds on $A_{\max}, B_{\max}, C_{\max}, D_{\max}$, the motion is relatively limited, as the body lacks sufficient curvature and torsional flexibility to generate strong propulsive patterns. This leads to weaker deformations and consequently reduced locomotion efficiency. As the bounds are increased, the optimizer can explore a richer set of deformation patterns, leading to more pronounced body undulations and improved locomotion performance. However excessively large bounds can lead to overly aggressive or less smooth deformations, which may reduce physical realism or introduce instability in the motion. Overall, these observations indicate that an intermediate range of coefficient bounds strikes the best balance, providing sufficient control authority for efficient locomotion while maintaining smooth, physically consistent body configurations. In the present study, this balance is achieved at $\max(A_{ij}) = \max(B_{ij}) = 2$ and $\max(C_{ij}) = \max(D_{ij}) = 0.2$.

\subsubsection{Role of Torsional Amplitudes (Figure~\ref{fig:Img3})}
In this experiment, we investigate the influence of torsional amplitudes by increasing the bounds $C_{\max}$ and $D_{\max}$. Unlike earlier cases where torsion is already present at lower amplitudes ($C_{\max}=D_{\max}=0.1$), this study focuses on understanding the effect of allowing stronger torsional deformation. The objective is to examine whether increasing torsional flexibility enhances the optimizer’s ability to explore the solution space and identify improved locomotion strategies. Since torsion governs out-of-plane deformation, higher bounds may allow the system to better utilize three-dimensional configurations. The results indicate higher lift from the ground with larger torsional coefficient values.

\subsubsection{Effect of Mass Distribution (Figures~\ref{fig:Img4} and \ref{fig:Img5})}
While many models assume uniform mass for simplicity, real biological systems often exhibit non-uniform mass distributions. To study this, we introduce a controlled non-uniform mass profile derived from a quadratic function, ensuring only mild variations along the body. The objective is to determine whether such variations affect the optimization outcome and whether they introduce additional complexity in the dynamics.
Figure~\ref{fig:Img4} considers this setup without any bending energy regularization, while Figure~\ref{fig:Img5} includes bending energy. This allows us to isolate the combined effects of mass distribution and bending energy regularization. Non-uniform mass distribution subtly improves the smoothness and stability of the motion.


\subsubsection{Role of Bending and Torsion Energy Regularization (Figures~\ref{fig:Img5} and~\ref{fig:Img6})}
The purpose of this experiment is to evaluate the necessity of incorporating both bending and torsion energy regularization terms in the cost function to obtain physically consistent locomotion. In the absence of such regularization, the optimization process can lead to geometrically unrealistic deformations. To systematically analyze this effect, we consider two complementary experiments. Figure~\ref{fig:Img5} illustrates the influence of increasing the amplitude bounds of the curvature and torsion parameterization. Two cases are shown --- a lower-bound configuration and a higher-bound configuration. Both correspond to an FDSA-optimized gait under anisotropic friction characterized by $\mu_n/\mu_f = 1$ and $\mu_b/\mu_f = 5$, incorporating a non-uniform mass distribution along with the inclusion of both the bending and torsion energy regularization terms in the cost function. It is observed that increasing the allowable parameter range enables richer deformation patterns. Figure~\ref{fig:Img6} presents a direct comparison of locomotion patterns under three different energy configurations: (i) without bending energy, (ii) without torsion energy, and (iii) with both bending and torsion energy terms included. All cases correspond to the same experimental setup, namely $\mu_n/\mu_f = 1$, $\mu_b/\mu_f = 5$, $\max(A_{jk}) = \max(B_{jk}) = 2$, $\max(C_{jk}) = \max(D_{jk}) = 0.2$, non-uniform mass distribution, using the FDSA optimizer. In the absence of bending energy (\emph{Case 1}), the body undergoes self-intersection, and in the absence of torsion energy (\emph{Case 2}), the body exhibits excessive and non-physical twisting, resulting in sharp rotational distortions along the curve. These behaviors arise because there is no penalty for rapid change in curvature and torsion, respectively. Thus, when both bending and torsion energy terms are included (\emph{Case 3}), these artifacts are eliminated. The resulting motion is smooth, geometrically consistent, and physically plausible.

\begin{figure*}[htbp]
\centering

\begin{subfigure}[t]{0.18\textwidth}
    \centering
    \includegraphics[width=\linewidth]{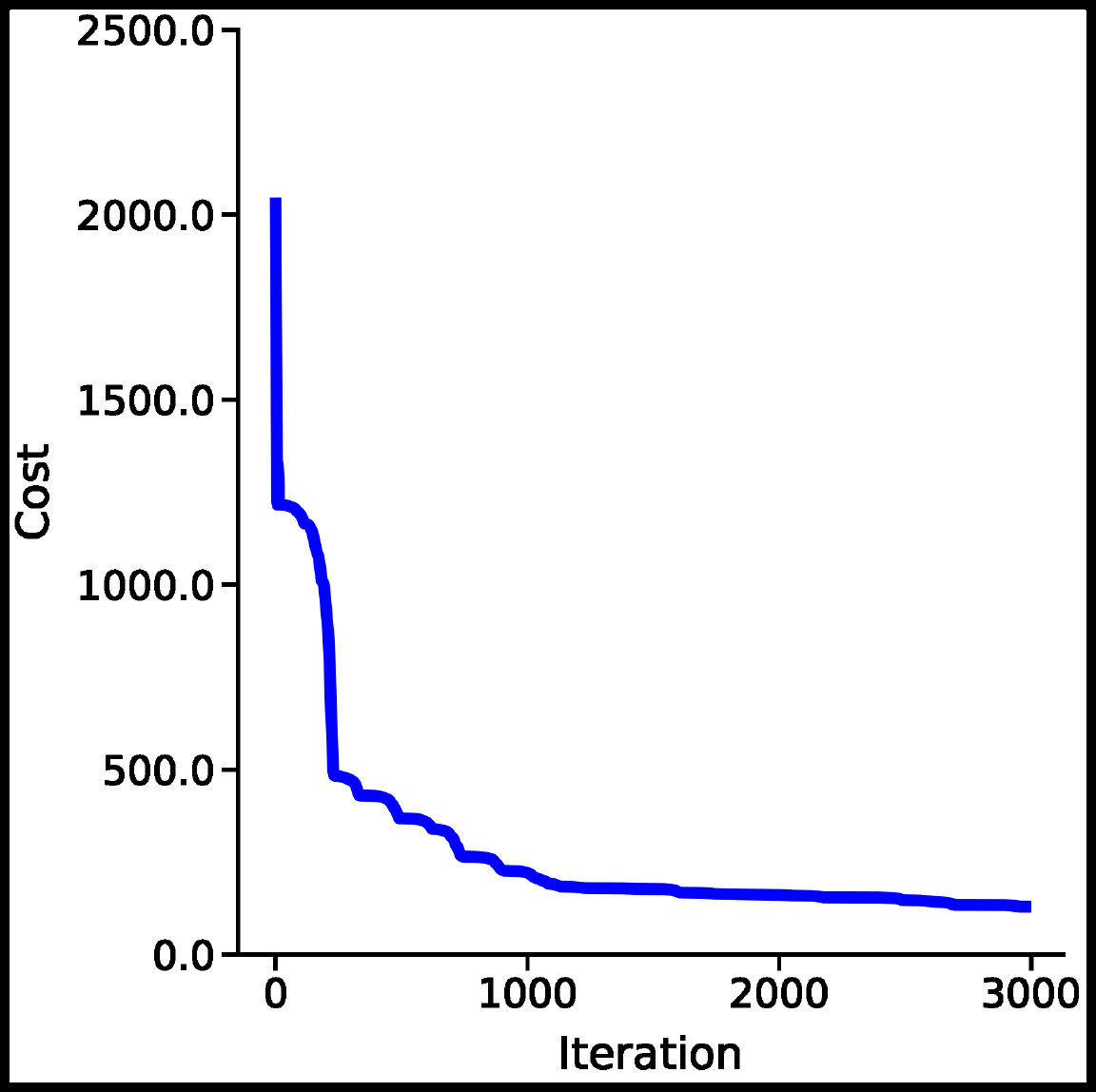}
    \caption{}
\end{subfigure}\hfill
\begin{subfigure}[t]{0.18\textwidth}
    \centering
    \includegraphics[width=\linewidth]{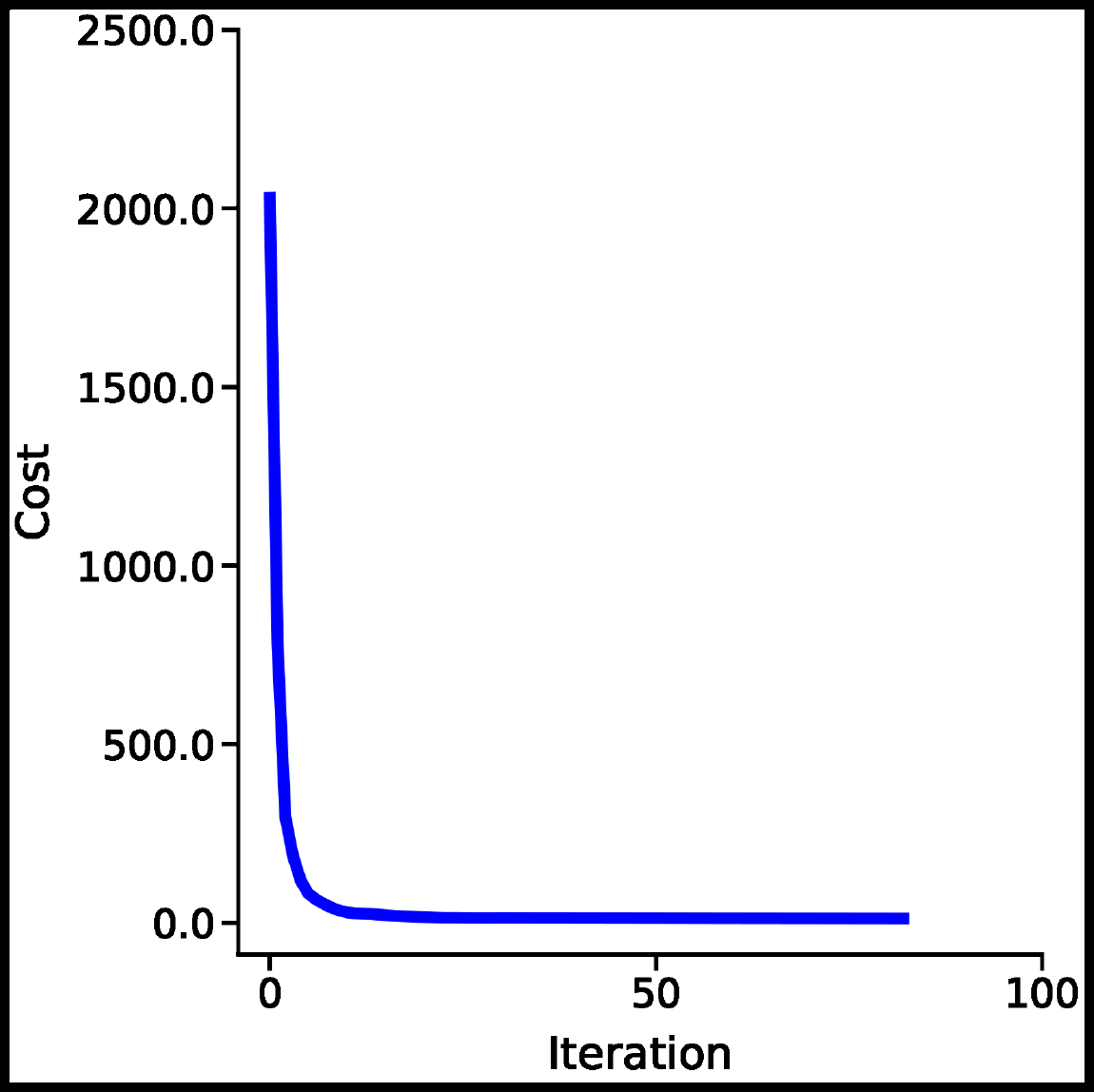}
    \caption{}
\end{subfigure}\hfill
\begin{subfigure}[t]{0.18\textwidth}
    \centering
    \includegraphics[width=\linewidth]{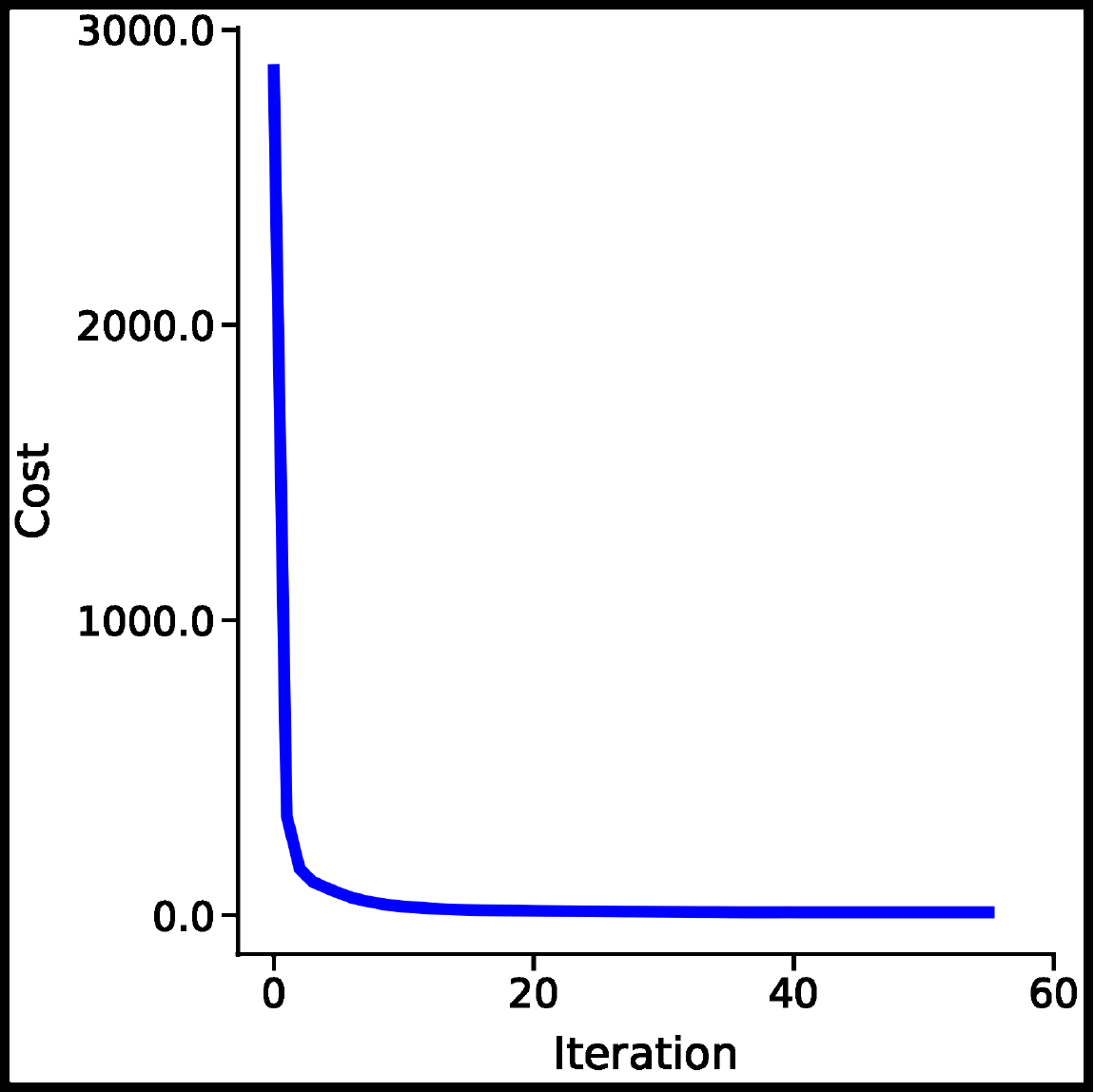}
    \caption{}
\end{subfigure}\hfill
\begin{subfigure}[t]{0.18\textwidth}
    \centering
    \includegraphics[width=\linewidth]{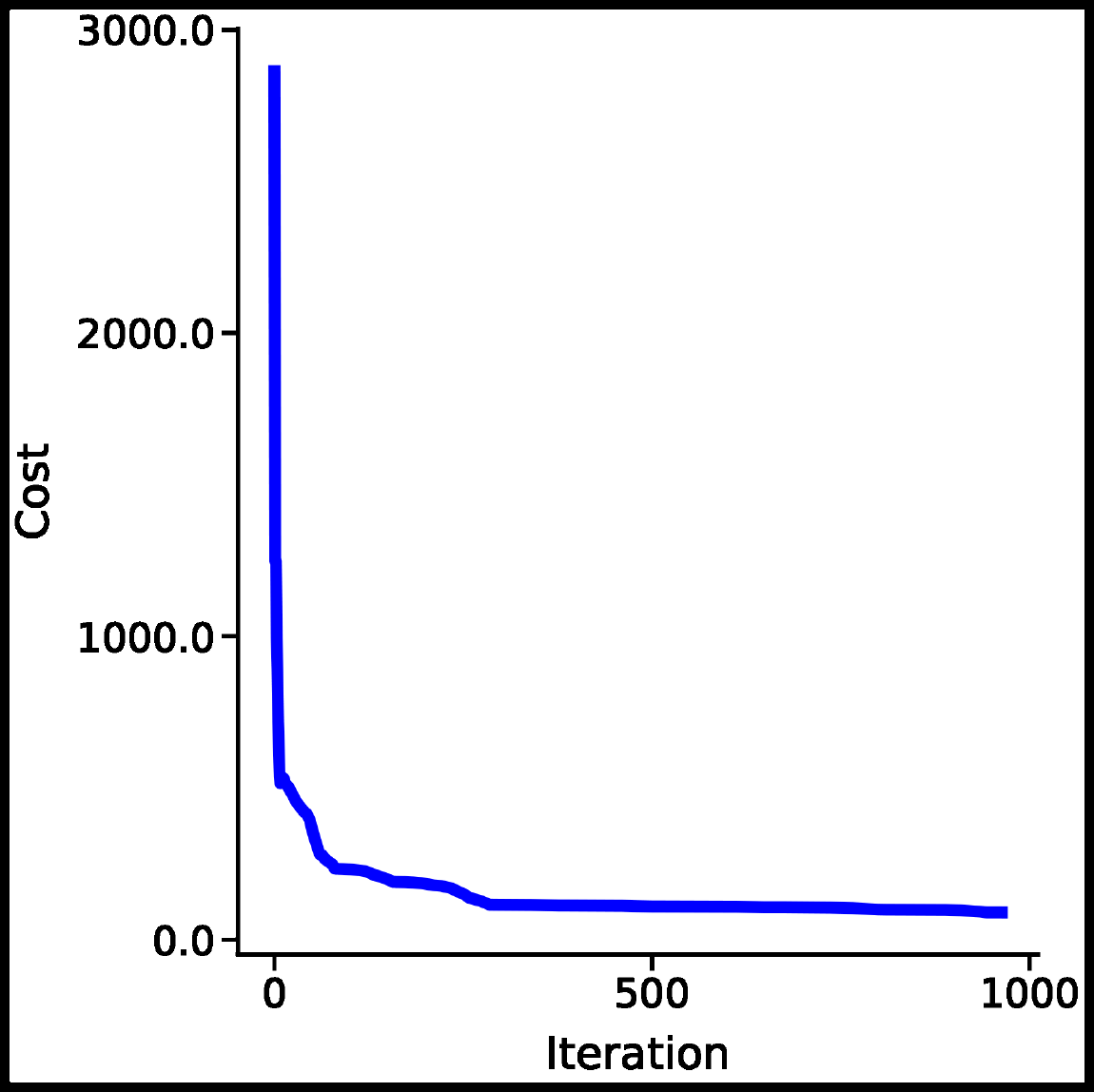}
    \caption{}
\end{subfigure}\hfill
\begin{subfigure}[t]{0.18\textwidth}
    \centering
    \includegraphics[width=\linewidth]{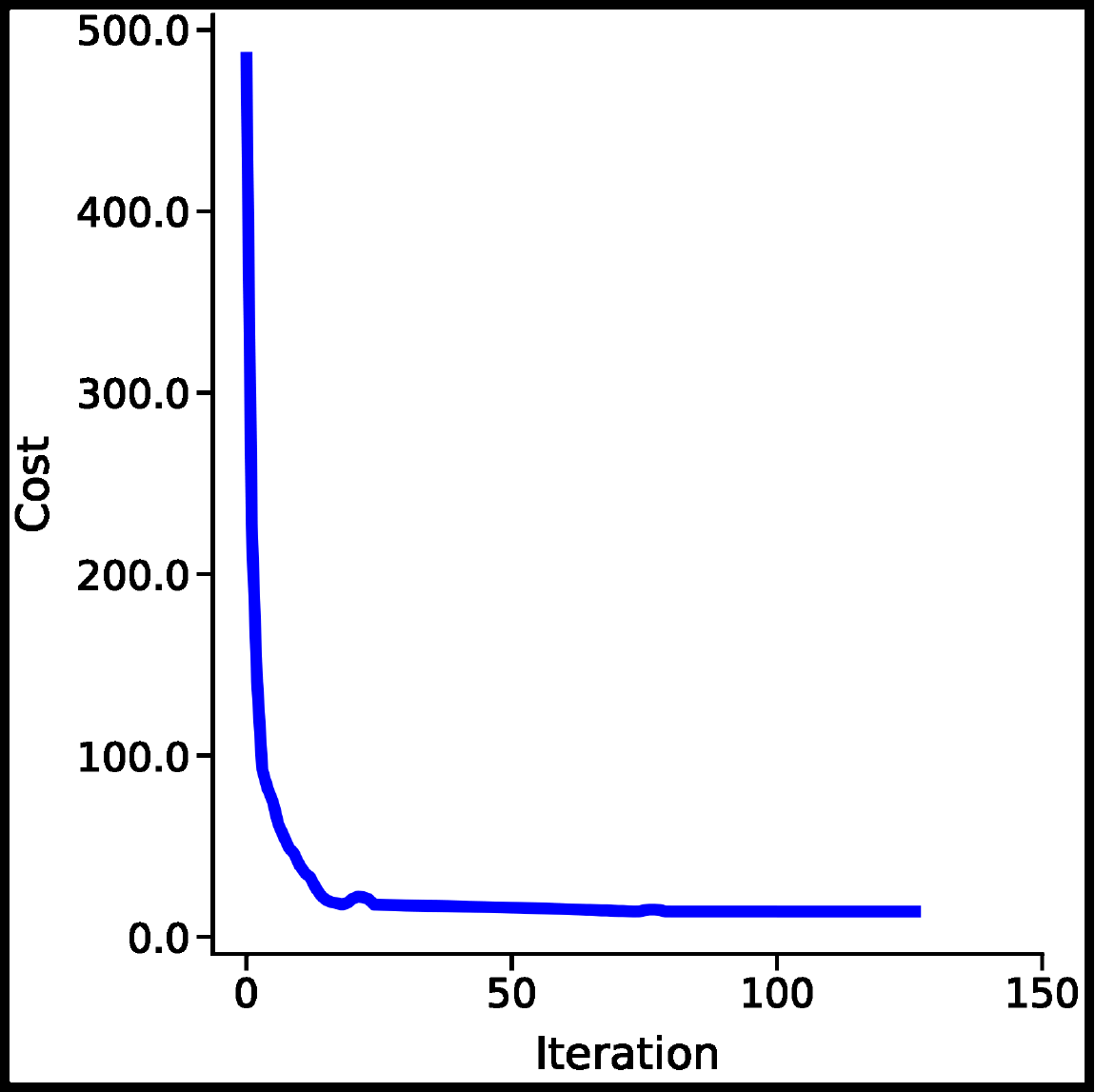}
    \caption{}
\end{subfigure}

\begin{subfigure}[t]{0.18\textwidth}
    \centering
    \includegraphics[width=\linewidth]{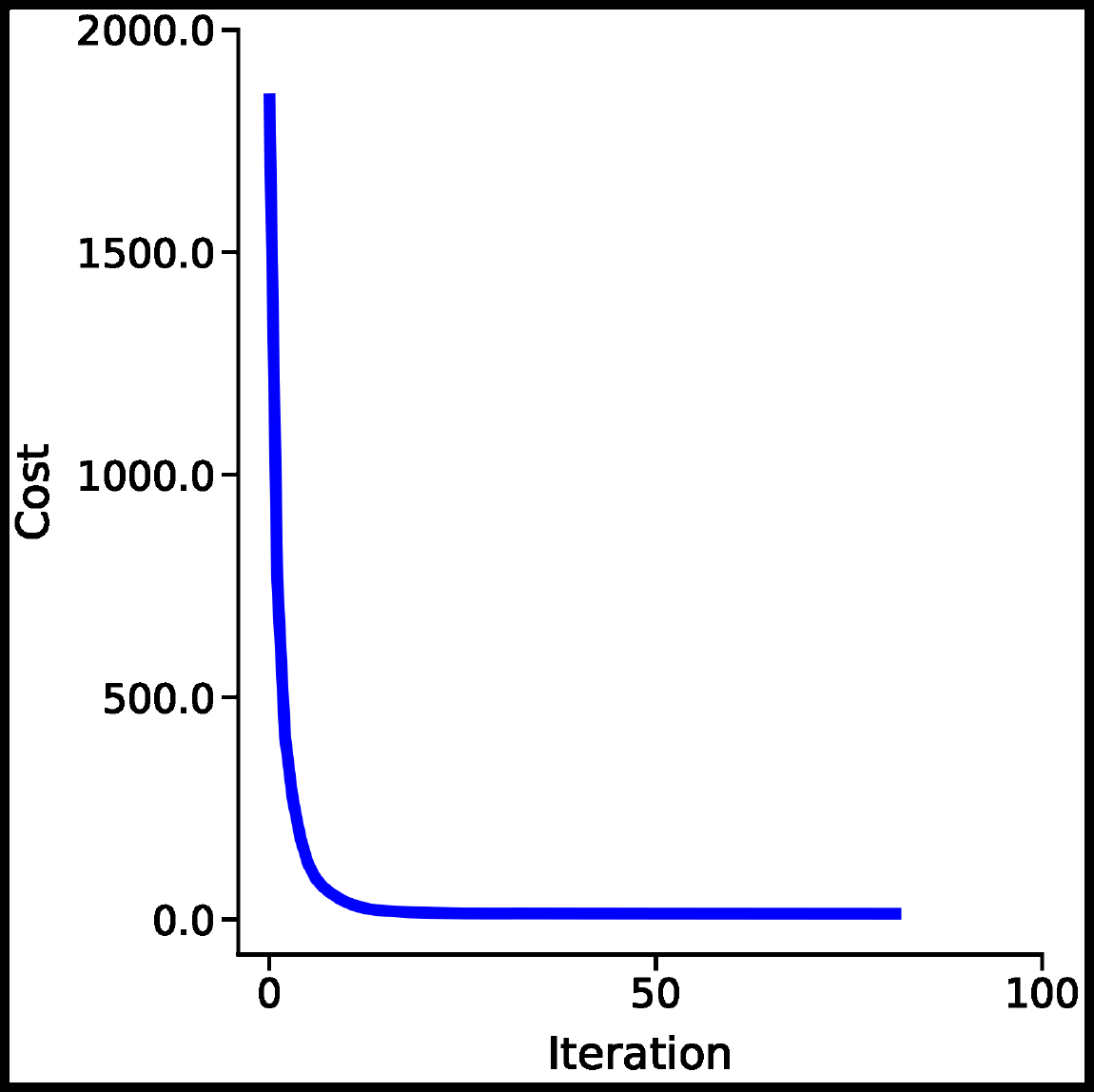}
    \caption{}
\end{subfigure}\hfill
\begin{subfigure}[t]{0.18\textwidth}
    \centering
    \includegraphics[width=\linewidth]{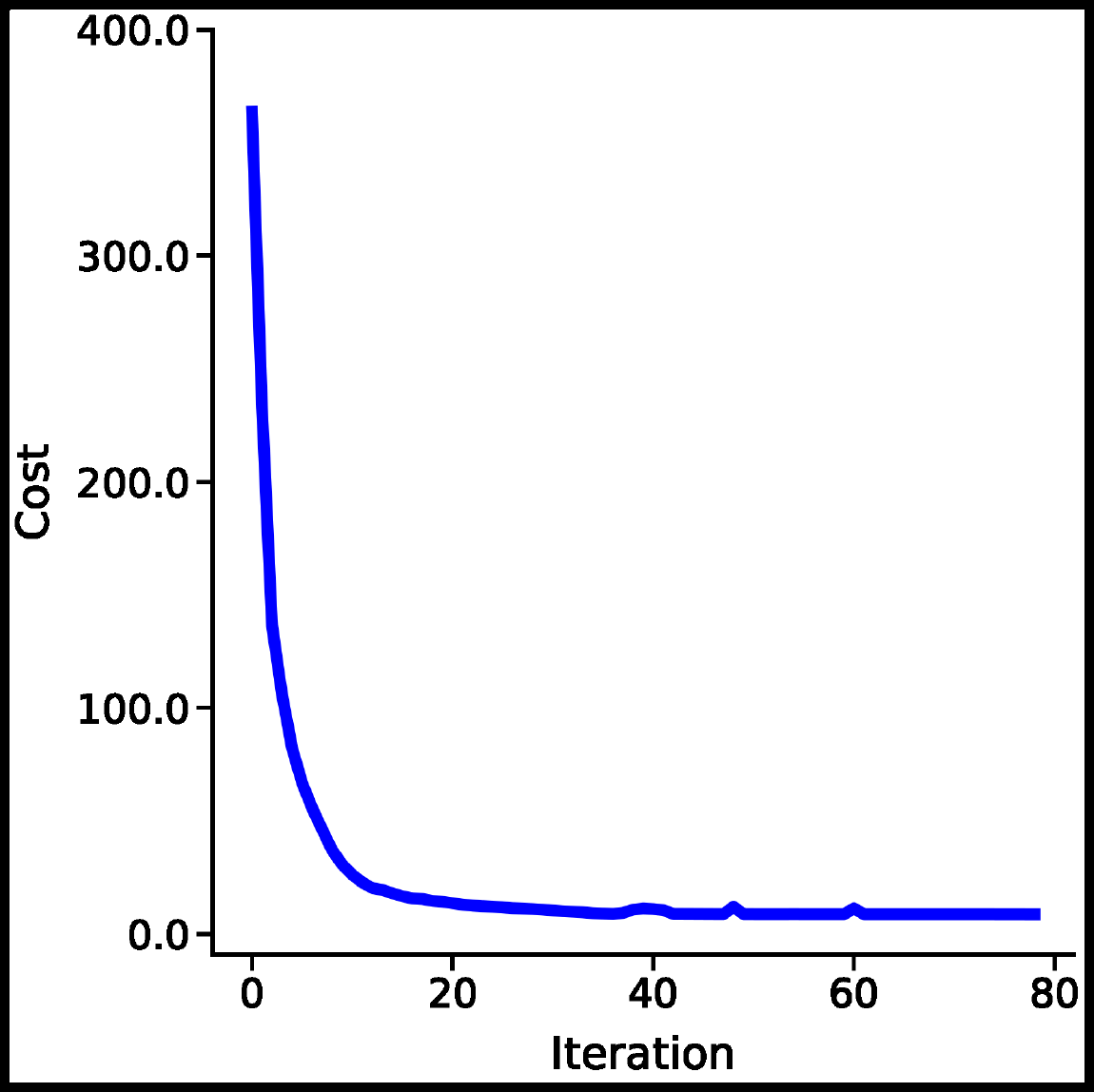}
    \caption{}
\end{subfigure}\hfill
\begin{subfigure}[t]{0.18\textwidth}
    \centering
    \includegraphics[width=\linewidth]{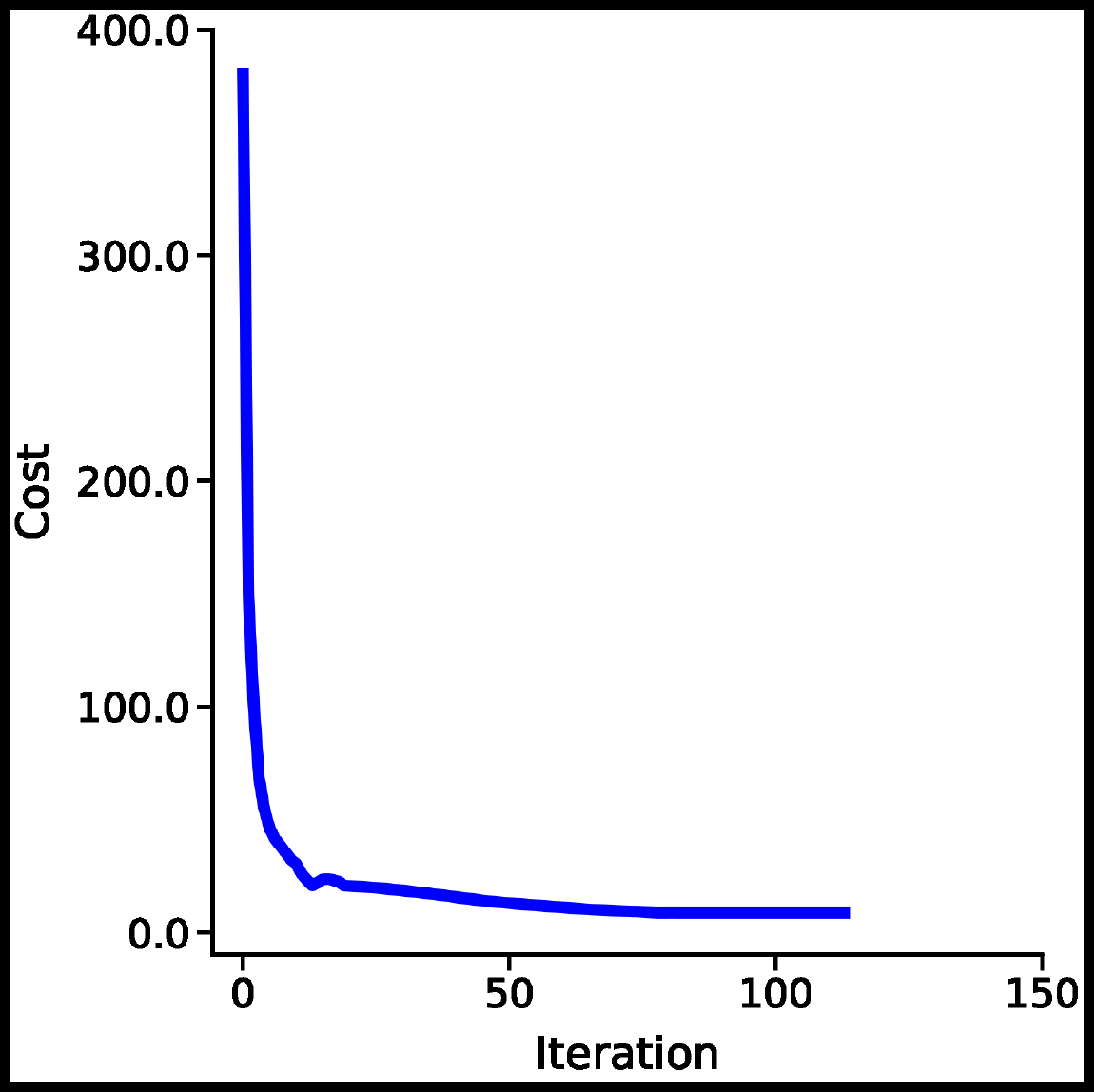}
    \caption{}
\end{subfigure}\hfill
\begin{subfigure}[t]{0.18\textwidth}
    \centering
    \includegraphics[width=\linewidth]{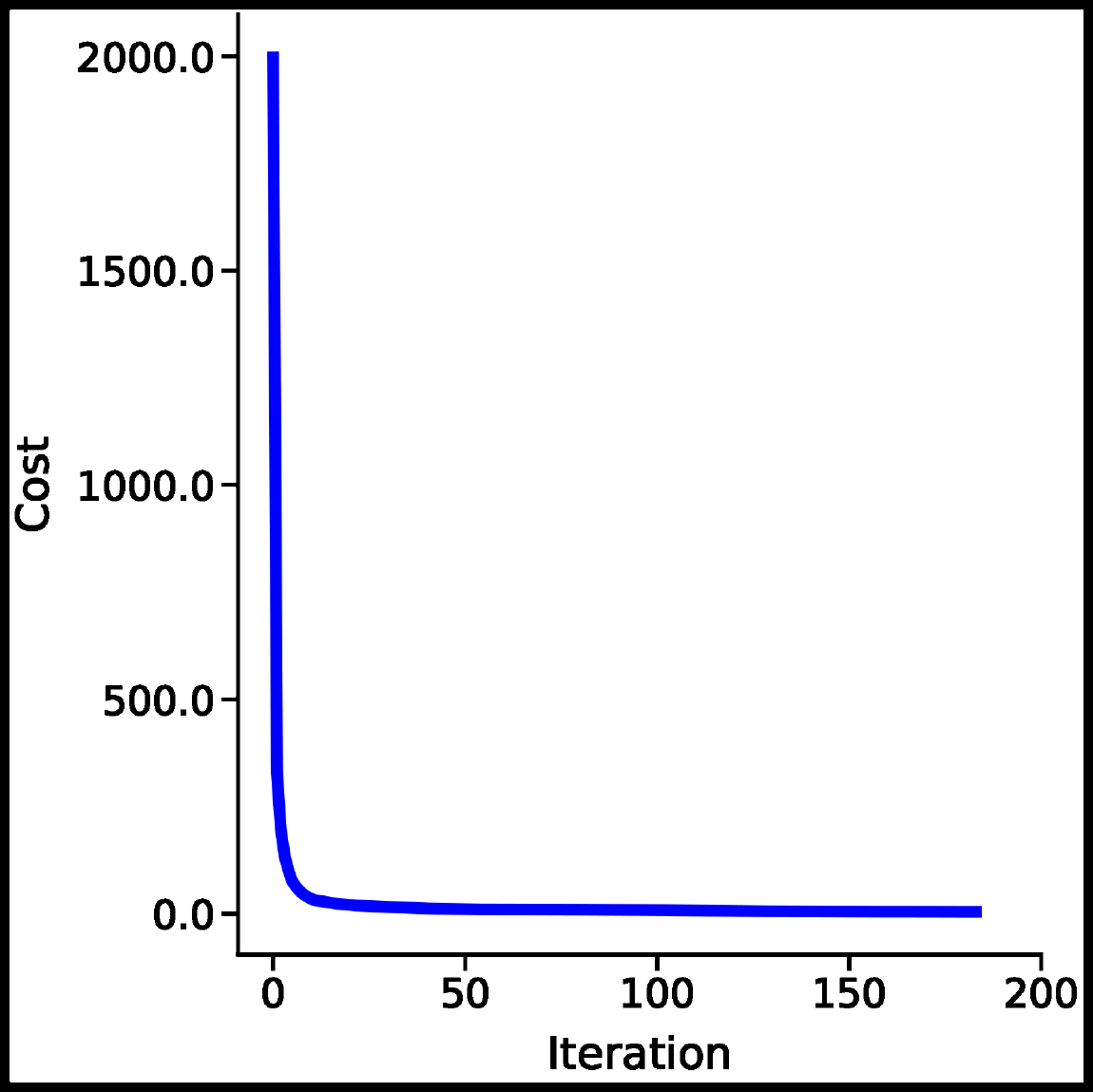}
    \caption{}
\end{subfigure}\hfill
\begin{subfigure}[t]{0.18\textwidth}
    \centering
    \includegraphics[width=\linewidth]{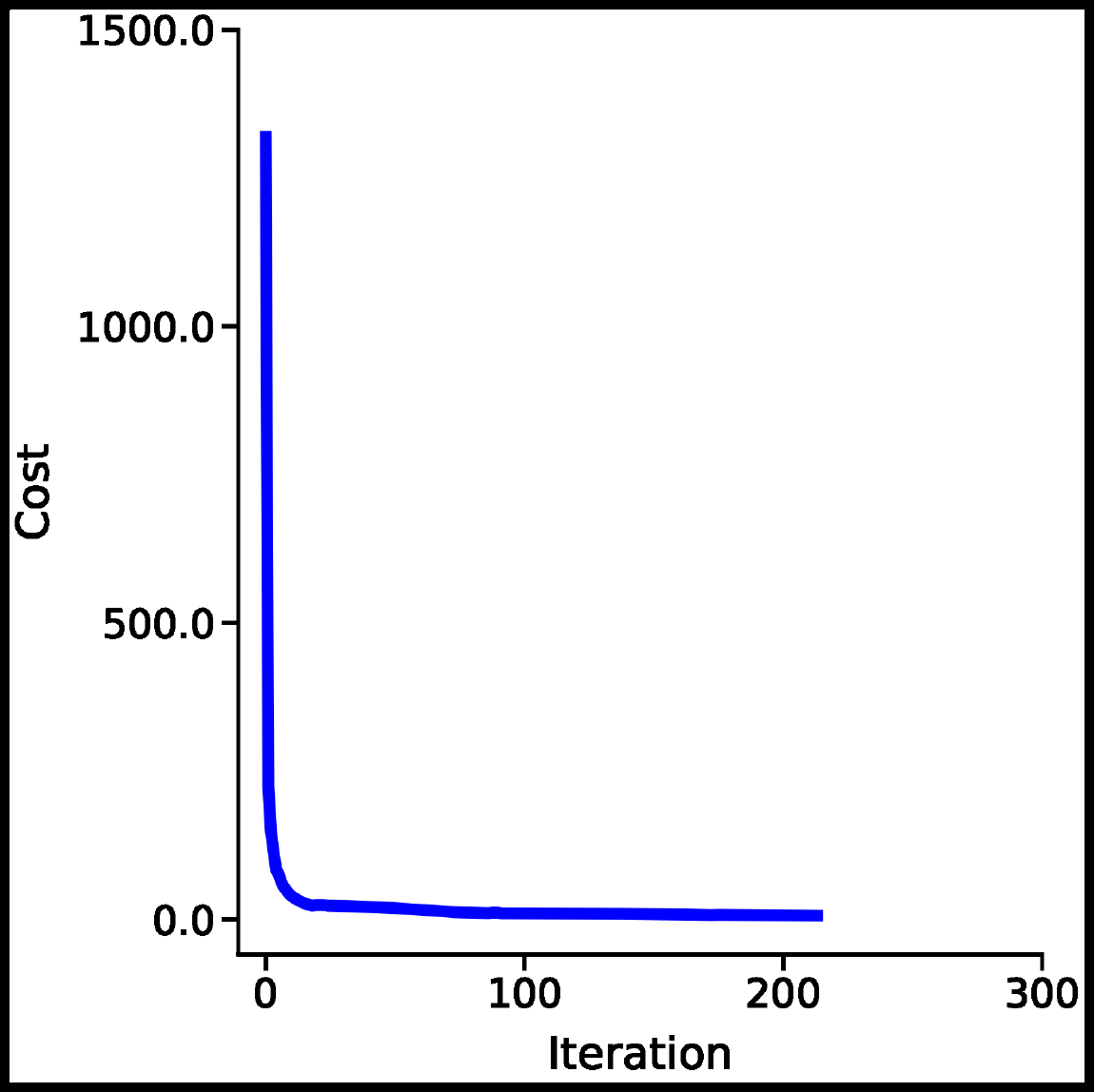}
    \caption{}
\end{subfigure}

\caption{
Convergence behavior of the optimization process for different parameter settings, where the x-axis represents the number of iterations and the y-axis denotes the normalized cost value. In all cases, the friction ratios are fixed. (a) SPSA with $A_{\max}=B_{\max}=1$, $C_{\max}=D_{\max}=0.1$, uniform mass distribution, without bending energy. (b) FDSA with the same parameters as (a). (c) FDSA with $A_{\max}=B_{\max}=2$, $C_{\max}=D_{\max}=0.1$, uniform mass distribution, without bending energy. (d) SPSA with the same parameters as (c). (e) FDSA with $A_{\max}=B_{\max}=3$, $C_{\max}=D_{\max}=0.1$, uniform mass distribution, without bending energy. (f) FDSA with $A_{\max}=B_{\max}=1$, $C_{\max}=D_{\max}=0.1$, non-uniform mass distribution, without bending energy. (g) FDSA with $A_{\max}=B_{\max}=2$, $C_{\max}=D_{\max}=0.2$, non-uniform mass distribution, without bending energy. (h) FDSA with $A_{\max}=B_{\max}=3$, $C_{\max}=D_{\max}=0.3$, non-uniform mass distribution, without bending energy. (i) FDSA with $A_{\max}=B_{\max}=1$, $C_{\max}=D_{\max}=0.1$, non-uniform mass distribution, with bending energy. (j) FDSA with $A_{\max}=B_{\max}=2$, $C_{\max}=D_{\max}=0.2$, non-uniform mass distribution, with bending energy. Across all cases, the cost decreases rapidly during early iterations and stabilizes, indicating convergence, with FDSA exhibiting smoother and more stable convergence compared to SPSA.
}
\label{fig:Graph1}
\end{figure*}

\subsubsection{Comparison of Optimization Methods (Figure~\ref{fig:Img2},Figure~\ref{fig:Img7} and Figure~\ref{fig:Graph1})}
To evaluate the performance of different optimization strategies, we compare FDSA(see Figure~\ref{fig:Img2}) and SPSA(see Figure~\ref{fig:Img7}) under similar experimental conditions. Figure~\ref{fig:Graph1} presents the evolution of the cost function over iterations. We examine how quickly each method reduces the cost and how stable the convergence trajectory is. The comparison indicates that SPSA exhibits faster initial convergence than FDSA, as it achieves a rapid reduction in the cost function during the early iterations. However, this comes at the expense of increased variability, with the convergence trajectory showing noticeable fluctuations due to the stochastic nature of the gradient approximation. In contrast, FDSA exhibits more stable, smoother convergence behavior, with a more consistent reduction in cost across iterations. Although FDSA may converge more slowly in the initial stages, it tends to achieve a lower final cost and provides more reliable convergence. Overall, SPSA is advantageous in terms of convergence speed, while FDSA offers better stability and solution quality, making it more suitable for obtaining physically consistent and optimal locomotion patterns.

\subsubsection{Side-winding Motion on Inclined Plane}
The simulation on a $10^\circ$ inclined plane (see Figure~\ref{fig:Img7-1}) shows that the proposed framework can generate stable and effective motion even when moving against gravity. The optimized gait clearly exhibits a side-winding motion, where different parts of the body alternately lift off and come into contact with the ground. This pattern allows some segments of the body to maintain traction while others move forward, thereby reducing backward slip. Using bounded curvature and torsion coefficients with $\max(A_{ij}) = \max(B_{ij}) = 2 \;\&\; \max(C_{ij}) = \max(D_{ij}) = 0.1$, the model produces smooth 3D deformations that enable uphill motion. The simulations are performed under anisotropic friction conditions with $\mu_n/\mu_f = 1 \;\&\; \mu_b/\mu_f = 5$. Locomotive behavior remains stable and smooth with time.

\subsubsection{Motion in Fluid with High Bending \& Torsion Energy}
The motion in a fluid-like setting is approximated by applying the same curvature-torsion-based framework to an eel-shaped body (see Figure~\ref{fig:Eel}). In this experiment, higher-order bending and torsion energy terms are included to promote smoother, more physically consistent deformations, which are typical of swimming organisms. The parameters are bounded as $\max(A_{ij}) = \max(B_{ij}) = 1 \;\&\; \max(C_{ij}) = \max(D_{ij}) = 0.1$, with a uniform mass distribution. Anisotropic friction ratios, $\mu_n/\mu_f = 1 \;\&\; \mu_b/\mu_f = 5$, are used to be consistent with the ground-based simulations. The resulting motion shows eel-like undulatory behavior, in which wave-like patterns travel along the body, generating forward motion. The simulation qualitatively captures the key features of eel locomotion. The inclusion of higher bending and torsion energy helps produce smoother body shapes and avoids unrealistic deformations. Thus, the proposed framework is flexible enough to approximate fluid-like motion, even though it is primarily designed for ground-based locomotion.

\begin{figure}[h!tb]
    \centering
    \includegraphics[width=0.485\columnwidth]{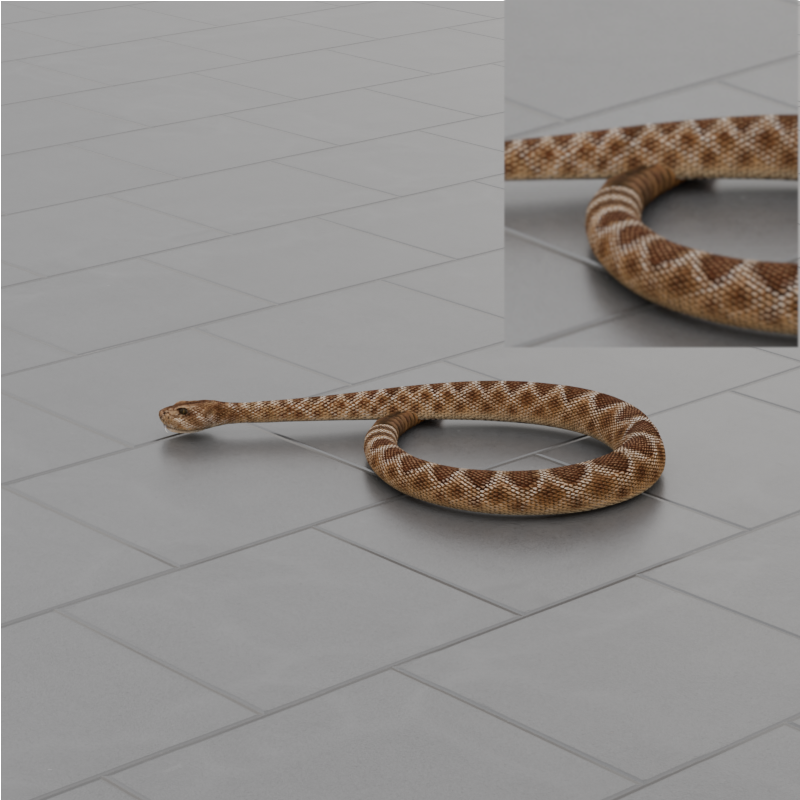}~%
    \includegraphics[width=0.485\columnwidth]{Figures/m4.png}
    \caption{\textit{Bending Energy}: The image illustrates that, in the absence of bending energy, the body undergoes self-intersection (left), but inclusion of bending energy removes it (right). The inset provides a zoomed view of the intersection region, highlighting the geometric overlap.}
    \label{fig:comparison_zoom}
\end{figure}

\subsubsection{Quantitative Evaluation (Tables~\ref{tab:Table1} and \ref{tab:Table2})}
The experiments are further quantified through Tables~\ref{tab:Table1} and~\ref{tab:Table2}, which summarize the convergence characteristics and locomotion performance metrics, respectively. These tables are designed to provide a systematic comparison across all configurations.
Table~\ref{tab:Table1} focuses on optimization efficiency, while Table~\ref{tab:Table2} evaluates the effectiveness of the resulting locomotion in terms of body displacement.

\subsection{Comparison Study}
\subsubsection{Study of Locomotion With and Without Bending and Torsional Energy Terms in cost function}
To further emphasize the role of bending energy regularization, we present a focused comparison based on the configurations shown in Figure~\ref{fig:Img6}. All cases correspond to the same experimental setup, namely $\mu_n/\mu_f = 1$, $\mu_b/\mu_f = 5$, $\max(A_{jk}) = \max(B_{jk}) = 2$, $\max(C_{jk}) = \max(D_{jk}) = 0.2$, non-uniform mass distribution, using the FDSA optimizer. The difference between \emph{Case 1} and \emph{Case 3} is the inclusion of the bending energy term in the cost function.
As observed in Figure~\ref{fig:Img6} (top row), the absence of the bending energy term leads to configurations where the body undergoes self-intersection. This occurs because the optimization is driven solely by frictional work, with no penalty for excessive curvature. In contrast, when the bending energy term is included, such configurations are suppressed, and the resulting motion remains geometrically consistent as observed in Figure~\ref{fig:Img6} (bottom row). To highlight this distinction more clearly, Figure~\ref{fig:comparison_zoom} presents a representative configuration along with a zoomed-in view of the region where self-intersection occurs. 

\begin{figure}[h!tb]
    \centering
    \includegraphics[width=0.485\columnwidth]{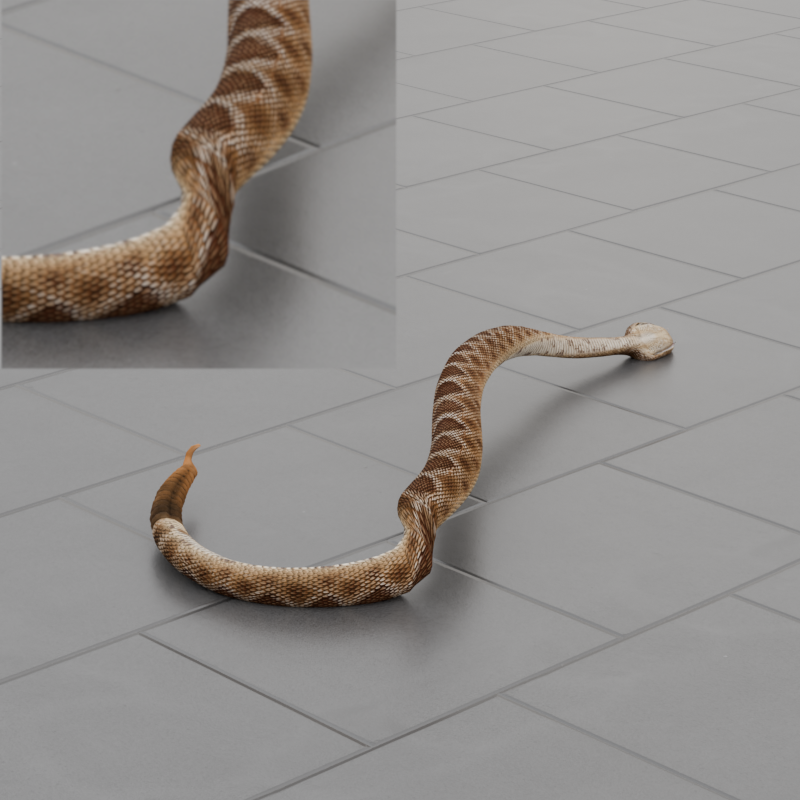}~%
    \includegraphics[width=0.485\columnwidth]{Figures/m5.png}
    \caption{\textit{Torsion Energy}: The image illustrates that, in the absence of torsion energy, the body undergoes a twist (left), but inclusion of torsion energy removes it (right). The inset provides a zoomed view of the twisted region, highlighting the geometric distortion.}
    \label{fig:comparison_zoom_twist}
\end{figure}

In addition to bending energy, torsional energy regularization is equally important for ensuring physically meaningful locomotion. As shown in Figure~\ref{fig:Img6} (middle row), in the absence of torsion energy, the body exhibits excessive, non-physical twisting, resulting in sharp rotational distortions along the curve. This behavior arises because the optimization process is driven primarily by frictional work, without any constraint penalizing rapid variations in torsion. Consequently, the solution may exploit large twisting deformations to reduce contact forces, resulting in geometrically inconsistent configurations. Thus, when the torsion energy term is included, such abrupt twisting is effectively penalized, promoting smooth and physically realistic rotational deformations as observed in Figure~\ref{fig:Img6} (bottom row). To further illustrate this effect, Figure~\ref{fig:comparison_zoom_twist} presents both configurations along with a zoomed-in view of the twisted region, clearly demonstrating how torsion regularization eliminates geometric distortion.

\subsubsection{Genetic Algorithm vs Our Method}
The motion obtained using the genetic algorithm (see Figure~\ref{fig:Genetic})~\cite{Alben2022} appears less smooth, with noticeable irregularities while the body bends and moves over time. The wave pattern along the body is not very consistent, and the curvature does not propagate evenly from head to tail. Because of this, different parts of the snake seem to move slightly out of sync, making the motion look somewhat jerky and non-smooth. In some regions, the body deformation appears uneven, which gives the impression of small discontinuities in the motion. Moreover, the movements generated with the genetic algorithm frequently display unrealistic bending and twisting as depicted in Figure~\ref{fig:Genetic}. In contrast, the motion generated using FDSA and SPSA (see Figure~\ref{fig:Img2} and Figure~\ref{fig:Img7}) is much smoother and more structured. The wave-like deformation propagates uniformly along the body, and the shape changes gradually without abrupt shifts. The body segments move in a well-coordinated way, with better synchronization among the snake's parts. This leads to a continuous, flowing motion that appears more stable over time.


\begin{table*}[h!tb]
\centering
\footnotesize
\resizebox{\textwidth}{!}{%
\begin{tabular}{l l c c c c c c c c c c}
\toprule
\multicolumn{12}{c}{\textbf{Convergence Characteristics of Optimization Methods}} \\
\midrule
\textit{\text{Exp.}}
& \textit{\text{Opt.}}
& \textit{\text{$\mu_n/\mu_f$}}
& \textit{\text{$\mu_b/\mu_f$}}
& \textit{\text{$A_{\max}$}}
& \textit{\text{$B_{\max}$}}
& \textit{\text{$C_{\max}$}}
& \textit{\text{$D_{\max}$}}
& \textit{\text{Mass}}
& \textit{\text{B.E.}}
& \textit{\text{$N_c$}}
& \textit{\text{$\bar P_{\mathrm{sat}}$} (norm.)}\\
\midrule

Fig.~\ref{fig:Img2} (T) & FDSA & 1 & 5 & 1 & 1 & 0.1 & 0.1 & U & No & 31 & $1.42$ \\
Fig.~\ref{fig:Img2} (M) & FDSA & 1 & 5 & 2 & 2 & 0.1 & 0.1 & U & No & 36 & $1.06$ \\
Fig.~\ref{fig:Img2} (B) & FDSA & 1 & 5 & 3 & 3 & 0.1 & 0.1 & U & No & 24 & $1.80$ \\
Fig.~\ref{fig:Img3} (M) & FDSA & 1 & 5 & 2 & 2 & 0.2 & 0.2 & U & No & 43 & $0.927$ \\
Fig.~\ref{fig:Img3} (B) & FDSA & 1 & 5 & 3 & 3 & 0.3 & 0.3 & U & No & 21 & $1.78$ \\
Fig.~\ref{fig:Img4} (T) & FDSA & 1 & 5 & 1 & 1 & 0.1 & 0.1 & NU & No & 32 & $1.40$ \\
Fig.~\ref{fig:Img4} (B) & FDSA & 1 & 5 & 2 & 2 & 0.2 & 0.2 & NU & No & 42 & $8.90$ \\
Fig.~\ref{fig:Img5} (T) & FDSA & 1 & 5 & 1 & 1 & 0.1 & 0.1 & NU & Yes & 51 & $1.07$ \\
Fig.~\ref{fig:Img5} (B) & FDSA & 1 & 5 & 2 & 2 & 0.2 & 0.2 & NU & Yes & 24 & $2.33$ \\
Fig.~\ref{fig:Img7} (T) & SPSA & 1 & 5 & 1 & 1 & 0.1 & 0.1 & U & No & 2925 & $12.9$ \\
Fig.~\ref{fig:Img7} (M) & SPSA & 1 & 5 & 2 & 2 & 0.1 & 0.1 & U & No & 940 & $9.06 $ \\
Fig.~\ref{fig:Img7} (B) & SPSA & 1 & 5 & 3 & 3 & 0.3 & 0.3 & NU & No & 960 & $8.99$ \\
\bottomrule
\end{tabular}%
}
\caption{
Convergence characteristics of the optimization methods across different experimental configurations. 
Each experiment (``Exp.'') corresponds to a specific set of friction ratios, coefficient bounds, mass distribution, and bending energy inclusion, where T, M, and B denote Top, Middle, and Bottom figures, respectively. 
``Opt.'' denotes the optimizer used (FDSA or SPSA). 
The ratios $\mu_n/\mu_f$ and $\mu_b/\mu_f$ represent the normal and backward friction relative to forward friction. 
$A_{\max}$, $B_{\max}$, $C_{\max}$, and $D_{\max}$ denote bounds on curvature and torsion coefficients. 
``Mass'' indicates the mass distribution, where U denotes uniform and NU denotes non-uniform mass distribution, and ``B.E.'' specifies the inclusion of bending energy. 
The $N_c$ is the corresponding number of iterations required to reach a steady state, and $\bar P_{\mathrm{sat}}$ is the cost when it reaches the steady state.
}
\label{tab:Table1}
\end{table*}

\begin{table*}[h!tb]
\footnotesize
\centering
\resizebox{\textwidth}{!}{%
\begin{tabular}{l l c c c c c c c c c}
\toprule
\multicolumn{11}{c}{\textbf{Locomotion Performance Metrics}} \\
\midrule
\textit{\text{Exp.}}
& \textit{\text{Opt.}}
& \textit{\text{$\mu_n/\mu_f$}}
& \textit{\text{$\mu_b/\mu_f$}}
& \textit{\text{$A_{\max}$}}
& \textit{\text{$B_{\max}$}}
& \textit{\text{$C_{\max}$}}
& \textit{\text{$D_{\max}$}}
& \textit{\text{Mass}}
& \textit{\text{B.E.}}
& \textit{\text{Disp.}}\\
\midrule

Fig.~\ref{fig:Img2} (T) & FDSA & 1 & 5 & 1 & 1 & 0.1 & 0.1 & U & No & 1.74 \\
Fig.~\ref{fig:Img2} (M) & FDSA & 1 & 5 & 2 & 2 & 0.1 & 0.1 & U & No & 3.43 \\
Fig.~\ref{fig:Img2} (B) & FDSA & 1 & 5 & 3 & 3 & 0.1 & 0.1 & U & No & 3.20 \\
Fig.~\ref{fig:Img3} (M) & FDSA & 1 & 5 & 2 & 2 & 0.2 & 0.2 & U & No & 3.70 \\
Fig.~\ref{fig:Img3} (B) & FDSA & 1 & 5 & 3 & 3 & 0.3 & 0.3 & U & No & 3.91 \\
Fig.~\ref{fig:Img4} (T) & FDSA & 1 & 5 & 1 & 1 & 0.1 & 0.1 & NU & No & 1.97 \\
Fig.~\ref{fig:Img4} (B) & FDSA & 1 & 5 & 2 & 2 & 0.2 & 0.2 & NU & No & 4.08 \\
Fig.~\ref{fig:Img5} (T) & FDSA & 1 & 5 & 1 & 1 & 0.1 & 0.1 & NU & Yes & 1.15 \\
Fig.~\ref{fig:Img5} (B) & FDSA & 1 & 5 & 2 & 2 & 0.2 & 0.2 & NU & Yes & 3.07 \\
Fig.~\ref{fig:Img7} (T) & SPSA & 1 & 5 & 1 & 1 & 0.1 & 0.1 & U & No & 0.42 \\
Fig.~\ref{fig:Img7} (M) & SPSA & 1 & 5 & 2 & 2 & 0.1 & 0.1 & U & No & 0.93 \\
Fig.~\ref{fig:Img7} (B) & SPSA & 1 & 5 & 3 & 3 & 0.3 & 0.3 & NU & No & 4.08 \\
\bottomrule
\end{tabular}%
}
\caption{
Locomotion performance metrics across different experimental configurations. 
Each experiment (``Exp.'') corresponds to a specific set of friction ratios, coefficient bounds, mass distribution, and bending energy inclusion, where T, M, and B denote Top, Middle, and Bottom figures, respectively. 
``Opt.'' indicates the optimizer used (FDSA or SPSA). 
The ratios $\mu_n/\mu_f$ and $\mu_b/\mu_f$ represent the normal and backward friction relative to forward friction. 
$A_{\max}$, $B_{\max}$, $C_{\max}$, and $D_{\max}$ denote bounds on curvature and torsion coefficients. 
``Mass'' indicates the mass distribution, where U denotes uniform and NU denotes non-uniform mass distribution, and ``B.E.'' specifies the inclusion of bending energy. 
``Disp.'' represents the net center-of-mass displacement.
These metrics collectively capture the mobility, directional stability, and energetic performance of the optimized locomotion.
}
\label{tab:Table2}
\end{table*}


\section{Conclusion and Limitations}\label{sec:conclusion}
This work presents a unified and physically consistent framework for modeling and optimizing three-dimensional, limbless soft-body locomotion using a differential-geometric formulation. By representing the limbless soft body as a space curve governed by curvature and torsion, the proposed approach provides a compact yet expressive description of complex body deformations. This intrinsic representation, combined with a force-balanced dynamic model and anisotropic frictional interactions, enables the emergence of realistic locomotion patterns without prescribing motion a priori.

The results demonstrate that efficient locomotion can be systematically discovered by optimizing curvature-torsion parameter spaces. The inclusion of torsion introduces critical out-of-plane deformation, allowing the system to modulate ground contact and reduce frictional losses. This highlights the importance of three-dimensional modeling in capturing realistic locomotion strategies that are not achievable in planar formulations. Furthermore, incorporating bending and torsional energy regularization ensures physically meaningful, smooth configurations, preventing non-physical artifacts such as excessive twisting or self-intersection. 

The optimization framework, based on stochastic approximation methods (FDSA and SPSA) with momentum and adaptive learning, proves effective at navigating the high-dimensional, non-linear search space. It consistently converges to stable, periodic gaits that balance frictional dissipation and internal deformation costs. The formulation also demonstrates flexibility, accommodating variations in frictional anisotropy, mass distribution, and control bounds, making it a general tool for studying soft-body locomotion across different physical settings.

The proposed framework offers a scalable and extensible foundation for future research in soft robotics, bio-inspired locomotion, and control of deformable systems. Future directions may include incorporating more complex environmental interactions, learning-based control strategies, and real-world robotic implementations to further enhance the applicability of the proposed method. The proposed method has some limitations in its current form. In the underwater eel simulation, while the body is fully submerged, its motion is influenced by complex hydrodynamic effects such as drag and lift. Those effects are not included in the present model. Moreover, although the inclusion of bending and torsion energies eliminates unrealistic configurations such as self-intersection and twisting, it fails in some extreme cases. These could be interesting potential avenues for future research. 


\bibliographystyle{ACM-Reference-Format}
\bibliography{bibliography}

\end{document}